\documentclass[12pt]{article}
\usepackage{graphicx}
\usepackage{latexsym}
\makeatletter
\@addtoreset{equation}{section}
\makeatother

\def\cone{R^\mu{}_{\nu\rho\sigma}}
\def\gcone{\hat R^\mu{}_{\nu\rho\sigma}}

   \let\d=\delta

   \let\U=\Upsilon  
\let\C=\Chi 
\let\la=\label  
  
 \def\bd{\begin{document}} \def\ed{\end{document}}
\def\ds{\documentstyle} \let\fr=\frac \let\bl=\bigl \let\br=\bigr
\let\Br=\Bigr \let\Bl=\Bigl
\let\bm=\bibitem
\let\na=\nabla
\let\pa=\partial \let\ov=\overline
\newcommand{\be}{\begin{equation}}
\newcommand{\ee}{\end{equation}}
\def\ba{\begin{array}}
\def\ea{\end{array}}
\newcommand{\ho}[1]{$\, ^{#1}$}
\newcommand{\hoch}[1]{$\, ^{#1}$}
\newcommand{\bea}{\begin{eqnarray}}
\newcommand{\eea}{\end{eqnarray}}
\newcommand{\ra}{\rightarrow}
\newcommand{\lra}{\longrightarrow}
\newcommand{\Lra}{\Leftrightarrow}
\newcommand{\ap}{\alpha^\prime}
\newcommand{\bp}{\tilde \beta^\prime}
\newcommand{\tr}{{\rm tr} }
\newcommand{\Tr}{{\rm Tr} }
\newcommand{\NP}{Nucl. Phys. }
\newcommand{\A}{A}
\newcommand{\R}{{\rm R}}
\newcommand{\C}{{\rm C}}
\newcommand{\SO}{{\rm SO}}
\newcommand{\SL}{{\rm SL}}
\newcommand{\Sp}{{\rm Sp}}
\newcommand{\ISO}{{\rm ISO}}
\newcommand{\SU}{{\rm SU}}
\newcommand{\E}{{\rm E}}
\newcommand{\Hol}{{\mathcal H}}
\newcommand{\Spin}{{\rm Spin}}
\newcommand{\Kappa}{\kappa}
\renewcommand{\d}{{\rm d}}
\newcommand{\Det}{\mbox{{\rm Det}}}
\newcommand{\beq}{\begin{equation}}
\newcommand{\eeq}[1]{\label{#1}\end{equation}}
\renewcommand{\Re}{\mathrm{Re}\,}
\renewcommand{\Im}{\mathrm{Im}\,}
\newcommand{\fft}[2]{{\frac{#1}{#2}}}

\newcommand{\tamphys}{\it
Institute for Quantum Science and Engineering and Hagler Institute for Advanced Study, Texas A\&M University, College Station, TX, 77840, USA
\&\\
Theoretical Physics, Blackett Laboratory, Imperial College London,\\
 London SW7 2AZ, United Kingdom

\&\\
Mathematical Institute, Andrew Wiles Building, University of Oxford,
 \\
Oxford OX2 6GG, United Kingdom\\}
\newcommand{\auth}{M. J. Duff}
\vspace{24pt}

\begin{document}
\begin{flushright}

\hfill\ \ \ { IMPERIAL-TP-2018-MJD-03}\ \ \\\
\end{flushright}

\hfill{}

\hfill{}

\vspace{24pt}

\begin{center}
\large{{\bf Thirty years of Erice on the brane}\footnote{Based on lectures at the International Schools of Subnuclear Physics 1987-2017 and the International Symposium {\it 60 Years of Subnuclear Physics at Bologna}, University of Bologna, November 2018.}}
\vspace{20pt}

\auth

\vspace{10pt}

{\tamphys}

\end{center}
 \abstract{After initially meeting with fierce resistance,  {\it branes}, $p$-dimensional extended objects which go beyond particles ($p=0$) and strings ($p=1$), now occupy centre stage in theoretical physics as microscopic components of M-theory, as the seeds of the AdS/CFT correspondence, as a branch of particle phenomenology, as the higher-dimensional progenitors of black holes and, via the {\it brane-world}, as entire universes in their own right.   Notwithstanding this early opposition, Nino Zichichi invited me to to talk about supermembranes and eleven dimensions at the 1987 School on Subnuclear Physics and has continued to keep Erice on the brane ever since. Here I provide a distillation of my Erice brane lectures and some personal recollections.
%\end{document}
\begin{center}
\end{center}
%{\vfill\leftline{}\vfill}
%\pagebreak
\setcounter{page}{1}
\tableofcontents
\newpage
\vspace{24pt}
\noindent

{\it So are we quarks, strings, branes or what?}
\medskip\\
\indent
New York Times, September 22, 1998
\bigskip
%{\it I know of no better way of teaching science than through its history}

%Steven Weinberg
%
\section{Introduction}
 \label{Introduction}

\subsection{Geneva and Erice: a tale of two cities}

 In 1987 I was a staff member in the Theory Division at CERN, on leave of absence from Imperial College London. Inspired by supergravity \cite{Freedman:1976xh,Deser:1976eh}, I spent the early 1980s advocating  spacetime dimensions greater than four \cite{Duff:1986hr} and the late 1980s advocating worldvolume dimensions greater than two \cite{Khuristring}.  The latter struggle was by far the harder. See for example \cite{Duff:2015yra}.  At this time CERN was playing a prominent part in the development of branes and the 11-dimensional foundations of what was later to be called M-theory.  See, for example, CERN TH-4124-85 \cite{Duff:1985bv}, CERN-TH-4664-87 \cite{Howe}, CERN-TH-4731-87 \cite{Inami}, CERN-TH-4749-87 \cite{Duffnot}, CERN-TH-4779-87 \cite{Savvidy}, CERN-TH-4797-87 \cite{Duff:1987qa},  CERN-TH-4818-87 \cite {Bergshoeffsing}, CERN-TH-4820/87 \cite{Biran}, CERN-TH-4924/87 \cite{Curtright}.  As a matter of fact, the Oxford English Dictionary attributes first usage of the word {\it brane} to  the May 1987 CERN preprint \cite{Inami} by Duff, Inami, Pope, Sezgin and Stelle, published the following year in Nuclear Physics B\footnote{ Paul Townsend's lecture at the Trieste Spring School in April 1987 was intended to be entitled ``P-branes for pea-brains'', but organizer Ergin Sezgin baulked (at pea-brains, not $p$-branes).}.  See Fig \ref{braneoed.jpeg}. Since then, according to INSPIRE there have been 46,192 papers on branes garnering 1,786,998 citations as of November 2018.  According to \cite{He:2018dlv}, {\it brane} ranks 13th in the list of most frequent words in hep-th titles\footnote{The top 20 are {\it model, theory, black-hole, quantum, gravity, string, susy, solution, field, equation, symmetry, brane, inflation, gauge-theory, system, geometry, sugra, new, generalized.}}.

 \begin{figure}[h]
\centering
\includegraphics[scale=0.35]{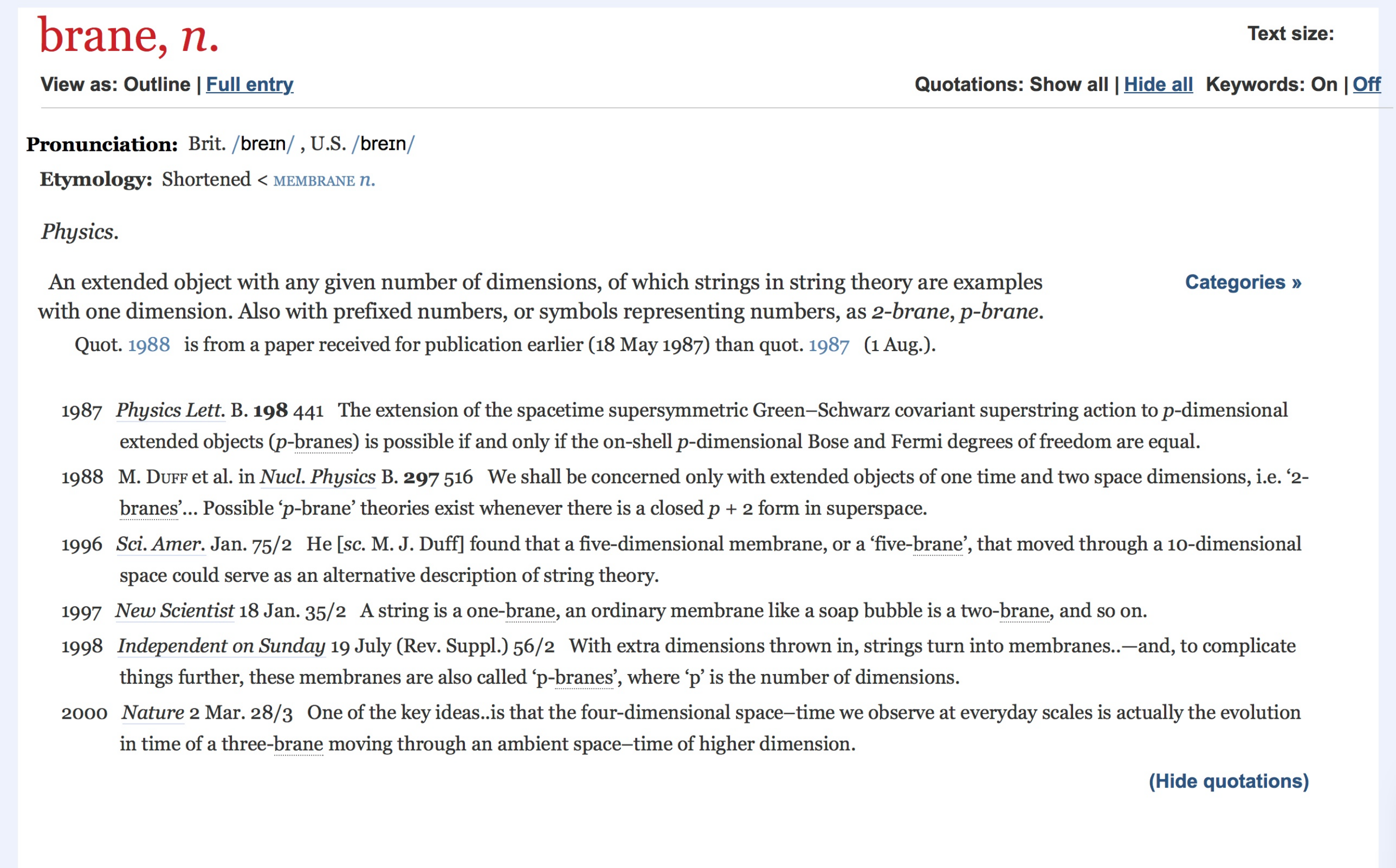}
\caption{Oxford English Dictionary: the word {\it brane}}
\label{braneoed.jpeg}                     
\end{figure}
\begin{figure}[h]
\centering
\includegraphics[scale=0.4, angle =270]{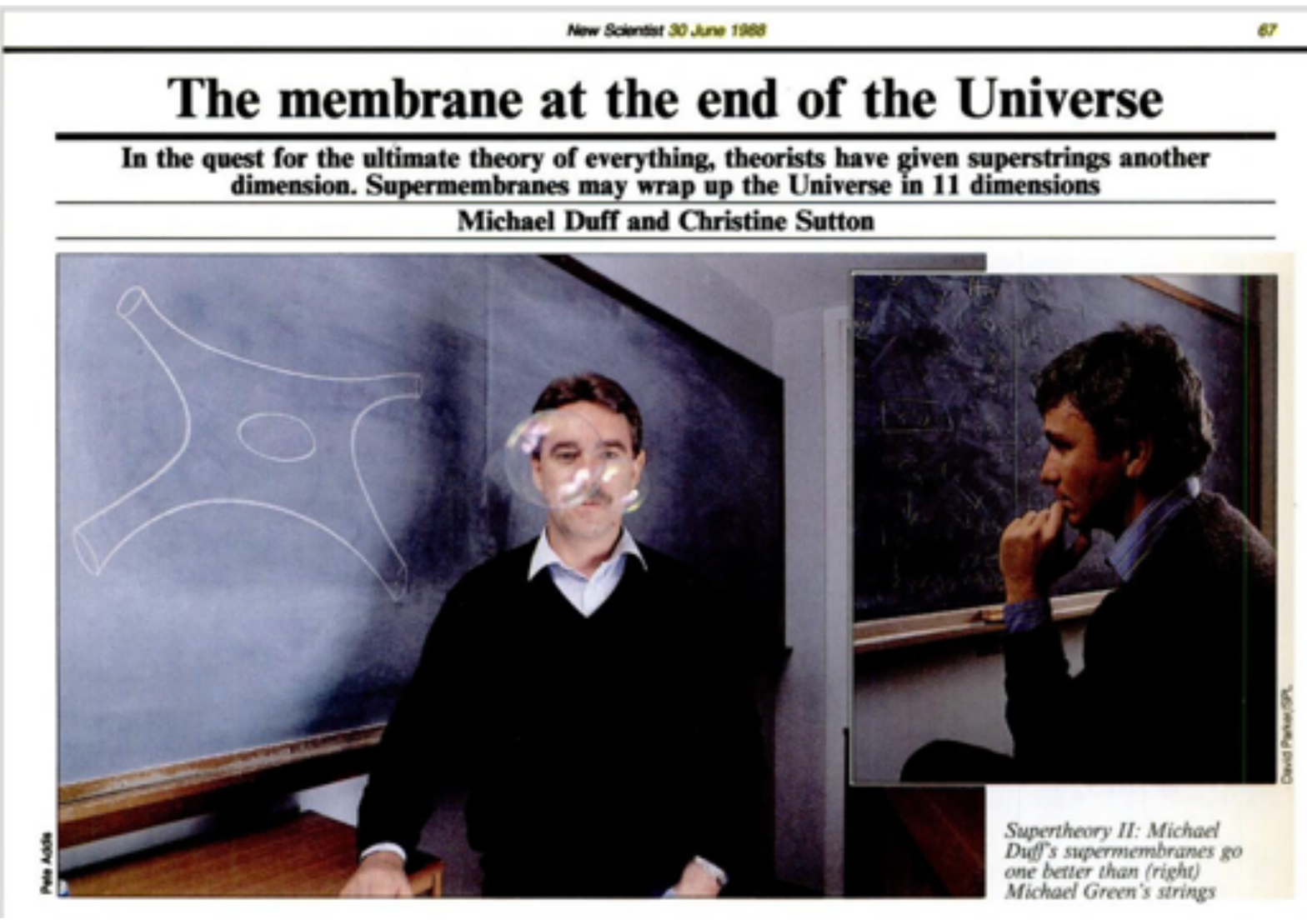}
\caption{\it 1987 article in New Scientist }. 
\label{Screenshot}
\end{figure}

 The 1987 Annual Report of the CERN Theory Division was upbeat:

{\it Finally there were a few papers that are highly critical of string theory and its prospects, and a few that started a heroic study of more complicated objects, namely supermembranes. During 1987 the CERN theory group became the leading research centre  for this subject, which is still in its infancy. The main goal is to understand why there exists an elegant and unique eleven dimensional supergravity, while string theory seems to be restricted to ten dimensions.}

That year I also co-authored an article \cite{Sutton} for New Scientist with Christine Sutton, former editor of the CERN Courier, entitled {\it The Membrane at the End of the Universe}, describing  conformal field theories arising from branes living on the boundary of anti-de Sitter space (AdS) \cite{Blencowe2}, a theme later to play a part in the AdS/CFT correspondence \cite{Maldacena,Gubser,Wittenads}. See Fig \ref{Screenshot}. By the way, I apologized to Mike Green for the caption inserted by New Scientist without my knowledge.  Mike reminded me recently that at the 1983 High Energy Physics  Conference in Brighton, he and I played a game of crazy golf on the promenade in order to decide whether spacetime had ten or eleven dimensions. I won (the golf that is). My excuse for needing a reminder about the golf was that later that same evening I met my future wife.

In 1988, together with fellow brane enthusiasts Chris Pope and Ergin Sezgin, I accepted an invitation from Dick Arnowitt to take up a faculty position in the Physics Department at Texas A\&M and was also lucky enough to have Jianxin Lu assigned to me as a graduate student. He and I were to co-author 20 braney papers. There was  less brane activity at CERN\footnote{But more Type IIA\&M-theory in Texas.}, see for example, CERN-TH-4970/88 \cite{Antoniadis}, CERN-TH-5048-88 \cite{Ruiz-alt}, CERN-TH-5188-88 \cite{Floratos}, CERN-TH-5239-88 \cite{Floratos}, as reflected in the tone of the Annual Report for 1988:

{\it During 1988 string theory has continued to thread much of the formal work in the TH division, renewed attention has been given to the development of conformal field theory and, relative to the previous year, work on supermembranes has somewhat shrunk.}

For those wishing to follow its past and present activities, the CERN Theory Division has three recommendations shown in Fig. \ref{CERN}. 
\begin{figure}[h]
\centering
\includegraphics[scale=0.41]{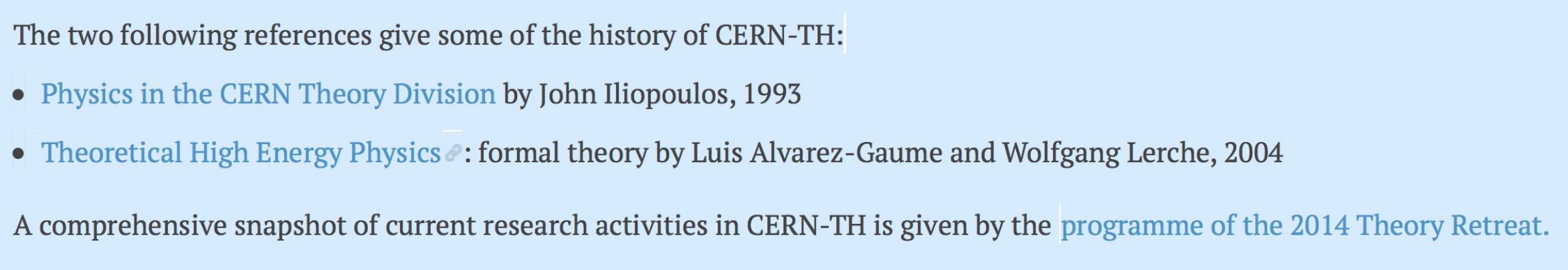}
\caption{ Homepage of the CERN Theory Division http://th-dep.web.cern.ch/cern-theoretical-physics}
\label{CERN}
\end{figure}The  two historical references seem rather coy about the above activities in supermembranes and we find only

{\bf Physics in the CERN Theory Division}: {\it Strings are the simplest extended objects. Although theories of higher dimensional objects have been studied (membranes, etc.), only strings seem to yield consistent theories.}

{\bf Theoretical High Energy Physics}: {\it There was also some activity in the study of the theory of Supermembranes, and in particular in \cite{Howe} it was shown how to extract 10-dimensional superstrings from 11-dimensional supermembranes.}

Whereas branes are at the forefront of current activities:
 
{\bf Programme of the 2014 Theory Retreat}: includes {\it Brane wrapping 3-cycles, D5-brane effective action, Intersecting 7-branes, D5/M5-brane superpotentials, dS-vacua in Type IIB with branes on singularities,  Non-perturbative 7-branes, T-branes/gluing branes, Unoriented D-branes, etc etc}

Such ambivalence towards branes was not unique to CERN\footnote{Later CERN contributions to branes include CERN-TH-6675-93 \cite{Lublack}, CERN-TH-7542-94 \cite{Khuristring}.}.  There are no superstrings in eleven dimensions but there are supermembranes \cite{Bergshoeff1,Bergshoeff1,Duffstelle} which is why between the 1984 Superstring Revolution and the 1995 M-theory Revolution many string theorists were opposed to eleven dimensions. Membrane-related grant proposals tended to attract hostile referee reports during that period and papers with titles like {\it Supermembranes: a fond farewell} and {\it Eleven dimensions (Ugh!)} did not help. One string theorist announced that ``I want to cover up my ears every time I hear the word membrane'' and some organisers of the annual superstring conferences even banned the use of the ``M-word''. My colleague Paul Townsend, one of the membrane pioneers, compared this with the theatrical superstition of calling Macbeth the ``M-Play''. This opposition continued even after it was shown in 1987 \cite{Howe} that one of the five consistent ten-dimensional superstring theories, the Type IIA string, was just the limiting case of the eleven-dimensional supermembrane  \footnote{In his recent book {\it Why string theory?'}, Joseph Conlon \cite{Conlon} writes ``When I first read this paper I was quite shocked by its existence; according to the supposed history of string theory I had `learned', such a paper could not have been written for almost another decade'.} .

An exception to this negativity was Nino Zichichi and in 1987 he invited me to give two lectures on branes at the School on Subnuclear Physics in Erice.  Ironically, an experimentalist could see what many theorists could not: since supermembranes are not forbidden by supersymmetry they must be compulsory.  See Section \ref{compulsory}. He has not only continued to welcome me and others to speak about branes at Erice in the intervening 30 years (together more recently with his co-organizer Gerard 't Hooft) but has also promoted them himself. See \cite{40} for a recent example. I should also mention that another Erice visitor, CERN theorist Sergio Ferrara, was always very supportive \cite{FerraraSagnotti}.

Our purpose here is to give a personal account of these previous lectures and their place in the scheme of things as seen from a 2017 perspective\footnote{I have omitted recollections of lectures I gave at the 2006, 2012, 2013, 2014 and 2015 Schools since branes were not their main focus.}. Accordingly, the Section assigned to each lecture also contains a Subsection devoted to subsequent developments. Of course this means that important topics not anticipated in the lectures will not be discussed as thoroughly as those that were.  Other Erice lectures devoted to branes include those of Khuri \cite{Khurierice}, Witten \cite{Wittenerice}, Polchinski \cite{Polchinskierice},  Bachas  \cite{Bachaserice},  Antoniadis \cite{Antoniadiserice}, Randall \cite{Randallerice} and  Sagnotti \cite{Sagnottierice}.  Two other historical accounts which are well worth reading are those of Witten \cite{Wittenkyoto} and Polchinski \cite{Polchinski:2017vik}.

\subsection{Co-authors}
 
Thanks to my braney collaborators:  Alex Anastasiou,  Alex Batrachenko, Eric Bergshoeff, Miles Blencowe,  Leron Borsten,  Duminda Dahanayake, John Dixon, Sergio Ferrara,  Gary Gibbons,  Paul Howe, Mia Hughes, Haja Ibrahim, Takeo Inami, Jussi Kalkkinen, Ramzi Khuri, James T. Liu, Hong Lu, Jianxin Lu, Alessio Marrani, Rubin Minasian, Silvia Nagy, Roberto Percacci, Jan Plefka, Chris Pope, Joachim Rahmfeld,  William Rubens, Henning Samtleben, Hisham Sati, Ergin Sezgin, Kelly Stelle, Christine Sutton, Paul Townsend, W. Y. Wen  and Edward Witten.

\subsection{Nomenclature}

The names given to various branes have evolved as their place in the
scheme of things has become clearer.  For example, {\it M-theory} is an eleven-dimensional unified theory \cite{Wittenvarious,Comments} incorporating  \cite{Wittenkyoto} earlier ideas on duality \cite{Sen:1992fr,Duffkhuri,Schwarzsen,Duffstrong} and 
on supersymmetric branes \cite{Bergshoeff1,Callan1,Callan2,Duffstelle,Hulltownsend} that subsumes $D=11$ supergravity and the five $D=10$ superstring theories \cite{Duffworld1}. See Section \ref{M} for the etymology of {\it M-theory}}. Following its discovery, the $D=11$ supermembrane and super 5-brane
became known as the {\it M2-brane} and {\it M5-brane} respectively.
(Discrete subgroups of) the Cremmer-Julia symmetries \cite{Cremmerjulia}, conjectured to be
brane analogue of string T-dualities \cite{Luduality2} became known as {\it U-dualities} of M-theory \cite{Hulltownsend}.
Similarly the Type II $p$-branes, which appear as CFTs on the boundary of AdS \cite{Blencowe2} and as closed string
solitons carrying Ramond-Ramond charge \cite{Horowitz1,Callan1,Callan2,Luscan}, are now known as {\it
D-branes} following the realization by Polchinski \cite{Polchinski} that
they admit the dual open string interpretation of {\it
Dirichlet-branes} \cite{strings89,Dai}: surfaces of dimension $p$ on which the
open strings can end: 

{\it No talk at Texas A\&M would be
complete without mention of supermembranes. If one compactifies the
Type I SO(32) superstring, which is unoriented, and sends $r\rightarrow
0$, one obtains a theory with a super-D-brane...}

 J. Polchinski, Strings 89, Texas A\&M, March 1989 \cite{strings89}. 
 
 At the same time the heterotic and Type II 5-branes carrying Neveu-Schwarz 
 charge were renamed {\it NS-branes} and the fundamental string the {\it F-string} to distinguish it from
the D-string.  The D=6 dyonic string became known as the D1-D5-brane
system.  In charting the history of these various branes we shall adopt
the convention in this lecture of using their modern names unless we are 
quoting verbatim an earlier lecture.  Moreover, we reserve the
name {\it D-brane} for the 1/2 BPS Type II branes whose mass equals their
charge and use the name {\it black branes} for those whose mass exceeds
their charge.

Just as the scalar multiplet CFT that occupies the boundary of $AdS_4$ is called the {\it singleton}, so we call the vector supermultiplet that occupies the boundary of $AdS_5 $ the {\it doubleton} and the tensor supermultiplet that occupies the boundary of $AdS_7$ the {\it tripleton}. This nomenclature is based on the rank of $AdS_{p+2}$ and differs from \cite{Gunaydin}.

\section{1987 Not the Standard Superstring Review }

%INTERNATIONAL SCHOOL OF SUBNUCLEAR PHYSICS - Director:  A. ZICHICHI
%26th Course:  The Super-World-III 
%7 - 15 August 1988  \cite{Classical5}
1987 INTERNATIONAL SCHOOL OF SUBNUCLEAR PHYSICS - Director: A.
ZICHICHI 25th Course: The Super World - II 6 - 14 August 1987 \cite{Duffnot}

The first of my lectures at the School on Subnuclear Physics,  {\it Not the standard superstring review} \cite{Duffnot}, was an appraisal of the current state of superstrings which differed from the superstring orthodoxy in those heady days following the 1984 Superstring revolution. Specifically I focussed on the vacuum degeneracy problem and supermembranes.  However, I tempered my scepticism by saying:

{\it In order not to be misunderstood, let me say straight away that I share the conviction that superstrings are the most exciting development in theoretical physics for many years, and that they offer the best promise to date of achieving the twin goals of a consistent quantum gravity and a unification of all the forces and particles of Nature. Where I differ is the degree of emphasis that I would place on the unresolved problems of superstrings, and the likely time scales involved before superstrings (or something like superstrings) make contact with experimental reality.}
\subsection{Vacuum degeneracy and the multiverse} 

{\it In the absence of an exhaustive classification, we do not know how many (consistent compactifications to four-dimensions) there are\footnote{It had already been noted in \cite{Duff:1986hr}  that there are an infinite number of compact Einstein manifolds in seven dimensions and hence an infinite number of compactifications of $D=11$  down to $D=4$.} but it surely runs into billions \cite{Schellekens}. For the time being, therefore, the phrase ``superstring-inspired phenomenology'' can only mean sifting through these billions of heterotic models in the hope of finding one that is realistic. The trouble wth this needle-in-a-haystack approach is that even if we found one with good phenomenology, we would be left wondering in what sense this could be called a``prediction'' of string theory.} 

{\it Some cosmologists, on the other hand, accept vacuum degeneracy as a fact of life. They argue that the Universe has billions of different vacua and we just happen to be living in one of them with $SU(3) \times SU(2) \times U(1)$, three families etc. In which case, as Murray Gell-Mann puts it, physics will have been reduced to an environmental science like botany.}

\subsection{Supermembranes} 

Membrane theory has a strange history which goes back even further than
strings \cite{Duff:2004nh}. The idea that the elementary particles might correspond to
modes of a vibrating membrane was put forward originally in 1962 by 
Dirac \cite{Dirac1}.  When string theory came along in the 1970s, there were 
some attempts to revive the membrane idea but things did not change much until 1986 when Hughes, Liu and Polchinski
\cite{Hughes} showed that it was possible to combine membranes with
supersymmetry: the {\it supermembrane} was born.
\begin{figure}[h]
\centering\includegraphics[scale=1]{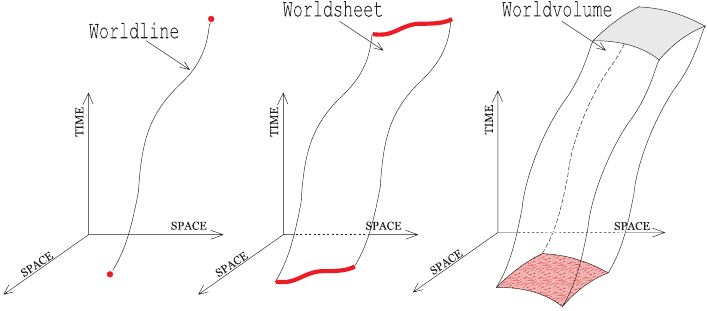}
\caption{Particles, strings and membranes.}
\label{psm}
\end{figure}
Consequently, while all the progress in string theory was going on, a small
splinter group was posing the question: Once you have given up
$0$-dimensional particles in favor of $1$-dimensional strings, why not
$2$-dimensional membranes or in general $p$-dimensional objects (inevitably
dubbed {\it $p$-branes})?  Just as a $0$-dimensional particle sweeps out a
$1$-dimensional {\it worldline} as it evolves in time, so a $1$-dimensional
string sweeps out a $2$-dimensional {\it worldsheet} and a $p$-brane sweeps
out a $d$-dimensional {\it worldvolume}, where $d=p+1$. See Fig. \ref{psm}.  Of course, there
must be enough room for the $p$-brane to move about in spacetime, so $d$
must be less than or equal to the number of spacetime dimensions $D$.  In
fact, as we shall see in Section \ref{ravioli}, supersymmetry places further
severe restrictions both on the dimension of the extended object and the
dimension of spacetime in which it lives \cite{Achucarro}. One can
represent these as points on a graph where we plot spacetime dimension $D$
vertically and the $p$-brane dimension $d=p+1$ horizontally. This graph is
called the {\it brane-scan} \cite{Duff:1990na}. See Table \ref{brane-scan}.  In the early
eighties Green and Schwarz \cite{Greenschwarz} had shown that spacetime
supersymmetry allows classical superstrings moving in spacetime dimensions
$3,4,6$ and $10$. (Quantum considerations rule out all but the
ten-dimensional case as being truly fundamental. Of course some of these
ten dimensions could be curled up to a very tiny size in the way suggested
by Kaluza and Klein \cite{Duff1}. Ideally six would be compactified
in this way so as to yield the four spacetime dimensions with which we are
familiar.) It was now realized, however, that these $1$-branes in
$D=3,4,6$ and $10$ should now be viewed as but special cases of this more
general class of supersymmetric extended object.  

Curiously enough, the maximum spacetime dimension permitted is eleven, 
where Bergshoeff, Sezgin and Townsend found their supermembrane
\cite{Bergshoeff1,Bergshoeff2} which couples to eleven-dimensional
supergravity \cite{Cremmerjuliascherk}. (The $3$-form gauge field of $D=11$
supergravity had long been suggestive of a membrane interpretation
\cite{Julia:1979fw}). Moreover, it was then possible to show \cite{Howe} by
simultaneous dimensional reduction of the spacetime and worldvolume 
that the membrane looks like a string in ten dimensions. In fact, it yields
precisely the Type $IIA$ superstring:

{\it We do not yet know whether this ``supermembrane'' is consistent at the quantum level but the orthodox claim that only strings can be quantum consistent now looks much less certain.}
\subsection{Subsequent developments}
\begin{itemize}
\item{The multiverse}

The  loss of uniqueness in going from ten dimensions to four, is nowadays called the {\it Landscape} problem \cite{Susskind,BoussoPolchinski}. The many universes are known collectively as the {\it Multiverse}. See, for example, Linde's {\it A brief history of the multiverse} \cite{linde}. though some might find it too brief.

\item{M-theory}

Branes now play vital role in M-theory.  Reviews on branes may be found in
\cite{Duff:1987qa,Classical5,Khuristring,Duffsupermembranes5,Stelle5,
Polchinskistrings0,Dbranes,nlab,Westbranes}.
Reviews of $M$-theory may be found in
\cite{Schwarzpower,DuffM,TownsendM,Duffworld1,Kaku1,Kaku2,Ortin,Huerta:2018xyh,nlab}.

%\item{Brane-scan}

\end{itemize}
%\newpage
 \section{1987 From super-spaghetti to super-ravioli }
 \label{ravioli}
 
 1987 INTERNATIONAL SCHOOL OF SUBNUCLEAR PHYSICS - Director: A.
ZICHICHI 25th Course: The Super World - II 6 - 14 August 1987 \cite{Duff:1990na}
 
 Since my second lecture attempted to justifiy this passage from strings to membranes and bearing in mind its location, I called it {\it From super-spaghetti \footnote{What better place to recall this than Bologna?} to super-ravioli}. It began:
 
{\it Many of the supergravity theories that we used to study a few years ago are now known to be merely the field theory limit of an underlying string theory. For example, N=2a supergravity in 10 dimensions is just the field theory limit of the Type IIA superstring.  What are we to make, therefore, of supergravity theories which cannot be obtained from strings such as N = 1 supergravity in eleven dimensions? This is a particularly puzzling example since it is well known that upon dimensional reduction to 10 dimensions, it yields the above-mentioned N = 2a theory. Indeed, if supersymmetry allows $D \leq 11$, why do strings stop at $D = 10$? }

%We now show how to derive  the Type IIA superstring in ten dimensions by a simultaneous dimensional reduction of the world-volume and the space-time.

\begin{table}[h]
$
\begin{array}{ccccccccccccccc}
~&D\uparrow&&&&&&&&&&&~\\
~&11&.&~&&s&&&t&&&&&~\\
~&10&.&v&s/v&v&v&v&s/v&v&v&v&v&~\\
~&9&.&s&&&&s&&&&&&~\\
~&8&.&~&&&s&&&&&&&~\\
~&7&.&~&&s&&&t&&&&&~\\
~&6&.&v&s/v&v&s/v&v&v&&&&&~\\
~&5&.&s&&s&&&&&&&&~\\
~&4&.&v&s/v&s/v&v&&&&&&&~\\
~&3&.&s/v&s/v&v&&&&&&&&~\\
~&2&.&s&&&&&&&&&&~\\
~&1&.&~&~&~&~&~&~&~&~&~&~&~\\
~&0&.&.&.&.&.&.&.&.&.&.&.&.~\\
~&~&0&1&2&3&4&5&6&7&8&9&10&11&d\rightarrow
\end{array}
$
\caption{ The old brane-scan involves only scalar multiplets $s$ on the worldvolume; the new one includes vector multiplets $v$ and antisymmetric tensor multiplets $t$.}
\label{brane-scan}
\end{table}

\subsection{The old brane-scan}
\label{fermi}

It is ironic that although one of the motivations for the original
supermembrane paper \cite{Hughes} was precisely to find the superthreebrane
as a {\it topological defect} of a supersymmetric field theory in $D=6$; 
the discovery of the other supermembranes proceeded in the opposite
direction. Hughes et al. showed that kappa symmetry could be generalized
to $d > 2$ and proceeded to construct a threebrane displaying an explicit
$D=6, N=1$ spacetime supersymmetry and kappa invariance on the
worldvolume. It was these kappa symmetric Green-Schwarz actions, rather
then the soliton interpretation which was to dominate the early work on the
subject\footnote{Strangely enough in Yau's version of the history, it was the other way around \cite{Yau}}. First of all, Bergshoeff, Sezgin and Townsend \cite{Bergshoeff1}
found corresponding Green-Schwarz actions for other values of $d$ and $D$,
in particular the eleven-dimensional supermembrane. 

Let us introduce the coordinates $Z^M$ of a curved superspace
\begin{equation}
Z^M=(x^{\mu}, \theta^{\alpha})
\end{equation}
and the supervielbein $E_M{}^A (Z)$ where $M = \mu,\alpha$ are world
indices and $A = a,\alpha$ are tangent space indices.  We also define the
pull-back
\begin{equation}
E_i{}^A = \partial_iZ^ME_M{}^A
\end{equation}
We also need the super-$d$-form $B_{A_{d}\ldots A_{1}}(Z)$. 
Then the supermembrane action ihas a kinetic term, a worldvolume cosmological term, and
a Wess-Zumino term
\be
S= T_d\int
d^d\xi\biggl\lbrack -\frac{1}{2}{\sqrt -\gamma}\gamma^{ij}
E_i{}^aE_j{}^b \eta_{ab}+\frac{1}{2} (d-2){\sqrt -\gamma}
+\frac{1}{d!}\epsilon^{i_1\ldots
i_d}E_{i_{1}}{}^{A_1}\cdots E_{i_{d}}{}^{A_d}B_{A_{d}\ldots A_{1}}
\biggr\rbrack.
\label{supermembrane}
\ee
  This action has the virtue that it reduces to
the Green-Schwarz superstring action when $d = 2$.

The target-space symmetries are superdiffeomorphisms,  Lorentz invariance and
$d$-form gauge invariance.  The worldvolume symmetries are ordinary
diffeomorphisms and kappa invariance referred to earlier which is known
to be crucial for superstrings, so let us examine it in more detail.  The
transformation rules are

\begin{equation}
\delta Z^M E^a{}_M = 0, ~~~\delta Z^ME^{\alpha}{}_M =
\kappa^{\beta}(1+\Gamma)^{\alpha}{}_{\beta}
\label{transf}
\end{equation}
where $\kappa^{\beta}(\xi)$ is an anticommuting spacetime spinor but
worldvolume scalar, and where

\begin{equation}
\Gamma^{\alpha}{}_{\beta} = \frac{(-1)^{d(d-3)/4}}{d!{\sqrt
-\gamma}}\epsilon^{i_{1}..i_{d}}E_{i_{1}}{}^{a_{1}}
E_{i_{2}}{}^{a_{2}}\ldots E_{i_{d}}{}^{a_{d}}\Gamma_{a_{1..a{_d}}}~~.
\end{equation}
Here $\Gamma_a$ are the Dirac matrices in spacetime and 
\begin{equation}
\Gamma_{a_{1}..a_{d}}=\Gamma_{[a_1{\cdots}a_{d}]}~~.
\end{equation}
This kappa symmetry has the following important consequences:

\noindent 
1)~~The symmetry is achieved only if certain constraints on the antisymmetric
tensor field strength $F_{MNP..Q}$(Z) and the supertorsion are
satisfied.  In particular the Bianchi identity $dF = 0$ then requires the
$\Gamma$ matrix identity
\begin{equation}
\biggl ( d{\bar \theta} \Gamma_ad\theta\biggr ) \biggl
(d{\bar \theta} \Gamma^{ab_{1}\ldots b_{d-2}} d\theta\biggr )~=0
\end{equation}
for a commuting spinor d$\theta$.  As shown by Achucarro, Evans, Townsend 
and Wiltshire \cite{Achucarro} this is satisfied only for certain values of d
and D.  Specifically, for $d\geq 2$
\begin{eqnarray}
d&=&2:~~~D = 3, 4, 6, 10\nonumber\\
d&=&3:~~~D = 4, 5, 7, 11\nonumber\\
d&=&4:~~~D = 6, 8\nonumber\\
d&=&5:~~~D = 9\nonumber\\
d&=&6:~~~D = 10~~.
\end{eqnarray}
Note that we recover as a special case the well-known result that 
Green-Schwarz superstrings exist {\it classically} only for $D = 3, 4,
6,$ and $10$.  Note also $d_{max} = 6$ and $D_{max} = 11$.  The upper limit
of $D = 11$ is already known in supergravity \cite{Freedman:1976xh,Deser:1976eh} but there it is necessary to
make extra assumptions concerning the absence of consistent higher spin
interactions.  In this formulation of supermembranes, it follows automatically. \\

\noindent
2)~~The matrix $\Gamma$ of (1.20) is traceless and satisfies

\begin{equation}
\Gamma^2 = 1
\end{equation}
\noindent
when the equations of motion are satisfied and hence the matrices
$(1\pm\Gamma)/2$ act as projection operators.  The transformation rule
(1.19) therefore permits us to gauge away one half on the fermion degrees of
freedom.  As described below, this gives rise to a matching of physical boson
and fermion degrees of freedom on the worldvolume.  \\

\noindent
3)~~In the case of the eleven-dimensional supermembrane, it has been shown [16]
that the constraints on the background fields $E_M{}^A$ and $B_{MNP}$
are nothing but the equations of motion of eleven-dimensional
supergravity \cite{Bergshoeff1,Bergshoeff2}.  

\subsection{Type $IIA$ superstring in $D=10$ from supermembrane in $D=11$}
\label{reduce}
We begin with the bosonic sector of the $d=3$ worldvolume of the $D=11$
supermembrane:
\begin{eqnarray}
S_3&=&T_3\int d^3\xi\biggl[-{1\over2}\sqrt{-\gamma}\gamma^{ij}
\partial_i X^M\partial_j X^N G_{MN}(X) +{1\over2}\sqrt{-\gamma}\nonumber\\
&&\qquad\qquad
+{1\over3!}\epsilon^{ijk}\partial_i X^M\partial_j X^N\partial_k X^P
A_{MNP}(X)\biggr]\ ,
\label{membrane}
\end{eqnarray}
where $T_3$ is the membrane tension, $\xi^i$ ($i=1,2,3$) are the
worldvolume coordinates, $\gamma^{ij}$ is the worldvolume metric and
$X^M(\xi)$ are the spacetime coordinates $(M=0,1,\ldots,10)$.  Kappa
symmetry \cite{Bergshoeff1,Bergshoeff2} then demands that the
background metric $G_{MN}$ and background 3-form potential $A_{MNP}$
obey the classical field equations of $D=11$ supergravity, whose
bosonic action is
\begin{equation}
I_{11}=\frac{1}{2\kappa_{11}{}^2}\int d^{11}x\sqrt{-G}
\left[R_G-\frac{1}{2\cdot4!}F_{\scriptscriptstyle MNPQ}^2\right]
-\frac{1}{12\kappa_{11}{}^2} \int A_3\wedge F_4 \wedge F_4 \ ,
\label{supergravity11}
\end{equation}
where $F_4=dA_3$ is the 4-form field strength. In particular, $F_4$
obeys the field equation
\begin{equation}
d*F_4=-{1\over2}F_4{}^2
\label{equation4}
\end{equation}
and the Bianchi identity
\begin{equation}
dF_4=0\ .
\label{Bianchi4}
\end{equation}
To see how a double worldvolume/spacetime compactification of the
$D=11$ supermembrane theory on $S^1$ leads to the Type $IIA$ string in
$D=10$ \cite{Howe}, let us denote all $(d=3, D=11 )$ quantities by a hat
and all $(d=2, D=10)$ quantities without.  We then make a ten-one split
of the spacetime coordinates
\be
{\hat X}^{\hat M}=(X^M,Y)\qquad M=0,1,\ldots,9
\ee
and a two-one split of the worldvolume coordinates
\begin{equation}
{\hat \xi}^{\hat i}= (\xi^i,\rho)\qquad i=1,2
\end{equation}
in order to make the partial gauge choice
\be
\rho=Y\ ,
\ee
which identifies the eleventh dimension of spacetime with the third
dimension of the worldvolume. In other words, the membrane is wrapped around the $S^1$ (See  \cite{russo} for subtleties concerning zero modes).
The dimensional reduction is then
effected by taking the background fields ${\hat G}_{{\hat
M}{\hat N}}$ and ${\hat A}_{{\hat M}{\hat N}{\hat P}}$ to be independent of
$Y$.  The string backgrounds of dilaton $\Phi$, string $\sigma$-model metric
$G_{MN}$, $1$-form $A_M$, $2$-form $B_{MN}$ and $3$-form $A_{MNP}$ are given
by% %
\footnote{The choice of dilaton prefactor, $e^{-\Phi/3}$, is dictated by the
requirement that $G_{MN}$ be the $D=10$ string $\sigma$-model metric.  To
obtain the $D=10$ fivebrane $\sigma$-model metric, the prefactor is unity
because the reduction is then spacetime only and not simultaneous
worldvolume/spacetime.  This explains the remarkable ``coincidence''
\cite{Lublack} between $\hat G_{MN}$ and the $D=10$ fivebrane $\sigma$-model
metric.}
\begin{eqnarray}
{\hat G}_{MN}&=& e^{-\Phi/3}\left(
\begin{array}{cc}
G_{MN}+e^\Phi A_MA_N&e^{\Phi}A_M\\
e^{\Phi}A_N&e^{\Phi}
\end{array}
\right)\nonumber\\
{\hat A}_{MNP}&=&A_{MNP}\nonumber\\
{\hat A}_{MNY}&=&B_{MN}\ .
\end{eqnarray}

The actions (\ref{membrane}) and (\ref{supergravity11}) now reduce to
\begin{eqnarray}
S_2=T_2\int d^2\xi\biggl[&-&{1\over2}\sqrt{-\gamma}\gamma^{ij}
\partial_i X^M\partial_j X^N G_{MN}(X) 
\nonumber \\
&-& {1\over2!}\epsilon^{ij}\partial_i X^M\partial_j X^N
B_{MN}(X)+\cdots\biggr] 
\end{eqnarray}
and
\begin{eqnarray}
I_{10}&=&\frac{1}{2\kappa_{10}{}^2}\int d^{10}x\sqrt{-G}e^{-\Phi} \Big[
R_G+(\partial_{\scriptscriptstyle M}\Phi)^2
-\frac{1}{2\cdot3!}H_{\scriptscriptstyle MNP}^2
-\frac{1}{2\cdot2!}e^\Phi F_{\scriptscriptstyle MN}^2
 \nonumber\\
&&~~~~~~~~~~-\frac{1}{2\cdot4!}e^\Phi J_{\scriptscriptstyle MNPQ}^2 \Big]
-{1\over2\kappa_{10}{}^2}\int{1\over2}F_4\wedge F_4\wedge B_2\ ,
\label{eq:supergravity10}
\end{eqnarray}
where the field strengths are given by $J_4=F_4+A_1H_3$, $H_3=dB_2$ and
$F_2=dA_1$.

One may repeat the procedure in superspace to obtain
\begin{eqnarray}
S_2=T_2\int
d^2\xi\biggl[-{1\over2}\sqrt{-\gamma}\gamma^{ij}{E_i{}^aE_j{}^b\eta_{ab}} 
+{1\over2!}\epsilon^{ij}\partial_i X^M\partial_j X^N
B_{MN}(Z)\biggr] 
\label{string}
\end{eqnarray}
which is just the action of the Type $IIA$ superstring.

\subsection{Bose-fermi matching on the worldvolume}
\label{oldscan}

The matching of physical bose and fermi degrees of freedom on the
{\it worldvolume} may, at first sight, seem puzzling since we began with
only spacetime supersymmetry. The explanation is as follows. As the $p$-brane
moves through spacetime, its trajectory is described by the
functions $X^M (\xi)$ where $X^M$ are the spacetime coordinates ($M = 0, 1,
\ldots, D - 1$) and $\xi^i$ are the worldvolume coordinates ($i = 0, 1,
\ldots, d - 1$).  It is often convenient to make the so-called
 {\it static gauge choice} by making the $D = d + (D - d)$ split
\begin{equation}
X^M (\xi) = (X^{\mu} (\xi), Y^m (\xi)),
\end{equation}
where $\mu = 0, 1, \ldots, d - 1$~and~$m = d, \ldots, D - 1$,
and then setting
\begin{equation}
X^{\mu} (\xi) = \xi^{\mu}.
\end{equation}
Thus the only physical worldvolume degrees of freedom are given
 by the $(D - d)~Y^m (\xi)$.  So the number of on-shell bosonic degrees of
freedom is
\begin{equation}
{N_B = D - d.}
\end{equation}
To describe the super $p$-brane we augment the $D$ bosonic coordinates $X^M
(\xi)$ with anticommuting fermionic coordinates $\theta^{\alpha} (\xi)$.
Depending on $D$, this spinor could be Dirac, Weyl, Majorana or
Majorana-Weyl. The fermionic kappa symmetry means that half of the spinor
degrees of freedom are redundant and may be eliminated by a physical gauge
choice.  The net result is that the theory exhibits a {\it $d$-dimensional
worldvolume supersymmetry} \cite{Achucarro} where the number of fermionic
generators is exactly half of the generators in the original spacetime
supersymmetry.  This partial breaking of supersymmetry is a key idea.  Let
$M$ be the number of real components of the minimal spinor and $N$ the
number of supersymmetries in $D$ spacetime dimensions and let $m$~and~$n$
be the corresponding quantities in $d$ worldvolume dimensions.  Let us
first consider $d > 2$.  Since kappa symmetry always halves the number of
fermionic degrees of freedom and going on-shell halves it again, the
number of on-shell fermionic degrees of freedom is
\begin{equation}
{N_F = {1\over 2}mn = {1\over 4}MN.}
\end{equation}
Worldvolume supersymmetry demands $N_B = N_F$ and hence
\begin{equation}
{D - d = {1\over 2}mn = {1\over 4}MN.}
\label{bosefermi}
\end{equation}
A list of dimensions, number of real dimensions of the minimal spinor and
possible supersymmetries is given in Table \ref{mini}, from which we
see that there are only $8$ solutions of (\ref{bosefermi}) all with $N = 1$, as
shown in Table \ref{brane-scan}.  We note in particular that $D_{{\rm max}} =
11$ since $M \geq 64$ for $D \geq 12$ and hence (\ref{bosefermi}) cannot be
satisfied.  Similarly
 $d_{{\rm max}} = 6$ since $m \geq 16$ for $d \geq 7$.  The case $d = 2$ is
special because of the ability to treat left and right moving modes
independently.  If we require the sum of both left and right moving bosons
and fermions to be equal, then we again find the condition (\ref{bosefermi}). 
This provides a further $4$ solutions all with $N = 2$, corresponding to
Type $II$ superstrings in $D = 3, 4, 6$~and~$10$ (or 8 solutions
in all if we treat Type $IIA$ and Type $IIB$ separately).  Both the
gauge-fixed Type $IIA$ and Type $IIB$ superstrings will display $(8, 8)$
supersymmetry on the worldsheet. If we require only left (or right) matching,
then  (\ref{bosefermi}) replaced by
\begin{equation}
{D - 2 = n = {1\over 2}MN,}
\end{equation}
which allows another $4$ solutions in $D = 3, 4, 6$~and~$10$,
all with $N = 1$. The gauge-fixed theory will display $(8,0)$ worldsheet
supersymmetry.  The heterotic string falls into this category.  The results
\cite{Achucarro} are indicated by the points labelled $s$ in Table 
\ref{brane-scan}. Point particles with $d=1$ are usually omitted from the
brane-scan \cite{Achucarro,Luscan,Khuristring}, but in Table
\ref{brane-scan} we have included them.

An equivalent way to arrive at the above conclusions is to list
all scalar supermultiplets and to interpret
the dimension of the target space, $D$, by
\begin{equation}
{D - d =~{\rm number~of~scalars}.}
\label{scalars}
\end{equation}
Indeed, these scalars are the Goldstone bosons associated with the spontaneous breaking of the $D-d$ translations. 
A useful reference is \cite{Strathdee} which provides an exhaustive
classification of all unitary representations of supersymmetry with maximum
spin $2$.  In particular, we can understand $d_{{\rm max}} = 6$ from this
point of view since this is the upper limit for scalar supermultiplets.

\begin{table}
\halign{\indent #&\qquad\hfil# \hfil&\quad\hfil
#\hfil&\quad\hfil # \hfil &
\quad \hfil # \hfil &\quad \hfil # \hfil &\quad #\hfil\cr
&&&Dimension & Minimal Spinor& Supersymmetry&\cr
&&&($D$ or $d$) & ($M$ or $m$) & ($N$ or $n$)&\cr
&&&11 & 32 & 1&\cr
&&&10 & 16 & 2, 1&\cr
&&&9 & 16 & 2, 1&\cr
&&&8 & 16 &2, 1&\cr&&&6 & 8 & 4, 3, 2, 1&\cr
&&&5 & 8 & 4, 3, 2, 1&\cr
&&&4 & 4 & 8, $\ldots$, 1&\cr
&&&3 & 2 & 16, $\ldots$, 1&\cr
&&&2 & 1 & 32, $\ldots$, 1&\cr}
\caption{Minimal spinor components and supersymmetries.}
\label{mini}
\end{table}

There are four types of solution with  
$8 + 8 $, $4 + 4 $,  $2 + 2 $ or  $1 + 1 $ degrees of freedom respectively. 
Since the numbers  $1 $, $2 $,  $4 $ and  $8 $ are also the dimension of the
four division algebras, these four types of solution are referred to as
real, complex, quaternion and octonion respectively.  The connection with
the division algebras can in fact be made more precise \cite{Kugo,Evans,Baez,Duff:2010vy,Huerta2011,Anastasiou:2013cya,Huerta2014}. 

\subsection{A heterotic 5-brane?}
\label{five}

Of particular interest was the $D=10$ fivebrane, whose Wess-Zumino term
coupled to a rank six antisymmetric tensor potential $A_{MNPQRS}$ just as the
Wess-Zumino term of the string coupled to a rank two potential $B_{MN}$.
Spacetime supersymmetry therefore demanded that the fivebrane coupled to
the $7$-form field strength formulation of $D=10$ supergravity
\cite{Chamseddine} just as the string coupled to the $3$-form version
\cite{deroo,Chapline}. These dual formulations of $D=10$ supergravity have long
been something of an enigma from the point of view of superstrings. As field
theories, each seems equally valid. In particular, provided we couple them to
$E_8 \times E_8$ or $SO(32)$ super-Yang-Mills \cite{Greenschwarz}, then
both are anomaly free \cite{Gates}. Since the $3$-form version
corresponds to the field theory limit of the heterotic string, we
conjectured  \cite{Duff:1987qa} that there ought to exist a {\it heterotic
fivebrane} which could be viewed as a fundamental anomaly-free theory in
its own right and whose field theory limit corresponds to the dual
$7$-form version. We shall refer to this as the {\it string/fivebrane
duality conjecture}. At this stage, however, the solitonic element had not
yet been introduced.

\subsection{E(8) x SO(16) in $D=11$?}

It is interesting to note that the three-eight split 
\be
SO(1,10) \supset SO(1,2) \times SO(8)
\ee
implied by the embedding of the three-dimensional worldvolume of the
supermembrane in  eleven-dimensional space-time had previously been invoked in \cite{Duff:1985bv} to exhibit the hidden $SO(16)$ symmetry of D = 11 supergravity, where the 128 bosonic degrees of freedom may be assigned to the coset $E_8/SO(16)$. We wondered what role $E_8$,  the Kac-Moody extension $E_9$ and the Lorentzian algebra $E_{10}$ will play for the supermembrane.

\subsection{Branes on the boundary of AdS}

Compactification of $D=11$ supergravity: $d=4$ anti-de Sitter space-time x $S^7$ yields  four-dimensional supergravity with maximum (N=8) supersymmetry and local SO(8) invariance \cite{Duff:1986hr}.  The vacuum symmetry is the AdS supergroup OSp(4/8) which admits the strange ``singleton" which have no analogue in the Poincare group and no immediate field theory interpretation. Owing to the N = 8 supersymmetry they form an ultrashort N = 8 supermultiplet consisting of eight spin-1/2 fermions and eight spin-0 bosons which transform according to
the $8_s$ and $8_v$ representations of $SO(8)$. Although we are dealing
with the four-dimensional anti de Sitter group $SO(2,3)$, we cannot
write down an action for these singletons living in $AdS_4$. However, as
discussed by Fronsdal \cite{Fronsdal}, we can write down an action living on its three-dimensional boundary $S^1 \times S^2$ with signature $(-,+,+)$.

But $8_s$ spin-1/2 and $8_v $ spin-0 on a 3-dimensional worldvolume with signature $(-,+,+)$ is just what we get from gauge-fixing the supermembrane! We noted that relativistic membranes and singletons have one more thing in common: they were both invented by Dirac at about the same time \cite{Dirac1,Dirac2}.

\subsection{Subsequent developments}
\begin{itemize}

\item{Role of D$=$11 supergravity}

Responding to my remark that $D=11$ supergravity hints at something beyond strings, Dean Rickles \cite{Rickles} finds it necessary to belittle the role of supergravity compared with superstrings in the historical development of M-theory, calling the years between the discovery of supergravity and the superstring revolution the Decade of Darkness. While it is true that eleven-dimensional quantum supergravity suffers from the ultraviolet divergences that ten-dimensional superstrings avoid, its very existence calls into question the notion that strings are the be-all-and-end-all of the final theory.   In his zeal to downgrade supergravity Rickles distorts the compliment to make it sound more like an insult:
``This became widely accepted, and one can find Michael Duff writing in 1988 that {\it Many of the supergravity theories that we used to study a few years ago are now known to be merely the field theory limit of an underlying string theory}.''

\item{M2 brane solutions of $D=11$ supergravity}

The eleven-dimensional supermembrane was subsequently seen to be a solution of the $D=11$ supergravity field equations \cite{Duffstelle} and now plays a vital role in M-theory where it is known as the  M2-brane.

\item{Type IIA string in $D=10$ from supermembrane in $D=11$}

Witten \cite{Wittenvarious,Comments} realises that the radius $R_{11}$ of the $S^1$
leads to the Type IIA string with coupling constant $g_{s}$ given by
\be
g_{s}=R_{11}^{3/2}   
\ee
and  we recover the weak coupling regime when $R_{11}\rightarrow 0$, 
which 
explains the earlier illusion that the theory is defined in 
$D=10$. 
\item {D-branes} 

Note that if Type $II$ $p$-branes exist for $p > 1$, they cannot be described by scalar multiplets \cite{Achucarro}.  In fact they are described by the vector multiplets that appear on the brane-scan. They subsequently acquired an interpretation as Dirichlet branes, or D-branes \cite{Polchinski}, surfaces of dimension $p$ on which open strings can end.

\item {Fivebrane as a soliton}

The heterotic 5-brane was found by Strominger to be a soliton solution of the heterotic string \cite{Strominger1}.

\item{Exceptional geometry}

Exceptional symmetries $E_8,E_9,E_{10}$, appearing not merely upon compactification but already in eleven dimensions, are now the subject of much investigation on the context of exceptional geometry. $E_{11}$ has taken this one stage further. See section \ref{objects}.

\item{AdS/CFT correspondence}

Branes on the boundary of AdS are a vital ingredient in the AdS/CFT correspondence \cite{Maldacena,Gubser,Wittenads}. Another vital ingredient, missing in these early days, was the non-abelian nature of the symmetries that appear when we stack N branes on top of one another \cite{Wittenbound}.  

\item{Brane-scan}

Further developments and elaborations on the brane-scan are summarized in Schreiber's n-lab \cite{nlab} and references  therein.

\end{itemize}
\begin{figure}[h]
\centering
\includegraphics[scale=0.5]{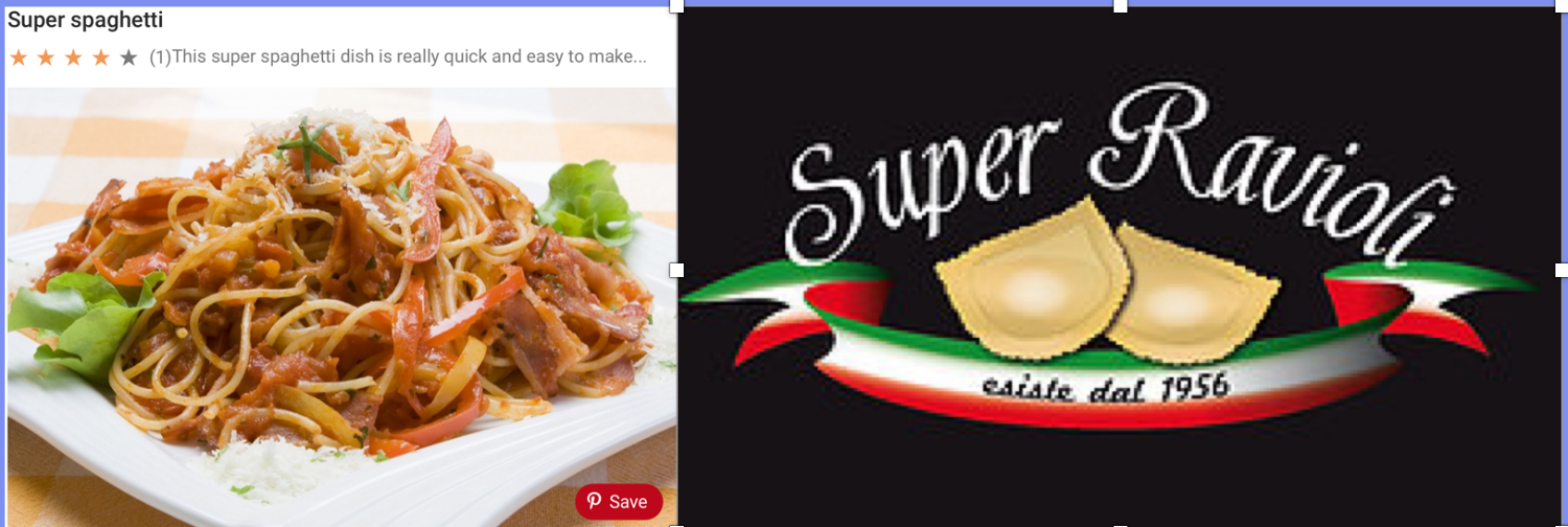}
\caption{\it From super spaghetti to super ravioli}. 
\label{Spag}
\end{figure}

\section{1988 Classical and Quantum Supermembranes}
INTERNATIONAL SCHOOL OF SUBNUCLEAR PHYSICS - Director:  A. ZICHICHI
26th Course:  The Super-World-III 
7 - 15 August 1988  \cite{Classical5}
\subsection{Conformal brane-scan:  predicts D3 and M5 in addition to M2}
\label{conformal}
 \begin{figure}[h]
\centering\includegraphics[scale=0.6 , angle =270]{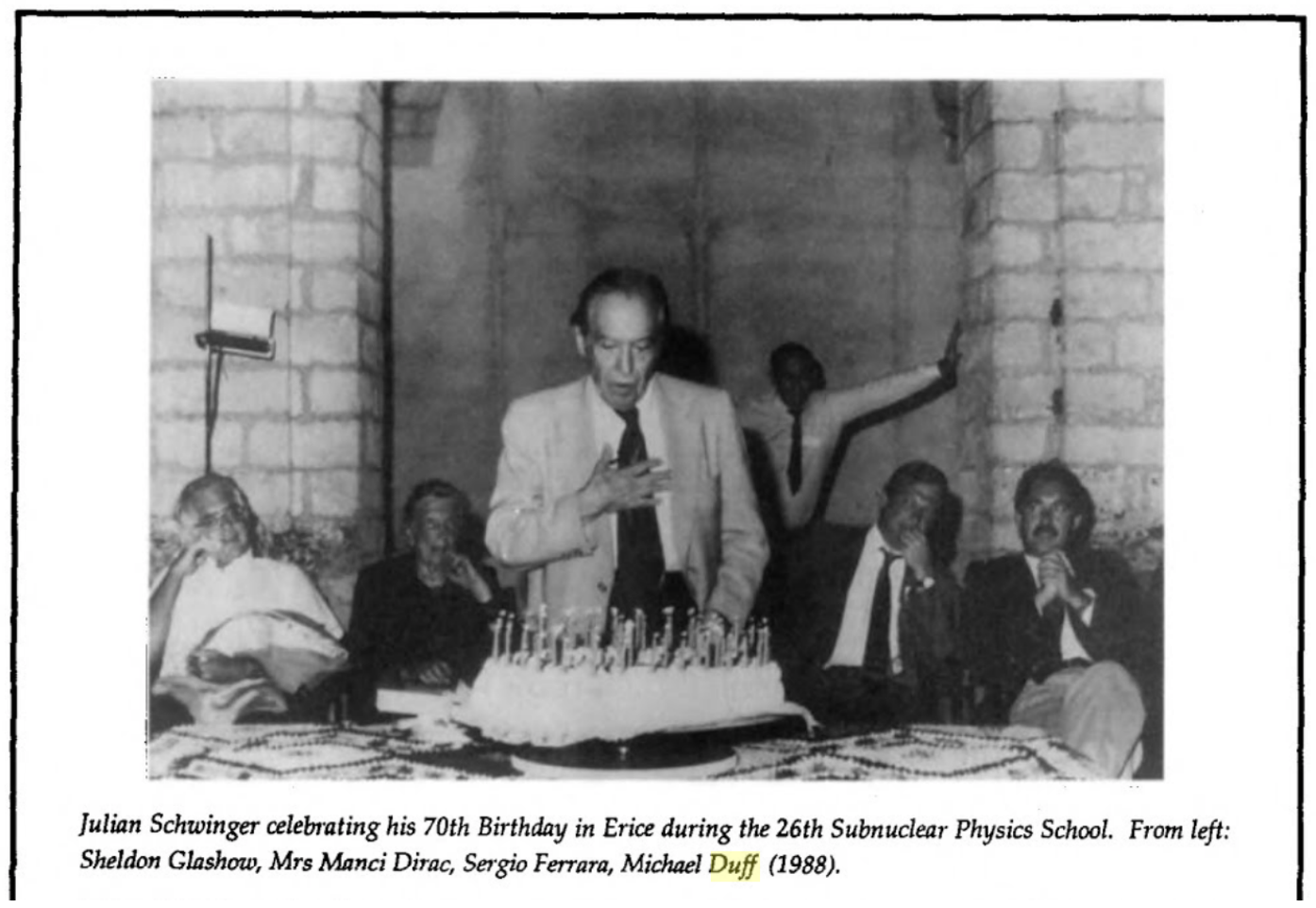}
\caption{In distinguished company}
\label{schwinger}
\end{figure}
%
%\subsection{Conformal brane-scan:  predicts D3 and M5 in addition to M2}
%\label{conformal}

In 1987 two versions of the brane-scan of $D$-dimensional super $p$-branes were put forward. The first by Achucarro, Evans, Townsend and Wiltshire \cite{Achucarro} pinpointed those twelve $(p,D)$ slots consistent with kappa-symmetric Green-Schwarz \cite{Greenschwarz} type actions for $p\geq 1$ . The results are the  slots labelled $s$ shown in Table \ref{brane-scan}.  Moving diagonally down the brane-scan corresponds to a simultaneous dimensional reduction of spacetime and worldvolume \cite{Duff:1987qa}.  Of course some of these $D$ dimensions could be compactified, in which case the double dimensional reduction may be interpreted as wrapping the brane around the compactified directions.  
\begin{table}
\begin{center}
\begin{tabular}{lcl}
\hline
Supergroup&AdS Dimension&$p$\\
&&\\
\hline
$OSp(p|2) \times OSp(q|2)$&3& 1 \\
$OSp(N|4)$& 4& 2 \\
$SU(2,2|N)$&5&3\\
$F(4)$&6&4 \\
$OSp(6,2|N)$&7&5\\
\hline
\end{tabular}
\caption{Supergroups admitting $p$-branes on the boundary of $AdS_{p+2} \times S^{D-p-2}$.}
\end{center}
\label{confbs}
\end{table}

The second brane-scan by Blencowe and the author \cite{Blencowe} generalized the membrane at the end of the universe idea [5, 6] to arbitrary $p$-branes with $p \geq 1$ by selecting those superconformal groups in Nahm's classification \cite{Nahm} with bosonic subgroups $SO(p+1, 2) \times SO(D-p-1) $ describing $p$-branes on the boundary of $AdS_{p+2} \times SO(D-p-2)$, as shown in Table \ref{conformal}. In each case the boundary CFT is described by the corresponding singleton (scalar), doubleton (scalar or vector) or tripleton (scalar or tensor) supermultiplet. The supersingleton lagrangian and transformation rules were also spelled out explicitly in this paper.  Interesting special cases of the conformal brane-scan of Table \ref{conformal} are $(p=2, D=11)$, $(p=3, D=10)$ and ($p=5, D=11$) which we now recognise as the M2, D3 and M5 branes. Although the M2 was known, this was the first appearance of D3 and M5. See Table \ref{confbs}.  As Steven Weinberg once remarked., the problem with theoretical physicists is not that they take themselves too seriously but that they don't take themselves seriously enough.

\begin{table}
\begin{center}
\begin{tabular}{lcl}
\hline
Supergroup&Spacetime&Brane\\
&&\\
\hline
$OSp(8|4)$& $AdS_4 \times S^7$& {\bf M2} \\
$SU(2,2|4)$&$AdS_5 \times S^5$&{\bf D3}\\
$OSp(6,2|4)$&$AdS_7 \times S^4$ &{\bf M5}\\
\hline
\end{tabular}
\caption{The conformal brane-scan predicts D3 and M5 in addition to M2} \end{center}
\label{253}
\end{table}
%\item
\subsection{Supermembranes and the signature of spacetime}
\label{signature}

If our senses are to be trusted, we live in a world with three space and
one time dimensions.  However, the revival of the Kaluza-Klein idea
\cite{Duff1}, brought about by supergravity, superstrings and M-theory, has
warned us that this may be only an illusion.  In any case, there is a
hope, so far unfulfilled, that the four-dimensional structure that we
apparently observe may actually be predicted by a {\it Theory of
Everything}.  Whatever the outcome, imagining a world with an arbitrary
number of space dimensions has certainly taught us a good deal about the
properties of our three-space-dimensional world.

In spite of all this activity, and in spite of the popularity of Euclidean
formulations of field theory, relatively little effort has been devoted to
imagining a work with more than one {\it time} dimension.  This
is no doubt due partially to the psychological difficulties we have in
treating space and time on the same footing.  As H. G. Wells reminds us in
{\it The Time Machine}: ``There is, however, a tendency to draw an unreal
distinction between the former three dimensions and the latter, because it
happens that our consciousness moves intermittently in one direction along
the latter from the beginning to the end of our lives."  There are also
more justifiable reasons associated with causality.  Nevertheless, one
might hope that a theory of everything should predict not only the
{\it dimensionality} of spacetime, but also its {\it signature}.

For example, quantum consistency of the superstring requires $10$ spacetime
dimensions, but not necessarily the usual $(9, 1)$ signature.  The
signature is not completely arbitrary, however, since spacetime
supersymmetry allows only $(9,1)$, $(5,5)$ or $(1,9)$.  Unfortunately,
superstrings have as yet no answer to the question of why our universe
appears to be four-dimensional, let alone why it appears to have
signature $(3,1)$.

In this 1989 lecture therefore I considered a world with an
arbitrary number $T$ of time dimensions and an arbitrary number $S$ of space
dimensions to see how far classical supermembranes restrict not only $S + T$
but $S$ and $T$ separately.  To this end I also allowed an $(s,t)$ signature
for the worldvolume of the membrane where $s \leq S$ and $t \leq T$ but are
otherwise arbitrary.  It is not difficult to show that there is once again a
matching of the bosonic and fermionic degrees of freedom as a consequence of
the kappa symmetry.  However severe constraints on possible supermembrane
theories will now follow by demanding spacetime supersymmetry \cite{Blencowe}.
\begin{table}
$
\begin{array}{cccccccccccccc}
S\uparrow&&&&&&&&&&&~\\
11&.&&&&&&&&&&&~\\
10&.&O&\tilde O&&&&&&&&&~\\
9&H&H/O&O&&&&&&&&&~\\
8&\tilde C&H&&&&&&&&&&~\\
7&\tilde C&H&&&&&&&&&&~\\
6&\tilde C/\tilde H&H&&&&O&\tilde O&&&&~\\
5&C&C/H&H&H&H&H/O&O&&&&&~\\
4&\tilde C&C&\tilde C&\tilde C&\tilde C&H&&&&&&~\\
3&.&R/C&\tilde R&\tilde R&\tilde C&H&&&&&&~\\
2&.&R&~&\tilde R&\tilde C&H&~&~&~&O&\tilde O&~\\
1&.&~&R&R/C&C&C/H&H&H&H&H/O&O&~\\
0&.&.&.&.&\tilde C&C&\tilde C/\tilde H&\tilde C&\tilde C&H&.&.~\\
~&0&1&2&3&4&5&6&7&8&9&10&11&T\rightarrow
\end{array}
$
\caption{The brane molecule}
\label{molecule}
\end{table}
The results are summarized by the {\it brane-molecule}
of Table \ref{molecule}.

Several comments are now in order:

\noindent
1) We see from Table \ref{molecule} that for every supermembrane with $(S,T)$
signature, there is  another with $(T,S)$.  Note the self-conjugate theories
that lie on the  $S = T$ line which passes through the $(5,5)$ superstring.

\noindent
2)~~There is, as yet, no restriction on the worldvolume signatures beyond
the original requirement that $s\leq S$ and $t\leq T$.

\noindent
3)~~If we were to redraw the $D/d$ brane-scan of Table \ref{brane-scan}
allowing now arbitrary signature, there would be no new $s$ points on the
plot, but rather the new solutions would be superimposed on the old ones. 
For example, there would now be six solutions occupying the $(d = 3, D =
11)$ slot instead of one. 

\noindent
4)~~Perhaps the most interesting aspect of the brane-molecule is the mod $8$
periodicity.  Suppose there exist signatures $(s,t)$ and $(S,T)$ which satisfy
both the requirements of bose-fermi matching and super-Poincare invariance. 
Now consider $(s',t')$ and $(S',T')$ for which

\begin{equation}
s'+t' = s +t~~~~
S'+T' = S+T
\end{equation}

As a consequence of the modulo $8$ periodicity theorem for real Clifford
algebras, the minimal condition on a spinor is modulo $8$ periodic 
e.g. $S - T =0$ mod $8$ for Majorana-Weyl.  So if, in addition we also have
\begin{equation}
S'-T'=S-T+8n~~~~n\epsilon Z~~~.
\end{equation}
then $(s',t')$ and $(S',T')$ satisfy bose-fermi matching.  (1.52) and
(1.53) imply
\begin{eqnarray}
S'&=& S + 4n\\
T'&=&T - 4n~~~.
\end{eqnarray}

Similarly, the membrane at the end of the universe admits a corresponding generalization to brane worldvolumes with $s$ space and $t $ time dimensions moving in a spacetime with $S \geq s$ space and $T \geq t$ time dimensions. The brane occupies the boundary of a universe of constant curvature so that the bosonic symmetry is $O(s + 1, t + 1) \times O(S- s, T -t)$. Supersymmetry restricts the values of $s, t, S, T$ to those for which this bosonic symmetry is a subgroup of a superconformal group, and the resulting superconformal theories have $(s + t) \leq 6$. For example, the possible signatures of M-theory are $(10,1), (9,2), (6,5), (5,6), (2,9), (1,10)$ and the possible $M2$-branes have worldvolume signatures $(3,0), (2,1), (1,2), (0,3)$.
\subsection{D$=$12?}
It is interesting to ask whether we have exhausted all possible 
theories of extended objects with Green-Schwarz type actions.  We demanded super-Poincare invariance but might there exist others for which the supergroup is not necessarily super-Poincare? Although the possibilities are richer, there are still severe constraints. Note, in particular, that the maximum spacetime dimension is now $D=12$ provided we have signatures $(10, 2)$, $(6, 6)$ or $(2, 10)$. These new cases are particularly interesting since they admit Majorana-Weyl spinors. In fact, twelve-dimensional supersymmetry algebras have been discussed before in the supergravity literature \cite{vanvan}. The RHS of the ${Q,Q} $ anticommutator yields not only a Lorentz generator but also a six index object so it is certainly not
super-Poincare. We conjectured (together with C. Hull and K. Stelle) that the (2, 2) extended object moving in $(10, 2)$ spacetime may (if it exists) be related by simultaneous dimensional reduction \cite{Howe} to the $(1, 1)$ Type IIB superstring in $(9,1)$.
\subsection{Area-preserving diffeomorphisms: Matrix models}

In string theory, the light cone gauge is convenient for quantization because it allows the elimination of all unphysical degrees of freedom and unitarity is guaranteed. Of course, one loses manifest Lorentz invariance
and one must be careful to check that it is not destroyed by quantization. In membrane theory, however, the lightcone gauge does not eliminate all unphysical degrees of freedom. Let us split
\[
X^{\mu}=  (X^{\pm},X^I)~~~~I=1,2, ...(D-2)
\]
\be
X^{\pm}=\frac{1}{\sqrt{2}}(X^0 \pm X^{D-1})
\ee
One can then solve for $X^-$ leaving the $( D - 2 )$ variables $X ^I$ . For membranes, however, only $( D- d)$ variables are physical. Thus the light-cone gauge must leave a residual gauge invariance \cite{Hoppe1,Hoppe2}.  This group is, in fact, the subgroup of the worldvolume diffeomorphism group that preserves the Lie bracket 
\be
\{f,g\}=\epsilon^{ab}\partial_af \partial_b g
\ee
 and is known as the group of area-preserving diffeomorphisms. For spherical membranes, this group is given  by $\lim_{N \rightarrow \infty} SU(N)$. Let us focus our attention on a $d = 3$ supermembrane in flat spacetime.  The light-cone action turns out to be
\be
S=\frac{1}{2}\int d\tau tr\{(D_0 A^I)^2-\frac{1}{2}[A^I,A^J][A^I,A^J]+i {\bar \lambda} D_0 \lambda+i{\bar \lambda}\gamma^I [A^I, \lambda]\}
\ee
where the fields are all in the adjoint representation of $SU(\infty)$. Remarkably, this looks like a $( D- 1)$-dimensional super-Yang-Mills theory dimensionally reduced to one time dimension.

One can generalise these results to the $d = 3$ supermembranes \cite{deWit1} in $D = 4, 5, 7$ and $11$, and one finds super-Yang-Mills quantum mechanical models corresponding to the dimensional reduction of super-Yang-Mills in $D = 3,4,6 $ and $10$, which provides yet another way of understanding the allowed values of $D$. (One might conjecture a similar relationship between the $d > 3$ membranes and quantum mechanical models, but this time the gauge symmetry could not be of the Yang-Mills type. It has been suggested \cite{{Bergshoeff1}} that they are given by infinite-dimensional non-Abelian antisymmetric tensor gauge theories.)

\subsection{Subsequent developments}
\begin{itemize}
\item{Black branes}

Following the {\it electric} M2-brane solution of $D=11$ supergravity \cite{Duffstelle}, the dual {\it magnetic} M5-brane solution was found by Gueven \cite{Gueven}.  The black $p$-brane solution of Types IIA and IIB supergravity were found by Horowitz and Strominger \cite{Horowitz1} and the extremal cases  were proven to be supersymmetric (1/2 BPS) in \cite{Luthree,Luscan}.  In his now famous paper, Polchinski \cite{Polchinski} provided an alternative derivation as Dirichlet-branes on which open strings may end.

\item{ Branes and the signature of spacetime}

Branes in exotic signatures were further studied by Hull \cite{Hull98}, Hull and Khuri \cite{Hull98c}, Batrachenko, Duff and Lu \cite{Bat}, Duff and Kalakinin \cite{kal2,kal1}.  Note. however, that, signature reversal $(S,T) \rightarrow (T,S)$ in general yields a different theory. The conditions for reversal invariance for both supergravities and branes are spelled out in  \cite{kal2,kal1}. A necesary but not sufficient conditon is that the Clifford algebra obey Cliff(S,T)$=$Cliff(T,S) which requires $S-T=0 ~mod ~4$.  Physics with more than one time has also been pursued by Bars \cite{Bars}.  Negative branes, supergroups, and the signature of spacetime was the subject of a recent paper by Dijkgraaf, Heidenreich, Jefferson and Vafa \cite{vdhj}.  A supergravity lagrangian in $(10,2)$ was recently proposed by Castellani \cite{Castellani:2017vbi}. 

\item{F-theory}

The idea of a 12-dimensional world was revived by Vafa in
the context of {\it $F$-theory} \cite{Vafa}, which involves Type $IIB$
compactification where the axion from the R-R sector
and dilaton from the NS-NS sector are allowed to vary on the internal manifold.  Given a manifold $M$ that has
the structure of a fiber bundle whose fiber is $T^2$ and whose base is some
manifold $B$, then  \be
F~ on ~M \equiv ~Type~IIB~on~B      
\ee

\item {Matrix models}

The $SU(\infty)$ Yang-Mills description of M2-branes was revived by Banks, Fischler, Shenker and Susskind in the matrix-model interpretation of M-theory \cite{BanksM}, which has received some recent attention by Maldacena and Milekhin \cite{Maldacena1}.
\item{Holographic duals}

Although the D3 worldvolume theory on the boundary of $AdS_5 \times S^5$ is the well-known $N=4$ Yang-Mills, the holographic duals of M-theory on $AdS_4 \times S^7$ and $AdS_7 \times S^4$ are more obscure.  ABJM (Aharony, Bergman, Jafferis, Maldacena) theory \cite{ABJM} is the favorite candidate for $M2$ but the $M5$ case is a $(2,0)$ CFT not describable by a lagrangian field theory. See also \cite{BL}. In this context and in the context of exotic signatures, it is worth bearing in mind that the (9,2) version of M-theory admits a doubly holographic  $AdS_4 \times AdS_7$ solution. This may be regarded either as a stack of M2 branes with conformal group SO(4,2) and R-symmetry SO(6,2), or as a stack of M5 branes for which the conformal and R-symmetries are interchanged \cite{Hull98b,Hull98c,Bat}. Perhaps the ABJM of one can throw light on the $(2,0)$ CFT of the other.

\end{itemize}
\section{\bf 1990 Symmetries of Extended Objects }
\label{objects}
INTERNATIONAL SCHOOL OF SUBNUCLEAR PHYSICS - Director: A. ZICHICHI
28th Course:  Physics up to 200 TeV 
16 - 24 July 1990 \cite{Duff90}

\subsection{ T-duality and double geometry: $Z^M=(x^\mu, y_\alpha)$}
%*\subsubsection*{Strings, T-duality and Double Geometry}
\indent
In this 1990 lecture, based on an earlier  paper \cite {Duffdual},  I pointed out that strings moving in an $n$-dimensional 
space $M^n$ with coordinates $X^{\mu}(\tau, \sigma)$, background metric $g_{\mu\nu}(X)$ and 
2-form $b_{\mu\nu}(X)$, could usefully be described by a doubled geometry with $2n$-dimensional 
coordinates
\be
Z^M=(X^{\mu},Y_{\sigma})
\ee
and doubled metric\footnote{$G_{MN}$ had previously appeared in \cite{Giveon} with a different 
physical interpretation as a metric on phase space.} 
%
%\be
%G_{MN}= \begin{pmatrix}
%g_{\mu\nu}-b_{\mu\alpha}g^{\alpha\beta}b_{\beta\nu}& b_{\mu\alpha}g^{\alpha\beta}\\
%-g^{\alpha\beta}b_{\beta\nu}&g^{\alpha\beta}\end{pmatrix}.
% \ee  
%
 \be
G_{MN}= \left(
\begin{array}{cc}
g_{\mu\nu}-b_{\mu\rho}\,g^{\rho\sigma}b_{\sigma\nu}& b_{\mu\rho}\,g^{\rho\sigma}\\
-g^{\mu\sigma}b_{\sigma\nu}&g^{\mu\nu}
\end{array}\,.
\right)
 \ee  
The motivation was twofold; worldsheet and spacetime:
\begin{enumerate}
\item{\bf Worldsheet}

 In the case when $M^n$ is the $n$-torus $T^n$, this renders manifest 
the $O(n,n)$ T-duality by combining worldsheet field equations and 
Bianchi identities via the constraint
\be
\Omega_{MN}\epsilon^{ij}\partial_{j}Z^N=
G_{MN}\sqrt{-\gamma}\gamma^{ij}\partial_{j}Z^N\,,
\label{dcon}
\ee
where
\be
\Omega_{MN}=\left(
\begin{array}{cc}
 0 & \delta_{\mu}{}^{\beta} \\
\delta^{\alpha}{}_{\nu} &0
\end{array}
\right)\,,
\ee
and $\gamma_{ij}$ is the worldsheet metric.  In components
\[
\epsilon^{ij}\partial_j Y_\nu=\sqrt{-\gamma}\gamma^{ij}\partial_{j}X^\mu g_{\mu\nu}+\epsilon^{ij}\partial_j X^{\mu }b_{\mu\nu}
\]

\[
\epsilon^{ij}\partial_j X^\nu=\sqrt{-\gamma}\gamma^{ij}\partial_{j}Y_\mu p^{\mu\nu}+\epsilon^{ij}\partial_j Y_{\nu }q^{\mu\nu}
\]
Here $Y_\nu(\tau,\sigma)$ are the coordinates of the T-dual string which interchanges field equations and bianchi identities. Its  background metric and 2-form are $p^{\mu\nu}(X)$ and $q^{\mu\nu}(X)$ where

\be
g=p^{-1}(1-qb)~~~b=-p^{-1}qg
\ee
\be
p^{-1}=g-bg^{-1}b~~~g^{-1}b=-pq^{-1}
\ee
so
\be
(p\pm q)(g\pm b)=1
\ee

An earlier alternative suggestion \cite{Duffhidden} was to use the non-symmetric metric $g_{\mu\nu}+b_{\mu\nu}$. The two alternatives are related by the two-vielbein (left L and right R) approach \cite {Siegeltwo}. 
\be
\left(
\begin{array}{cc}
e_\mu{}^a(L) e_\nu{}_a(R) & e_\mu{}^a(L) e^\nu{}_a(R)\\
e^\mu{}^a(L) e_\nu{}_a(R) &e^\mu{}^a(L) e^\nu{}_a(R)
\end{array}
\right)
=
\left(
\begin{array}{cc}
g_{\mu\nu}+b_{\mu\nu}& \delta_\mu{}^\nu\\
\delta^\mu{}_\nu&p^{\mu\nu}+q^{\mu\nu}
\end{array}
\right)
\ee
\medskip

\item{\bf Spacetime}

 In the case when $M^n$ is a generic manifold,  
the $2n$-dimensional diffeomorphisms with parameter 
$\xi^M=(\xi^{\mu}, \lambda_{\alpha})$
suggest a way of unifying $n$-dimensional diffeomorphisms 
\be
\delta g_{\mu\nu}=-\partial_{\mu}\xi^{\rho} g_{\rho\nu}-\partial_{\nu} 
\xi^{\rho} g_{\mu\rho}-\partial_{\rho}g_{\mu\nu}\xi^{\rho} \,,
\ee
and 2-form gauge invariance
\be
\delta B_{\mu\nu}=\partial_{\mu}\lambda_{\nu}-\partial_{\nu}\lambda_{\mu}\,.
\ee
After all, $G_{MN}$ is just the Kaluza-Klein metric with 
spacetime metric $g_{\mu\nu}$,
gauge field $A_{\mu}{}^a$ and internal metric $g_{ab}$
%\be
%G_{MN}= \begin{pmatrix}
%g_{\mu\nu}+A_{\mu}{}^{a}g_{ab}A_{\nu}{}^b& A_{\mu}{}^{a}g_{ab}\\
%g_{ab}A_{\nu}{}^b&g_{ab}\end{pmatrix}.
 %\ee  
 % \ee  
 \be
G_{MN}= \left(
\begin{array}{cc}
g_{\mu\nu}+A_{\mu}{}^{a}g_{ab}A_{\nu}{}^b& A_{\mu}{}^{a}g_{ab}\\
g_{ab}A_{\nu}{}^b&g_{ab}
\end{array}
\right)\,,
 \ee  
where the ``gauge field'' is $B_{\mu\alpha}$ and the ``internal'' 
metric is $g^{\alpha\beta}$.  If this programme were successful one 
would expect the $SL(n)/SO(n)$ coset of general relativity to be promoted 
to an $O(n,n) /(SO(n) \times SO(n))$, as conjectured in 
\cite{Duff:1985bv,Duffhidden}.
\end{enumerate} 

%
%for example \cite{Giveon:1994fu,Hull:2006va,Hull:2009mi,Hohm:2010jy,Hohm:2010pp,
%Thompson:2011uw,:2011jh,Grana:2012rr,West:2012qz,Hohm:2012gk,
%Hatsuda:2012vm,Hatsuda:2013dya}.
\subsection{ U-duality and ``exceptional'' geometry: $Z^M=(x^\mu,y_{\alpha\beta}, ...)$}
%\subsubsection*{Branes, U-duality and M-theory}
 \begin{figure}[h]
\centering
\includegraphics[scale=0.4]{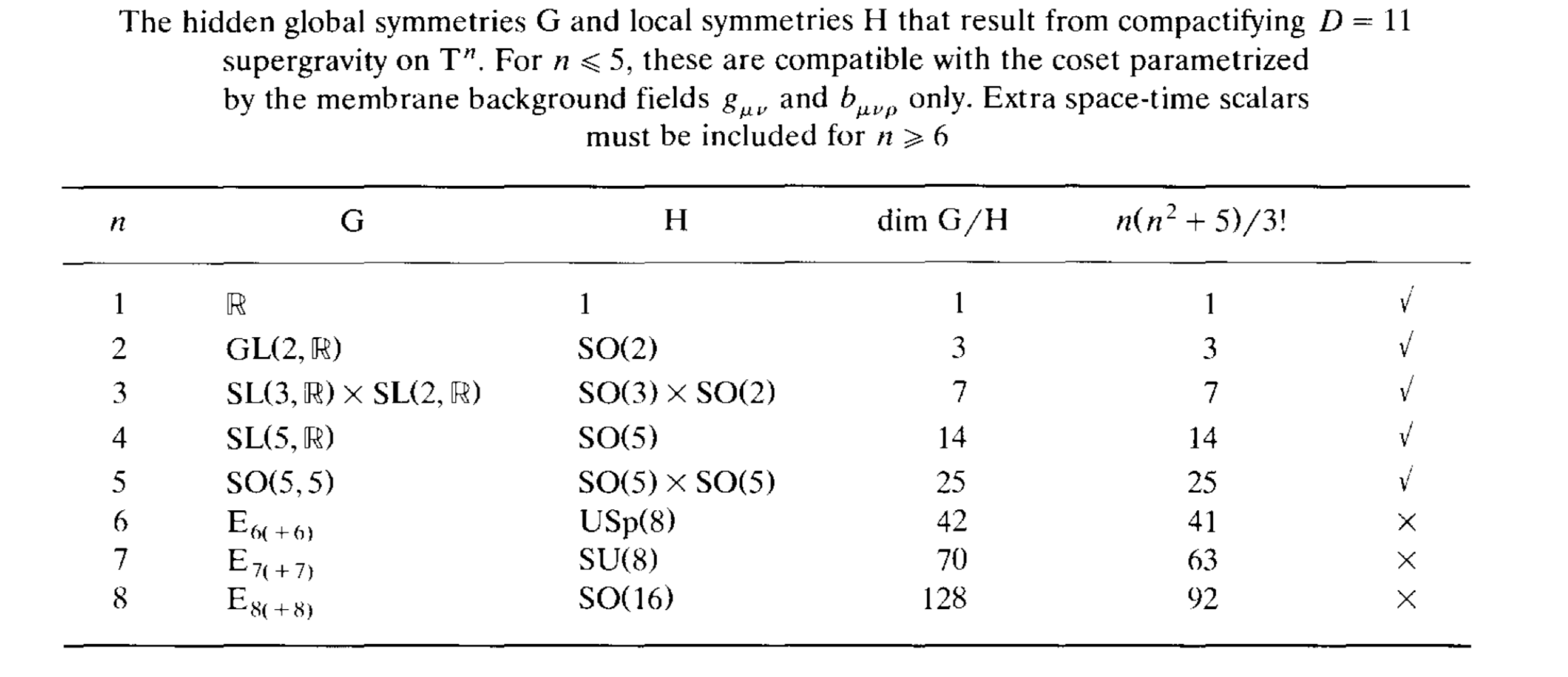}
\caption{U-dualities and branes}
\label{u.jpeg}
\end{figure}

\indent
Similarly, based on \cite{Luduality2,Luduality3}, I pointed out  that 
membranes moving in a ($n\leq 4$)-dimensional space $M^n$ with coordinates $X^\mu(\tau,\sigma,\rho)$, 
background metric $g_{\mu\nu}(X)$ and 3-form $B_{\mu\nu\rho}(X)$ could usefully be described by a geometry 
with $[n+n(n-1)/2]$-dimensional coordinates
\be
Z^M=(X^{\mu},Y_{\rho\sigma})
\label{a}
\ee
and generalized metric
\be
G_{MN}= 
\left(
\begin{array}{cc}
g_{\mu\nu}+b_{\mu\rho\sigma}\,g^{\rho\sigma\lambda\tau}b_{\lambda\tau\nu}& 
b_{\mu\rho\sigma}\,g^{\rho\sigma\lambda\tau}\\
\,g^{\mu\nu\rho\sigma} \, b_{\rho\sigma\nu}&
g^{\mu\nu\rho\sigma}\\
\end{array}
\right)\,,
\label{b}
 \ee  
  
where
\be
g^{\alpha\beta\gamma\delta}=\frac{1}{2}(g^{\alpha\gamma}g^{\beta\delta} 
-g^{\alpha\delta}g^{\beta\gamma} )\,.
\ee
\begin{figure}[h]
\centering
\includegraphics[scale=0.4]{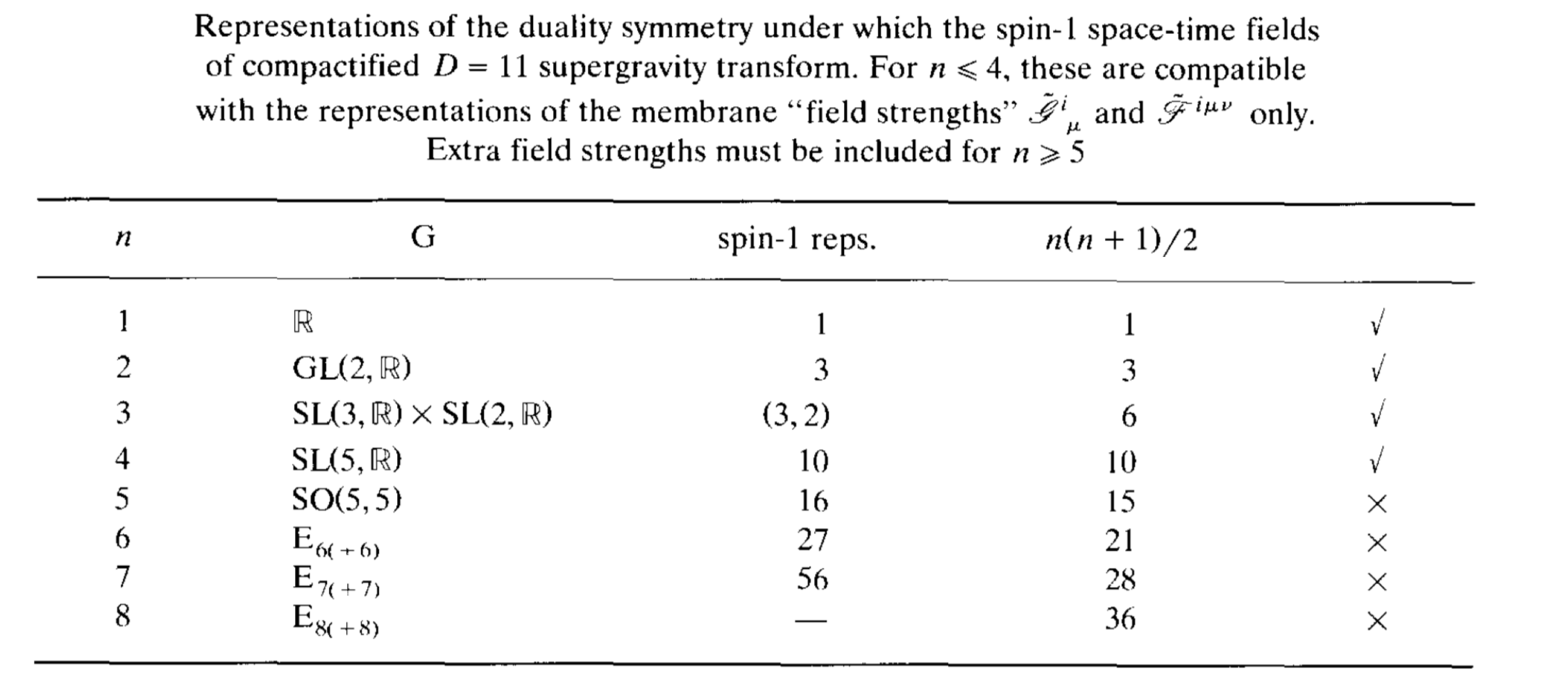}
\caption{U-dualities and branes}
\label{u.jpeg}
\end{figure}

Once again, the motivation was twofold; worldvolume and spacetime:
\begin{enumerate}
\item{\bf Worldvolume}

In the case when $M^n$ is the $n$-torus $T^n$, the hope was to render 
manifest the M-theory U-dualities (using modern parlance) by combining 
worldvolume field equations and Bianchi identities. For example, the 
U-duality would be $SL(5,R)$ in the case $n=4$. The restriction to 
$n\leq4$ arises because, just as the usual coordinates $X^\mu$ correspond 
to momentum in the supersymmetry algebra, so the extra coordinates 
$Y_{\mu\nu}$ correspond to the M2 central charge. But for $n \geq 5$, 
this is not enough, as shown in Fig.\ref{u.jpeg}. There is 
also the M5 central charge with corresponding coordinates 
$Y_{\mu\nu\rho\sigma\tau}$, which first appears in $n=5$. 
In general there are extra coordinates for all
central charges in the M-theory algebra for general $D$. For example, 
in the $n=7$ case  $X^\mu$, $Y_{\mu\nu}$, 
${\tilde Y}^{\mu\nu}\sim \epsilon^{\mu\nu\rho\sigma\tau\lambda\kappa}Y_{\rho\sigma\tau\lambda\kappa}$ 
and $\tilde X^\mu$ form a 56 of the U-duality symmetry $E_{7(7)}$. 

\item{\bf Spacetime}

If this programme were successful, one would expect the 
$SL(n)/SO(n)$ of general relativity to be promoted not merely to  
$O(n,n) /(SO(n) \times SO(n))$ but to $E_8/SO(16)$, with possible 
infinite-dimensional extensions involving $E_9$, $E_{10}$ 
as conjectured in \cite{Duff:1985bv,Duffhidden}
\end{enumerate} 
%\end{itemize}
\subsection{Subsequent developments}
\begin{itemize}
\item{Dual variables}

The variables $p$ and $q$ re-appear in the literature on noncommutative geometry \cite{noncom},  in non-geometric flux and $\beta$-supergravity where they are known as $ \tilde g$ and $\beta$ \cite{Andriot,Condescu,Cezar} and in Yang-Baxter equations \cite{yb}.

\item{Double field theory}

The worldsheet goal of rendering manifest the string 
T-duality $O(n,n)$ by doubling the coordinates was achieved successfully 
in \cite {Duffdual} and a T-dual worldsheet action using the 
doubled coordinates was then constructed in \cite{Tseytlin}.
There were missing ingredients in the spacetime approach:  
The generalized diffeomorphisms were subsequently supplied in 
\cite{Siegeltwo,Siegeltwo'}
\be
\delta G_{MN}=\xi^P\partial_PG_{MN}+(\partial_M\xi^P-
\partial^P\xi_M)G_{PN}+(\partial_N\xi^P-\partial^P\xi_N)G_{MP}\,,
\ee
and the section condition subsequently supplied in \cite{Hull1}
\be
\Omega^{MN}\partial_M\partial_N=0\,.
\ee
(The need for the section condition has, however, 
been called into question \cite{Grana,}.)  
Once these ingredients were included, it was possible also to 
build a generalised spacetime action for $G_{MN}$. This activity came to 
be known as ``Double Field Theory.'' \cite{Hull1,Hull2,Hull3,Hohm1,Hohm2}

\item{Anachronisms}

Although it is quite common for papers on Double Geometry to begin with its history, they often contain anachronisms, in the sense that results of paper A are said to extend those of paper B even though A preceded B. In particular, the chronology of the papers by Duff in August 1989 \cite{Duffdual}, by Tseytlin in February 1990 \cite{Tseytlin} and June 1990 \cite{Tseytlin'} and by Siegel in February 1993 \cite{Siegeltwo}, May 1993 \cite{Siegeltwo'} and August 1993 \cite{Siegeltwo''} is frequently reversed.
For example in  \cite{WWY} reference \cite{Duffdual} is described as a ``descendant'' of \cite{Tseytlin}, in the first version of \cite{BBMP} as an ``elaboration'' of  \cite{Tseytlin,Tseytlin'}, in \cite{Berman} as a ``development'' of  \cite{Siegeltwo,Siegeltwo'} and in \cite{Polyakov:2015wna} as ``recent''. The fact that \cite{Duffdual} is pre-arXiv may be a contributing factor to the ordering ambiguities.

\item{Exceptional field theory}

Similarly with the supermembrane, the generalised diffeomorphisms, section conditions and 
U-invariant actions came later. These activities involved  
 {\it Generalized Geometry} \cite{Coimbra}, {\it Exceptional Field Theory} \cite{Hull1,Hull2,Hull3,Hohm1,Hohm2,Blair,Deser,Hatsuda,Hohm3}
 and $E_{11}$ \cite{Weste11}.  For subsequent developments and variations 
on generalized geometry in M-theory and U-duality see, for example, 
\cite{Waldram,Hohm1,Hohm2,Berman1,Berman2,Hohm3} where the 5-brane and other extended objects were incorporated, as required for $n>4$. 
The $E_{11}$ approach \cite{Weste11} goes further with infinitely many coordinates 
of which those associated with the  M-theory central charges are but a subset.
	  
In summary, in contrast with strings where both the worldsheet and 
spacetime approaches have been successful, the brane worldvolume 
approach seems problematical \cite{Percacci,Sen95,Lukas97,perc} and, with the exception of 
\cite{Berman1,Hatsuda,perc}, recent developments have tended to 
focus on the spacetime approach where the extra coordinates (\ref{a}) and 
generalised metric (\ref{b}) have proved valuable. 
In any event, the need to include coordinates corresponding to central 
charges in the M-theory algebra exposes a major difference between 
U-duality in M-theory and T-duality in string theory. In string theory, 
T-duality takes strings into strings, but in M-theory U-duality mixes up 
$p$-branes with different $p$. It seems unlikely, therefore, that the 
M2-brane worldvolume alone is sufficient. Somehow the totality of $p$-brane 
worldvolumes must conspire to give the full U-duality. This remains an 
unsolved problem.

%Finally we note that the purpose of extra coordinates in both string and 
%M-theory is to render the T and U dualities manifest. If one is content 
%with non-manifest T-duality, one may invoke the Gaillard-Zumino (GZ) 
%approach \cite{Gaillard:1981rj}, as was done in \cite{Cecotti:1988zz}. 
%tThe GZ approach to U duality is discussed below.
\end{itemize}
%\begin{figure}[h]
%\centering
%\includegraphics[scale=0.4]{branes3.png}
%\caption{U-dualities and branes}
%\label{braneoed.jpeg}
%\end{figure}

\section{\bf 1991 A Duality between Strings and 5-branes }

INTERNATIONAL SCHOOL OF SUBNUCLEAR PHYSICS - Director:  A. ZICHICHI
29th Course:  Physics at the Highest Energy and Luminosity:  to Understand the Origin of Mass 
14 - 22 July 1991 \cite{Duff:1991tg}

%\subsection{String/5-brane duality}
\subsection{Elementary v solitonic branes}
The next development came when Townsend \cite{Townsend:1987yy} pointed out that
all the points on the ${\cal H,C,R}$
branescan sequences correspond
to topological defects of some globally supersymmetric field
theory which
break half the spacetime supersymmetries. He conjectured
that the
$p$-branes in the ${\cal O}$ sequence would also admit such a
solitonic
interpretation within the context of supergravity. 
The first hint in this direction came from  Dabholkar
{\it et al.} \cite{{Dabholkar}},
who presented a multi-string solution which in $D=10$ indeed
 breaks half the
supersymmetries. They obtained the solution by solving the
low-energy $3$-form supergravity equations of motion coupled to
a string
$\sigma$-model source and demonstrated that it saturated a
Bogomol'nyi bound
and satisfied an associated zero-force condition, these
properties being
intimately connected with the existence of unbroken spacetime
supersymmetry.  However, this $D=10$ string was clearly not the soliton
anticipated by
Townsend because it described a singular configuration with a
$\delta$-function source
at the string location. Moreover, its charge per unit length
 $e_2$ was an
``electric'' Noether charge associated with the equation of
 motion of the
antisymmetric tensor field rather than a ``magnetic''
topological charge
associated with the Bianchi identities. Consequently, in the
current
literature on the subject, this solution is now referred to as
the
``fundamental'' or ``elementary'' string. 

Similarly, the
supermembrane
solution of $D=11$ supergravity found in \cite{{Duffstelle}} was not
solitonic either
 because it was also obtained by coupling to a membrane
$\sigma$-model source. Curiously, however, the
curvature computed from its $\sigma$-model metric is finite at
the
location of the source, in contrast to
the case of the elementary string.

The next major breakthrough for $p$-branes as solitons came
 with the paper of
Strominger \cite{Strominger1}, who showed that $D=10$ supergravity coupled to
 super
Yang-Mills (without a $\sigma$-model source), which is the
field theory
limit
of the heterotic string \cite{Gross} , admits as a solution the
heterotic fivebrane.
In contrast to the elementary string, this fivebrane is a
genuine soliton,
being everywhere nonsingular and carrying a topological
magnetic charge $g_6$.
A crucial part of the construction was a Yang-Mills instanton
in the four
directions transverse to the fivebrane. He went on to suggest a
complete
strong/weak coupling duality with the strongly coupled string
corresponding
to the weakly coupled fivebrane and vice-versa, thus providing
a solitonic
interpretation of the string/fivebrane duality conjecture. In
this form,
string/fivebrane duality is in a certain sense an analog of the
Montonen-Olive \cite{Montonen} according to which the
magnetic monopole
states of four-dimensional spontaneously broken supersymmetric
Yang-Mills
theories may be viewed from a dual perspective as fundamental
in their own
right and in which the roles of the elementary and solitonic
states are
interchanged.

This strong/weak coupling theme was further developed in \cite {Luremarks} which also established a Dirac quantization rule
\be
{\kappa^2 T_2T_6=n\pi, \qquad\qquad n={\rm integer}}
\ee
relating the fivebrane tension $T_6$ to the string tension
$T_2$, which
 followed from the corresponding rule for the electric and
 magnetic charges
generalized to extended objects
\cite{Nepomechie,Teitelboim} $e_2 g_6=2n\pi$.

For the purposes of generalizing the Dirac quantization rule
for extended objects, we recall that just as a charged particle couples to an Abelian vector
potential $A_M$
displays a gauge invariance
\be{A_M \rightarrow A_M + \partial_M \Lambda}
\ee
and has a gauge invariant field strength
\be
{F_{MN} = 2\partial_{[M} A_{N]} \equiv \partial_M A_N
- \partial_N A_M
,}
\ee
a string couples to a rank-2 antisymmetric tensor potential
$A_{MN} = - A_{NM}$
with a gauge invariance
\be
A_{MN} \rightarrow A_{MN} + \partial_{[M}
\Lambda_{N]} ,
\ee
and field strength
\be
F_{MNP} = 3\partial_{[M} A_{NP]}.
\ee
In general, a $(d-1)$-brane couples to a $d$-form $A_{M_1 M_2
\cdots M_d}$ with
\be
A_{M_1 M_2 \cdots M_d} \rightarrow A_{M_1 M_2
 \cdots M_d} +
\partial_{[ M_1} \Lambda_{M_2 \cdots M_d ]},
\ee
and
\be
F_{M_1 M_2 \cdots M_{d+1}} = (d + 1) \partial_{[ M_1}
 A_{M_2
\cdots M_{d+1} ]}.
\ee
In the language of differential forms we may write for
 arbitrary $d$ and $D$
\be
{A_d \rightarrow A_d + d \Lambda_{d-1},}
\ee
and
\be
{F_{d+1} = dA_d ,}
\label{x}
\ee
from which the Bianchi identity
\be
{dF_{d+1} \equiv 0 }
\label{m}
\ee
follows immediately. In the absence of other interactions, the
equation of
motion for the $d$-form potential is
\be
d {^\ast} F_{D-d-1} = {^\ast}J_{D-d},
\label{e}
\ee
where the source $J$ is a $d$-form. Here we have introduced the
Hodge dual
operation $^\ast$ which converts a $d$-form into a $(D-d)$-form,
 e.g.
\be
{(^\ast J)^{M_1 M_2 \cdots M_{D-d}}
\equiv {1\over d!}
\varepsilon^{M_1 M_2 \cdots M_D} J_{M_{D-d+1} \cdots M_D},}
\ee
where $\varepsilon^{M_1 \cdots M_D}$ is the $D$-dimensional
 alternating symbol
with $\varepsilon^{01 \cdots D-1} = 1$.

In analogy with the usual Maxwell's equations, (\ref{e}) and (\ref{m}) imply the presence of an ``electric'' charge, i.e. a
 $(d-1)$-brane, but no
``magnetic'' charge, i.e. no $(D-d-3)$-brane. To restore the
duality symmetry by
introducing a $(D-d-3)$-brane we must modify (\ref{x}) to
\be F_{d+1} = dA_d + \omega_{d+1},\ee
so that the Bianchi identity  becomes
\be{dF_{d+1} = X_{d+2},}
\ee
with
\be{X_{d+2} = d\omega_{d+1}.}\ee
$X$ may be singular
\be
{X_{123 \cdots d+2} = g_{D-d-2} \delta^{d+2} (y),}
\ee
or may be smeared out so as
to be regular at the origin. We then have
\be{e_d = \int \limits_{S^{D-d-1}} {^\ast}F_{D-d-1} =
\int \limits_{M^{D-d}} {^\ast} J_{D-d},}
\ee
\be
{g_{D-d-2} = \int \limits_{S^{d+1}} F_{d+1} =
\int \limits_{M^{d+2}}
X_{d+2}.}
\ee
The Dirac quantization condition is again obtained by using the
generalization
of either the Dirac string \cite{Teitelboim} or Wu-Yang construction
\cite{Nepomechie} as
\be
\frac{e_d g_{D-d-2}}{4\pi} =\frac{1}{2}(n ={\rm integer})
\ee
Note that, $e_d$ and $g_{D- d -2}$ are not in general
dimensionless but rather
\be{[e_d] = - {1\over 2} (D-2d -2),
\qquad [g_{D-2d -2}] = {1\over 2}
(D - 2d -2).}\ee
They do become dimensionless when
\be{D= 2 (d + 1),}\ee
of which the point particle ($d = 1$) in $D = 4$ is the most
familiar special
case.

We shall now consider the elementary string where $X_{8}$ is singular, the solitonic fivebrane where $X_4$ is zero, the elementary fivebrane where $X_4$ is singular and the solitonic string where $X_8$ is zero. The solitons are regular in the sense that the curvature singularities are absent when written in terms of the corresponding dual frame sigma-model metrics given below \cite{sing}. Moreover a probe fivebrane takes infinite proper time to fall onto a heavy string and vice versa.

Then we allow for the presence of Yang-Mills fields and consider the solitonic string where $X_8$ is non-zero but regular and the solitonic fivebrane where $X_4$ is non-zero but regular.

\subsection{The elementary string and solitonic fivebrane}
\label{elementary}

We begin by recalling the elementary string solution of \cite{Dabholkar}.
We want to find a vacuum-like supersymmetric configuration with $D = 2$
super-Poincare symmetry from the 3-form version of $D = 10, N = 1$ supergravity
theory. As usual, the fermionic fields should vanish for this configuration.
We start by making an ansatz for the $D = 10$ metric $g_{MN}$, 2-form
$B_{MN}$ and dilaton $\phi$ ($M = 0, 1, \cdots,9$) corresponding to the
most general eight-two split invariant under $P_2 \times SO(8)$, where $P_2$ is
the $D = 2$ Poincare group. We split the indices
\begin{equation}
{x^M = (x^\mu, y_m),}
\label{split}
\end{equation}
where $\mu = 0,1$ and $m = 2, \cdots, 9$, and write the line element as
\begin{equation}
{ds^2 = e^{2A} \eta_{\mu\nu} dx^\mu dx^\nu + e^{2B}\delta^{mn} dy_mdy_n ,}
\label{element}
\end{equation}
and the two-form gauge field as
\begin{equation}
{B_{01} = - e^C .}
\end{equation}
All other components of $B_{MN}$ and all components of the gravitino $\psi_M$ and
dilatino $\lambda$ are set zero. $P_2$ invariance requires that the arbitrary
functions $A, B$ and $C$ depend only on $y^m$; $SO(8)$ invariance then requires
that this dependence be only through $y = \sqrt {\delta_{mn} y^m y^n}$.
Similarly, our ansatz for the dilaton is
\begin{equation}
{\phi = \phi(y).}
\end{equation}
As we shall now show, the four arbitrary functions $A, B, C,$ and $\phi$ are
reduced to one by the requirement that the above field configurations preserve
some unbroken supersymmetry. In other words, there must exist Killing spinors
$\varepsilon$ satisfying \cite{Dabholkar} 
\begin{equation}
{\delta \psi_M = D_M
\varepsilon + {1\over 96}\,e^{-\phi/
2}\big(\Gamma_M\,^{NPQ} - 9~\delta_M\,^N \Gamma^{PQ}\big)
 H_{NPQ}~
\varepsilon = 0,}
\end{equation}
\begin{equation}
{\delta \lambda = - {1\over2\sqrt{2}}~\Gamma^M
\partial_M \phi
\varepsilon + {1\over 24\sqrt{2}}~e^{-\phi/2}~\Gamma^{MNP}
H_{MNP}~\varepsilon = 0,}
\end{equation}
where
\begin{equation}
{H_{MNP} = 3\partial_{\lbrack M} A_{NP\rbrack}.}
\end{equation}
Here $\Gamma_A$ are the $D = 10$ Dirac matrices satisfying
\begin{equation}
{\lbrace \Gamma_A, \Gamma_B \rbrace = 2\eta_{AB}.}
\end{equation}
$A, B$ refer to the $D = 10$ tangent space, $\eta_{AB} = (-, +,\cdots, +)$, and
\begin{equation}
{\Gamma_{AB\cdots C} = \Gamma_{\lbrack A}
\Gamma_{B \cdots}
\Gamma_{C\rbrack},}
\end{equation}
thus $\Gamma_{AB} = {1\over 2}~(\Gamma_A \Gamma_B - \Gamma_B\Gamma_A)$, etc.
The $\Gamma$'s with world-indices $P, Q, R, \cdots$  have been converted using
vielbeins $e_M\,^A$. We make an eight-two split
\begin{equation}
{\Gamma_A = (\gamma_\alpha\otimes 1, \gamma_3\otimes \Sigma_a),}
\end{equation}
where $\gamma_\alpha$ and $\Sigma_a$ are the $D = 2$ and $D = 8$ Dirac
matrices, respectively. We also define 
\begin{equation}
{\gamma_3 = \gamma_0\gamma_1,} 
\end{equation}
so that $\gamma_3^2 = 1$ and 
\begin{equation}
{\Gamma_9 = \Sigma_2 \Sigma_3 \cdots \Sigma_9 ,}
\end{equation}
so that $\Gamma_9^2 = 1$. The most general spinor consistent
with $P_2 \times SO(8)$ invariance takes the form
\begin{equation}
{\varepsilon (x, y) = \epsilon \otimes \eta ,}
\end{equation}
where $\epsilon$ is a spinor of $SO(1,1)$ which may be further decomposed into
chiral eigenstates via the projection operators $(1\pm \gamma_3)$ and $\eta$ is
an $SO(8)$ spinor which may  further be decomposed into chiral eigenstates via
the projection operators $(1\pm\Gamma_9)$. The $N=1, D=10$ supersymmetry
parameter is, however, subject to the ten-dimensional chirality
condition
\begin{equation}
{\Gamma_{11}~\varepsilon = \varepsilon ,}
\end{equation}
where $\Gamma_{11} = \gamma_3\otimes\Gamma_9$ and so the $D=2$ and $D=8$
chiralities are correlated.

Substituting the ansatz into the supersymmetry transformation rules
leads to the solution \cite{Dabholkar}
\begin{equation}
{\varepsilon = e^{3\phi/8}\epsilon_0\otimes\eta_0,}
\end{equation}
 where $\epsilon_0$ and $\eta_0$ are constant spinors satisfying
\begin{equation} 
{(1 -\gamma_3)\epsilon_0 =0,\quad (1 - \Gamma_9) \eta_0 = 0,}
\label{chiral}
\end{equation}
and where
\begin{equation}
A= {3\phi\over 4} +c_A ,
\end{equation}
\begin{equation}
B= -{\phi\over 4} + c_B ,
\end{equation}
\begin{equation}
C= 2\phi +2 c_A ,
\end{equation}
where $c_A$ and $c_B$ are constants. If we insist that the
metric is asymptotically Minkowskian, then
\begin{equation}
{c_A = -~{3\phi_0\over 4},\quad c_B = ~{\phi_0\over 4},}
\end{equation}
where $\phi_0$ is the value of $\phi$ at infinity {\it i.e.} the dilaton
vev $\phi_0 =~<\phi>$. The condition (\ref{chiral}) means that one half of
the supersymmetries are broken.

At this stage the four unknown functions $A$, $B$, $C$ and
$\phi$ have been reduced to one by supersymmetry. To determine $\phi$, we must
substitute the ansatz into the field equations which follow from the action
$I_{10}({\rm string}) + S_2$ where $I_{10}({\rm string})$ is the bosonic sector
of the 3-form version of $D =10, N = 1$ supergravity given by
\begin{equation}
{I_{10}({\rm string}) = {1\over 2\kappa^2}
\int d^{10}x~\sqrt{-g}~\bigg(R - {1\over 2}
(\partial\phi)^2 - {1\over 2\cdot 3!}~e^{-\phi} H^2\bigg),}
\end{equation} 
and $S_2$ is the string $\sigma$-model action. Up until now we have employed
the canonical choice of metric for which the gravitational action is the
conventional Einstein-Hilbert action. This metric is related to the metric
appearing naturally in the string $\sigma$-model by 
\begin{equation}
{g_{MN}({\rm string}~\sigma{\rm\!-\!model}) =
e^{\phi/2} g_{MN}({\rm canonical}),}
\end{equation}
In canonical variables, therefore, the string $\sigma$-model action is
given by
\begin{eqnarray}
S_2 &=& - T_2 \int d^2 \xi \bigg({1\over 2}
\sqrt{-\gamma}~
\gamma^{ij}\partial_i X^M \partial_j X^N g_{MN}~
e^{\phi/2} - 2~\sqrt{-\gamma}
\nonumber \\
&& ~~~~~~~~~~~
+{1\over 2!}~\varepsilon^{ij} \partial_i X^M \partial_j X^N
 B_{MN}\bigg).
\end{eqnarray}
We have denoted the string tension by $T_2$. The supergravity field equations
are
\begin{eqnarray}
R^{MN} \!\!\!\! &-&\!\!\!\!
{1\over 2}~\bigg(\partial^M
\phi~\partial^N \phi - {1\over 2}~g^{MN} (\partial \phi)^2 \bigg)
- {1\over 2}~g^{MN} R
\nonumber \\
 \!\!\!\!&-&\!\!\!\! {1\over 2\cdot 2!} \bigg(H^M\,_{PQ} H^{NPQ} - {1\over6}
{}~g^{MN}
H^2\bigg) e^{-\phi}
=\kappa^2 T^{MN}(\rm string),
\end{eqnarray}
where
\begin{equation}
T^{MN}({\rm string}) = -T_2 \int d^2\xi\sqrt{-\gamma}
{}~\gamma^{ij}
\partial_i X^M \partial_j X^N e^{\phi/2}~{\delta^{10} (x - X)
\over \sqrt{-g}},
\end{equation}
\begin{equation}
\partial_M (\sqrt{-g}~e^{-\phi} H^{MNP}) =2\kappa^2 T_2 \int d^2
\xi~\varepsilon^{ij} \partial_i X^N \partial_j X^P
\delta^{10} (x - X),
\end{equation}
\begin{eqnarray}
 &&
\partial_M (\sqrt{-g}g^{MN}\partial_N\phi) + {1\over 2\cdot
3!}~e^{-\phi} H^2 =
\nonumber \\
&& ~~~~~~~~~~~= {\kappa^2 T_2\over 2}
\int d^6 \xi \sqrt{-\gamma}~\gamma^{ij}\partial_i X^M
\partial_j X^N g_{MN}
e^{\phi/2}{\delta^{10} (x - X)}.
\end{eqnarray}
Furthermore, the string field equations are
\[
\partial_i (\sqrt{-\gamma}~\gamma^{ij}
\partial_j X^N
g_{MN}~e^{\phi/2} ) - {1\over 2}~\sqrt {-\gamma}~\gamma^{ij}
\partial_i X^N
\partial_j X^P \partial_M (g_{NP}~e^{\phi/2} )-
\]
\begin{equation}
{1\over 2}~\varepsilon^{ij}
\partial_i X^N \partial_j X^P H_{MNP} = 0, 
\end{equation}
and
\begin{equation}
\gamma_{ij} = \partial_i X^M \partial_j X^N g_{MN}
e^{\phi/2}.
\end{equation}
To solve these coupled supergravity-string equations we make
the static
gauge choice
\begin{equation}
{X^\mu = \xi^\mu, \quad \mu = 0,1}
\end{equation}
and the ansatz
\begin{equation}
{X^m=Y^m = {\rm constant}, \qquad m=2,...,9.}
\end{equation}
As an example, let us now substitute the ansatz into and
the 2-form equation. We find
\begin{equation}
{\delta^{mn} \partial_m \partial_n e^{-2\phi} =
 - 2\kappa^2 T_2
e^{-\phi_0/2}\delta^8 (y),}
\end{equation}
and hence
\begin{equation}
{e^{-2\phi} = e^{-2\phi_0} \left(1 + {k_2\over y^6}
\right),}
\end{equation}
where the constant $k_2$ is given by
\begin{equation}{k_2 \equiv {\kappa^2 T_2\over 3\Omega_7}
{}~e^{3\phi_0/2},}
\label{single}
\end{equation}
and $\Omega_n$ is the volume of the unit $n$-sphere $S^n$. One
 may verify by using the expressions for the
Ricci tensor $R^{MN}$ and Ricci scalar $R$ in terms of $A$ and $B$
\cite{Khuristring} that all the field equations are reduced to a
 single equation (\ref{single}).

Having established that the supergravity configuration preserves half the
supersymmetries, we must also verify that the string configuration also
preserve these supersymmetries. As discussed in \cite{Bergshoeffgamma},
the criterion is that in addition to the existence of Killing spinors
 we must also have 

\begin{equation} {(1 - \Gamma)\varepsilon = 0,}
\end{equation} where the choice of sign is dictated by the choice of the
sign in the Wess-Zumino term in $S_2$, and where 
\begin{equation}
{\Gamma \equiv {1\over 2! \sqrt{-\gamma}}
{}~\varepsilon^{ij}
\partial_i X^M \partial_j X^N \Gamma_{MN}.}
\label{gamma}
\end{equation}
Since $\Gamma^2 = 1$ and tr $\Gamma = 0,\, {1\over 2} (1 \pm
\Gamma)$ act as projection operators. For our solution, we find  that
\begin{equation}
{\Gamma = \gamma_3 \otimes 1,}
\end{equation}
and hence (\ref{gamma}) is satisfied.  This explains, from a string point of
view, why the solutions we have found preserve just half
the supersymmetries. It originates from the fermionic kappa symmetry of
the superstring action. The fermionic zero-modes on the worldvolume are
just the Goldstone fermions associated with the broken supersymmetry.

As shown in \cite{Dabholkar}, the elementary string solution saturates a
Bogolmol'nyi bound for the mass per unit length
\begin{equation}
{{\cal M}_2 = \int d^8 y~\theta_{00},}
\end{equation}
where $\theta_{MN}$ is the total energy-momentum pseudotensor of the combined
gravity-matter system. One finds 
\begin{equation}
{\kappa {\cal M}_2 \geq {1\over\sqrt{2}} |e_2| e^{ \phi_0/2},}
\end{equation}
where $e_2$ is the Noether ``electric '' charge whose conservation follows the
equation of motion of the 2-form, namely
\begin{equation}
{e_2 = {1\over {\sqrt{2}\kappa}} \int\limits_{S^7}e^{-\phi}\,^{\ast} H,}
\end{equation} 
where $^{\ast}$ denotes the Hodge dual using the canonical metric
and the integral is over an asymptotic seven-sphere surrounding the string. We
find for our solution that
\begin{equation}
{{\cal M}_2 = e^{\phi_0/2}~T_2,}
\label{mass}
\end{equation}
and
\begin{equation}
{e_2 = \sqrt{2}\kappa~T_2.}
\label{charge}
\end{equation}
Hence the bound is saturated. This provides another way, in addition to
unbroken supersymmetry, to understand the stability of the solution.

The elementary string discussed above is a solution of the
coupled field-string system with action $I_{10}({\rm string})+ S_2$.  
As such it exhibits $\delta$-function singularities at $y = 0$.  It is
characterized by a non-vanishing Noether electric charge $e_2$.  By
contrast, we now wish to find a solitonic fivebrane, corresponding to a
solution of the source free equations resulting from $I_{10}({\rm string})$
alone and which will be characterized by a non-vanishing topological
``magnetic'' charge $g_6$.

To this end, we now make an ansatz invariant under $P_6 \times SO(4)$. 
Hence we write (\ref{split}) and (\ref{element}) as before where now $\mu =
0, 1 \ldots 5$ and $m = 6, 7, 8, 9$.  The ansatz for the antisymmetric
tensor, however, will now be made on the field strength rather than on the
potential.  From section (\ref{elementary}) we recall that a non-vanishing
electric charge corresponds to
\begin{equation}
{{1\over \sqrt{2} \kappa} e^{-\phi}{}^{\ast}H
 = e_2 \varepsilon_7/\Omega_7,}
\end{equation}
where $\varepsilon_n$ is the volume form on $S^n$. Accordingly,
to obtain a non-vanishing magnetic charge, we make the ansatz
\begin{equation}
{{1\over \sqrt{2} \kappa} H = g_6
\varepsilon_3/\Omega_3.}
\end{equation}
 Since this is an harmonic form, $H$ can no longer be written globally as
the curl of $B$, but it satisfies the Bianchi identity.  It is now not
difficult to show that all the field equations are
satisfied. The solution is given by
\begin{equation}
e^{2\phi}=e^{2\phi_0}\left(1+{k_6\over y^2}\right),
\end{equation}
\begin{equation}
ds^2=e^{-(\phi-\phi_0)/2}\eta_{\mu\nu}dx^\mu dx^\nu+e^{3(\phi-\phi_0)/2}
\delta_{mn}dy^mdy^n,
\end{equation}
\begin{equation}
 H=2k_6 e^{\phi_0/2}\varepsilon_3,
\end{equation}
where $\mu,\nu=0,1,...,5$, $m,n=6,7,8,9$ and where
\begin{equation}{k_6={\kappa g_6 \over \sqrt{2}\Omega_3}e^{-\phi_0/2}.}
\end{equation}
It follows that the mass per unit 5-volume now saturates a bound
involving the magnetic charge
\begin{equation}{{\kappa\cal M}_6={1\over \sqrt{2}} \mid g_6 \mid
e^{-\phi_0/2}.}\end{equation}
Note that the $\phi_0$ dependence is such that ${\cal M}_6$ is
large
for small ${\cal M}_2$ and vice-versa.

The electric charge of the elementary solution and the magnetic charge of
the soliton solution obey a Dirac quantization rule
\cite{Nepomechie,Teitelboim} 
\begin{equation}{e_2 g_6 = 2 \pi n, \qquad n =
{\rm integer},} \end{equation}
and hence 
\begin{equation}{g_6 = 2\pi n/\sqrt{2}\kappa T_2.}
\end{equation}
\subsection{The elementary fivebrane and solitonic string}
In keeping with the viewpoint that the fivebrane may be
regarded as
fundamental in its own right, Duff and Lu \cite {Luremarks} then
constructed the
{\it elementary fivebrane} solution by coupling the $7$-form
version of
supergravity to a fivebrane $\sigma$-model source in analogy
 with the
elementary string. This carries an electric charge $e_6$.
Thus the elementary fivebrane, as pointed out
by
Callan, Harvey and Strominger \cite{Callan1,Callan2}, could
 also be
regarded as a soliton when viewed from the dual perspective,
with
$g_6=e_6$. In other words, it provides a nonsingular
solution of the
source-free $3$-form equations.  By the same token, when viewed from the dual perspective,
the elementary
string provides a nonsingular solution of the source-free
$7$-form equations
 with $ g_2=e_2$ \cite{Lustrings}.

%Similar remarks apply to the elementary fivebrane \cite{Luelem} and the solitonic string for which
%\begin{equation}
%e_6 g_2 = 2 \pi n, \qquad n ={\rm integer}
%\end{equation}
%\begin{equation}
%\kappa {\cal M}_2={1\over \sqrt{2}} |g_2|e^{\phi_0/2}.
%\end{equation}
\subsection{Honey, I shrunk the instanton}

\begin{table}
\begin{center}
\begin{tabular}{lllllllllllllllllll}
\hline
&Elementary fivebrane
&Solitonic fivebrane\\
\hline
metric&$e^{-\phi/6}g_{MN}(canon)$&$e^{\phi/2}g_{MN}(canon)$&\\
action  
            &$I_{10}(fivebrane)+S_6(fivebrane)$
            &$I_{10}(string)+S_{YM}(string)$\\
            sources
&$X_4=e_6\epsilon_4\delta^4(y)$
&$X_4=g_6Tr F^2$\\
$T_{\mu\nu}$
&$-T_6e^{-\phi/2}g_{\mu\nu}\delta^4(y)$
&${\tilde T}_6g_{\mu\nu}e^{-\phi/2}TrF^2$\\

tension
            &$\sqrt{2}\kappa T_6=e_6$
            &$\sqrt{2}\kappa {\tilde T}_6=g_6$\\
charge 
            &electric 
            &magnetic \\
            &$\sqrt{2}\kappa e_6= \int\limits_{S^3}e^{\phi}{}^{\ast} H_7$
            &$\sqrt{2}\kappa g_6= \int\limits_{S^3}H_3$\\
mass  
           &$\sqrt{2}\kappa {\cal M}_6= |e_6| e^{-\phi_0/2}$
           &$\sqrt{2}\kappa {\cal M}_6=|g_6|e^{-\phi_0/2}$&\\

         %  Instanton &$X_8=\frac{1}{3(2\pi)^3T_6} \left(Tr F \wedge F \wedge F \wedge F-\frac{1}{7200}Tr F \wedge F \wedge Tr F \wedge F\right)$&
        %   &$X_4=\frac{1}{30\pi T_2} Tr F \wedge F$&&\\
\hline
\end{tabular}
\caption{Fivebranes: String/fivebrane duality implies $e_6g_2=2 \pi n=2\kappa^2 T_6 {\tilde T}_2$
 and identifying elementary and solitonic fivebranes yields $e_6=g_6$.}
\label{Fivebranes}
\end{center}
\end{table}
\begin{table}
\begin{center}
\begin{tabular}{lllllllllllllllllll}
\hline
&Elementary string 
&Solitonic string\\
\hline
metric
&$e^{\phi/2}g_{MN}(canon)$
&$e^{-\phi/6}g_{MN}(canon)$&\\
action  
&$I_{10}(string)+S_2(string)$
&$I_{10}(fivebrane)+S_{YM}(fivebrane)$\\
sources 
&$X_8=e_2\epsilon_8\delta^8(y)$
&$X_8=g_2Tr F^4$\\
$T_{\mu\nu}$&$-T_2e^{\phi/2}g_{\mu\nu}\delta^8(y)$
&${\tilde T_2}g_{\mu\nu}e^{\phi/2}TrF^4$\\
tension
&$\sqrt{2}\kappa T_2=e_2$
&$\sqrt{2}\kappa {\tilde T}_2=g_2$\\
charge 
&electric             
&magnetic  \\
&$\sqrt{2}\kappa e_2 = \int\limits_{S^7}e^{-\phi}{}^{\ast} H_3$
&$\sqrt{2}\kappa g_2 = \int\limits_{S^7}{H}_7$\\
mass   
&$\sqrt{2}\kappa {\cal M}_2 =  |e_2| e^{ \phi_0/2}$
&$\sqrt{2}\kappa {\cal M}_2= |g_2| e^{\phi_0/2}$\\

 % Instanton &$X_8=\frac{1}{3(2\pi)^3T_6} \left(Tr F \wedge F \wedge F \wedge F-\frac{1}{7200}Tr F \wedge F \wedge Tr F \wedge F\right)$&
  %   &$X_4=\frac{1}{30\pi T_2} Tr F \wedge F$&&\\
\hline
\end{tabular}
\caption{Strings: String/fivebrane duality implies $e_2g_6=2 \pi n=2\kappa^2 T_2 {\tilde T}_6$
 and identifying elementary and solitonic strings yields $e_2=g_2$.}
\end{center}
\label{Strings}
\end{table}

Now we incorporate the Yang-Mills fields.  In the case of Strominger's solitonic fivebrane
\begin{equation}
X_4=\frac{1}{30\pi T_2} Tr F \wedge F
\end{equation}
String/fivebrane duality then suggested that by coupling the
$7$-form version
of supergravity to super Yang-Mills (without a $\sigma$-model
source),
one ought to find a nonsingular heterotic string soliton
carrying a
topological magnetic charge $ g_2$. This was indeed the case \cite{Lustrings} , but scaling
arguments required
an unconventional Yang-Mills Lagrangian, quartic in the field
strengths with
\begin{equation}
X_8=\frac{1}{3(2\pi)^3T_6} \left(Tr F \wedge F \wedge F \wedge F-\frac{1}{7200}Tr F \wedge F \wedge Tr F \wedge F\right)
\end{equation}
The fivebrane equations admit a solution where $F_{mn}$ is a self-dual  $SO(4)$ instanton \cite{inst1} in the 4 directions orthogonal to the brane. Similarly, the string equations admit a solution where $F_{mn}$ is an $SO(8)$ instanton \cite{inst2} in the 8 directions orthogonal to the string.

Consistency demands that the sources, denoted generically by J, must be such that when we shrink the size of the instanton
\be
\lim \rho \rightarrow 0~~ J(quadratic~ Yang-Mills)=J (fivebrane~ sigma~model)
\ee
\be
\lim \rho \rightarrow 0~~ J(quartic ~Yang-Mills)=J (string~ sigma~model)
\ee
This is indeed the case. For example  the corresponding dilaton solutions (with $\phi_0=0$) are
%\begin{equation}
%ds^2=e^{4\phi/3}\eta_{\mu\nu} dx^\mu dx^\nu +e^{-2\phi/3} \delta_{mn} dy^m dy^n
%\end{equation}
%\begin{equation}
%B_{01}=-e^{2\phi}
%\end{equation}

\begin{equation}
e^{-2\phi} = 1 +\frac {k_6(y^2 + 2\rho^2)}{(y^2 + \rho^2)^2}  \rightarrow1 + \frac{ k_6}{y^2}
\end{equation}
for the 5-brane and
\begin{equation}
e^{-2\phi} = 1 + k_2 \frac{(y^6 + 6y^4\rho^2+ 15 y^2 \rho^4 + 20 \rho^6)}{(y^2 + \rho^2)^6} \rightarrow 1+\frac{k_2}{y^6}
\end{equation}
for the string.

\subsection{Message to the no-braners}

{\it Let me reiterate.This discovery of solitons means that 5-branes are there whether you like them or not. If you buy strings, and of course you are free not to, you have to buy 5-branes in the same package.}

\subsection{N=4 gauge theory on the D3-brane}

{\it We have recently constructed a self-dual Type IIB super 3-brane which represents a new point $(d=4, D=10)$ on the brane-scan. Earlier no-go theorems \cite{Achucarro} are circumvented because there are spin 1 fields on the worldvolume. In fact, the gauge-fixed theory on the worldvolume is described by a $(d=4, N=4)$ Maxwell multiplet.}

\subsection{Subsequent developments}
\begin{itemize}
\item{Branes and M-theory}

The realization that the equations of string theory admit branes as soliton solutions opened a new window on non-perturbative string theory and paved the way for M-theory.

\item{Fivebranes}

The 5-branes discussed in this section now feature prominently in M-theory, for example  heterotic/heterotic duality \cite{DMW}. They became known as Neveu-Schwarz 5-branes. 

\item{Dual frame metrics}

The string and fivebrane $\sigma$-model metrics in $D=10$ are special cases of $(d-1)$-brane metrics in $D$ dimensions and those of their $\tilde d =D-2-d$ duals.
\[
g_{MN}(d)=e^{a(d)\phi/d}g_{MN}(canon)
\]
\be
g_{MN}(\tilde d)=e^{a(d)\phi/{\tilde d}}g_{MN}(canon)
\ee
where $a(\tilde d)=-a(d)$ and
\be
a^2(d)=4-\frac{2d\tilde d}{d+\tilde d}
\ee

These dual frame metrics, for which by definition  the Nambu-Goto part of the $(d-1)$-brane sigma model is independent
of the dilaton, have found numerous applications, especially in the context of holography \cite{Gibbonsdufftownsend,Boonstra,Skenderis,Kan,Taylor}.

\item{Type II branes}

According to the classification of \cite{Achucarro} described in
Section \ref{ravioli}, no Type II $p$-branes with $p > 1$ could exist.
Moreover, the only brane allowed in $D=11$ was $p=2$. These  conclusions
were based on the assumption that the only fields propagating on the
worldvolume were scalars and spinors, so that, after gauge fixing, they fall
only into {\it scalar} supermultiplets, denoted by $s$ on the brane-scan of
Table \ref{brane-scan}. Indeed, these were the only kappa symmetric actions
known at the time. However, as we saw already in Section \ref{conformal}, there was evidence 
for a Type IIB self-dual 3-brane and an M5-brane \cite{Blencowe}. Moreover, using soliton arguments,  it was pointed out in
\cite{Callan1,Callan2} that both Type IIA and Type IIB superfivebranes
exist after all. Moreover, the Type IIB theory also admits the self-dual
superthreebrane as a soliton \cite{Luthree}. The no-go theorem is circumvented because in
addition to the superspace coordinates $X^M$ and $\theta^\alpha$ there are
also higher spin fields on the worldvolume: vectors or antisymmetric
tensors. This raised the question: are there other super $p$-branes and if
so, for what $p$ and $D$? In \cite{Luscan} an attempt was made to answer
this question by asking what new points on the brane-scan are permitted by
bose-fermi matching alone. Given that the gauge-fixed theories display
worldvolume supersymmetry, and given that we now wish to include the
possibility of vector and antisymmetric tensor fields, it is a
relatively straightforward exercise to repeat the bose-fermi matching
conditions of the Section \ref{ravioli} for vector and antisymmetric
tensor supermultiplets.  Let us begin with vector supermultiplets. Once again, we may proceed in one 
of two ways. First, given that a worldvolume vector has ($d - 2$) degrees of
freedom, the scalar multiplet condition (\ref{bosefermi}) gets replaced by 
\begin{equation} 
{D - 2 = {1\over 2}~mn ={1\over 4}~MN .}
\end{equation}
Alternatively, we may simply list all the vector supermultiplets in the
classification of \cite{Strathdee} and once again interpret $D$ via
(\ref{scalars}). The results \cite{Luscan,Khuristring} are shown by the
points labelled $v$ in Table \ref{brane-scan}.  Next we turn to antisymmetric tensor multiplets. In $d = 6$ there is a
supermultiplet with a second rank tensor whose field strength is self-dual: 
$(B_{\mu\nu}^-, \lambda^I, \phi^{[IJ]})$, $I = 1, \ldots, 4$.  This is has
chiral $d=6$ supersymmetry. Since there are five scalars, we have
$D=6+5=11$. There is thus a new point on the scan corresponding to the $D=11$
superfivebrane.  One may decompose this $(n_+,n_-)=(2,0)$ supermultiplet
under $(n_+,n_-)=(1,0)$ into a tensor multiplet with one scalar and a
hypermultiplet with four scalars.  Truncating to just the tensor multiplet
gives the zero modes of a fivebrane in $D=6+1=7$. These two tensor multiplets
are shown by the points labelled $t$ in Table \ref{brane-scan}. 

\item{D-branes}

Subsequently, all 
the v-branes 
were given a new interpretation as Dirichlet $p$-branes, called D-branes, 
surfaces of 
dimension$p$on which open strings can end and which carry R-R 
(Ramond-Ramond) charge \cite{Polchinski}. The IIA theory has D-branes with $p=0,2,4,6,8$ 
and the IIB theory has D-branes with $p=1,3,5,7,9$. They are related to one another by 
T-duality. In terms of how their tensions depend on the string 
coupling $g_{s}$, the D-branes are 
mid-way between the fundamental (F) strings and the solitonic (S) 
fivebranes: 

\be
T_{F1}\sim m_{s}^{2},~~~~T_{Dp}\sim \frac{m_{s}^{p+1}}{g_{s}}, 
~~~~T_{S5}\sim \frac{m_{s}^{6}}{g_{s}^{2}}
\ee

\item{M2-branes, M5-branes and the quantization of 4-form flux}

That 4-form flux of M-theory is quantized was implicit in the multimembrane solution of $D=11$ supergravity since the tension of a stack of N 2-branes is just N times that of a single brane $T_3$. It was spelled out explicitly in \cite{Duffliuminasian}
where we begin with the bosonic sector of the $d=3$ worldvolume of the $D=11$
supermembrane (\ref{membrane}) and the bosonic $D=11$ supergravity action (\ref{supergravity11})
While there are two dimensionful parameters, the membrane tension $T_3$
and the eleven-dimensional gravitational constant $\kappa_{11}$, they are
in fact not independent.  To see this, we note from (\ref{membrane}) that
$A_3$ has period $2\pi/T_3$ so that $F_4$ is quantized according to
\begin{equation}
\int F_4={2\pi n\over T_3}\qquad n={\rm integer}\ .
\label{eq:kquant}
\end{equation}
Consistency of such $A_3$ periods with the spacetime action,
(\ref{supergravity11}), gives the relation\footnote{A factor of 2 error in \cite{Duffliuminasian} was corrected in \cite{Schwarzpower,Dealwis}}
\begin{equation}
\frac{(2\pi)^2}{{2\kappa_{11}{}^2}T_3{}^3}\in Z\ .
\label{eq:k11t3}
\end{equation}

The $D=11$ classical field equations admit as a soliton a dual superfivebrane
\cite{Gueven,Lublack} which couples
to the dual field strength $\tilde F_7=*F_4$.  The fivebrane tension
${\tilde T}_6$ is given by the Dirac quantization rule \cite{Lublack}
\begin{equation}
2\kappa_{11}{}^2 T_3 {\tilde T}_6 =2\pi n \qquad n={\rm integer}\ .
\label{Dirac11}
\end{equation}
Using (\ref{eq:k11t3}), this may also be written as
\begin{equation}
2\pi {\tilde T}_6=T_3{}^2
\label{eq:newdirac}
\end{equation}
 Although Dirac quantization rules of
the type (\ref{Dirac11}) appear for other $p$-branes and their duals in
lower dimensions \cite{Lublack}, it is the absence of a dilaton in the
$D=11$ theory that allows us to fix both the gravitational constant and
the dual tension in terms of the fundamental tension.

From (\ref{equation4}), the fivebrane Bianchi identity reads
\begin{equation}
d\tilde F_7=-{1\over2}F_4{}^2\ .
\end{equation}
However, such a Bianchi identity will in general require gravitational
Chern-Simons corrections arising from a sigma-model anomaly on the
fivebrane worldvolume \cite{Dixon,Dixon:1992qd,Dixon:1991xz,Percacci,Bergshoeff1,Cederwall,Duffminasian}:
\begin{equation}
d\tilde F_7=-{1\over2}F_4{}^2 + (2\pi)^4{\tilde \beta}'{\tilde X}_8\ ,
\label{Bianchi7q}
\end{equation}
where ${\tilde \beta}'$ is related to the fivebrane tension by
$T_6=1/(2\pi)^3{\tilde \beta}'$ and where the $8$-form polynomial
${\tilde X}_8$, quartic in the gravitational curvature $R$,
describes the $d=6$ $\sigma$-model Lorentz anomaly of the $D=11$
fivebrane.
\begin{equation}
{\tilde X}_8={1\over(2\pi)^4}
\Bigl[-{1\over768}(\tr R^2)^2+{1\over192}\tr R^4\Bigr]\ .
\la{X}
\end{equation}
Thus membrane/fivebrane duality predicts a spacetime correction to the
$D=11$ supermembrane action
\begin{equation}
I_{11}({\rm Lorentz})=T_3\int A_3\wedge
{1\over(2\pi)^4}\Bigl[-{1\over768}(\tr R^2)^2+{1\over192}\tr R^4\Bigr] \ .
\end{equation}
By simultaneous dimensional reduction \cite{Howe} of $(d=3, D=11 )$ to
$(d=2, D=10)$ on $S^1$, this prediction translates into a corresponding
prediction for the Type IIA string:
\begin{equation}
I_{10}({\rm Lorentz})=T_2\int B_2\wedge
{1\over(2\pi)^4}\Bigl[-{1\over768}(\tr R^2)^2+{1\over192}\tr R^4\Bigr] \ ,
\end{equation}
where $B_2$ is the string $2$-form, $T_2$ is the string tension,
$T_2=1/{2\pi\alpha'}$, related to the membrane tension by
\begin{equation}
T_2=2\pi RT_3\ ,
\label{T}
\end{equation}
where $R$ is the $S^1$ radius.

Further elaboration of four-form flux quantization may be found in \cite{Wittenflux,Beckerseight}
\item{D-branes from M-branes}

In addition to M2 and M5 there are two other objects in $D=11$, the plane wave \cite{Hull} and the 
Kaluza-Klein monopole \cite{Han}, which though not branes are still 1/2 BPS. When 
spacetime is 
compactified a $p$-brane may remain a $p$-brane or else 
become a $(p-k)$-brane if it wraps around $k$ of the compactified 
directions.
For example, the Type IIA fundamental string emerges by wrapping 
the M2-brane around $S^{1}$ and shrinking its radius to zero, and the 
Type IIA 4-brane emerges in a similar way from the $M5$-brane. Several comments are now in order: (1) The number of scalars in a vector supermultiplet is such that, from
(\ref{scalars}), $D = 3, 4, 6$ or $10$ only, in accordance with
\cite{Strathdee}. (2) Vector supermultiplets exist for all $d \leq10$ \cite{Strathdee}, as may
be seen by dimensionally reducing the $(n=1, D=10)$  Maxwell
supermultiplet.  However, in $d = 2$ vectors have no degrees of freedom and
in $d = 3$ vectors have only one degree of freedom and are dual to scalars. 
In this sense, therefore, these multiplets will already have been included as
scalar multiplets in Section \ref{ravioli}. There is consequently some
arbitrariness in whether we count these as new points on the scan.  For example, it was recognized
\cite{Luscan} that by dualizing a vector into a scalar on the gauge-fixed
$d=3$ worldvolume of the Type IIA supermembrane, one increases the number
of worldvolume scalars, {\it i.e.} transverse dimensions, from $7$ to $8$
and hence obtains the corresponding worldvolume action of the $D=11$
supermembrane.  Thus the $D=10$ Type IIA theory contains a hidden $D=11$
Lorentz invariance \cite{Luscan,TownsendD}! More on D-branes from M-branes may be found in the paper by Townsend  \cite{{TownsendD}}.
(3) In listing vector multiplets, we have focussed only on the abelian
theories obtained by dimensionally reducing the Maxwell multiplet. One
might ask what role, if any, is played by non-abelian Yang-Mills
multiplets.

\item{Non-abelian gauge groups from stacked branes}

Since they are BPS, there is a no-force condition between the branes that allows 
us to have many branes of the same charge parallel to one another. The gauge group 
on a single D-brane is an abelian $U(1)$.  If we stack $N$ such branes on top of one 
another, the gauge group is the non-abelian $U(N)$. As we separate them this decomposes 
into its subgroups, so in fact there is a Higgs mechanism at work whereby the 
vacuum expectation values of the Higgs fields are related to the separation of the 
branes. For example the theory that lives on a stack of $N$ Type IIB $D3$ 
branes is a four-dimensional $U(N)$ $n=4$ super Yang-Mills theory. In 
the limit of large $N$ the geometry of this configuration tends to the 
product of five dimensional anti-de Sitter space and a five dimensional 
sphere, $AdS_{5}\times S^{5}$.   This provides the AdS/CFT correspondence.

 \item{The brane-world}
 
The 3-brane soliton of Type IIB 
supergravity was an early 
candidate for a `brane-world', firstly because of its 
dimensionality \cite{Horowitz1,Luthree} and secondly because gauge fields propagate 
on its worldvolume \cite{Luthree}.  The idea that our universe is a brane floating in a higher dimensional bulk is not new. See, for example \cite{Rubakov} and \cite{Gibbons:1986wg}. But the way in which the gauge fields are confined to the brane in the D-brane picture and in the Horava-Witten \cite{Horava1,Horava2} heterotic M-theory picture provided the impetus for a revival of the {\it braneworld} and large extra dimensions \cite{Arkani,Antoniadis:1998ig,Randall1,Randall2,Karch}.
\end{itemize}
\bigskip

%\end{document}

\section{1992 Four-dimensional string/string duality }

 26th Workshop of the Eloisatron Project, Erice, Italy, December 5-12,  1992
 \cite{Duffkhuri}
 
%\subsection{S-duality as T-duality} 

In this lecture we presented supersymmetric soliton solutions of the four-dimensional heterotic string corresponding to monopoles, strings and domain walls. These solutions admit the D = 10 interpretation of a fivebrane wrapped around 5, 4 or 3 of the 6 toroidally compactified dimensions and are arguably exact to all orders in $\alpha'$. The solitonic string solution exhibits an SL(2, Z) strong/weak coupling duality which however corresponds to an SL(2, Z) target space duality of the fundamental string.

\subsection{Fivebranes in D$=$10}

We first summarize the 't Hooft ansatz for the Yang-Mills instanton.
Consider the four-dimensional Euclidean action
\be S=-{1\over 2g^2}\int d^4x {\rm Tr} F_{\mu\nu}F^{\mu\nu},
\qquad\qquad \mu,\nu =1,2,3,4.
\ee
For gauge group $SU(2)$, the fields may be written as $A_\mu=(g/2i)
\sigma^a A_\mu^a$ and $F_{\mu\nu}=(g/2i)\sigma^a F_{\mu\nu}^a$\ \
(where $\sigma^a$, $a=1,2,3$ are the $2\times 2$ Pauli matrices).
A self-dual solution (but not the most general one) to the equation of motion
of this action is given by the 't Hooft ansatz
\be
A_\mu=i \overline{\Sigma}_{\mu\nu}\partial_\nu \ln f
\ee
where $\overline{\Sigma}_{\mu\nu}=\overline{\eta}^{i\mu\nu}(\sigma^i/2)$
for $i=1,2,3$, where
$
\overline{\eta}^{i\mu\nu}=-\overline{\eta}^{i\nu\mu}
=\epsilon^{i\mu\nu}$ for $ \mu,\nu=1,2,3,$ and $ \overline{\eta}^{i\mu\nu}=-\overline{\eta}^{i\nu\mu}
=-\delta^{i\mu} $ for $\nu=4$
and where $f^{-1}\Box\ f=0$. The ansatz for the anti-self-dual solution
is similar, with the $\delta$-term changing sign. From this ansatz, depending on how many of the four coordinates $f$ is
allowed to depend and depending on whether we compactify, we shall obtain
$D=10$ multi-fivebrane and $D=4$ multi-monopole,
multi-string and multi-domain wall solutions. We will discuss these four
cases in the next section. In this section, we do not specify the precise
form of $f$ or the dilaton function, but show that the derivation of the
solution and most of the arguments used to demonstrate
the exactness of the heterotic solution are equally valid for any $f$
satisfying $f^{-1}\Box\ f=0$.

It turns out that there is an analog to the 't Hooft ansatz for the Yang-Mills
instanton in the gravitational sector of the string, namely the axionic
instanton \cite{Rey}. In its simplest form, this instanton appears as a solution for the massless fields of the bosonic string. The identical instanton
structure arises in all supersymmetric multi-fivebrane solutions, in particular in the tree-level neutral solution
\[
g_{\mu\nu}=e^{2\phi}\delta_{\mu\nu}  \qquad \mu,\nu=1,2,3,4,
\]
\[
g_{ab}=\eta_{ab}  \qquad\quad   a,b=0,5,...,9,
\]
\be
H_{\mu\nu\lambda}=\pm 2\epsilon_{\mu\nu\lambda\sigma}\partial^\sigma\phi
\qquad \mu,\nu,\lambda,\sigma=1,2,3,4,
\label{ansatz}
\ee
with $e^{-2\phi}\Box\ e^{2\phi}=0$. The D'Alembertian refers to the
four-dimensional subspace $\mu,\nu,\lambda,\sigma=1,2,3,4$ and $\phi$ is taken
to be independent of $(x^0,x^5,x^6,x^7,x^8,x^9)$. For zero background fermionic
fields the above solution breaks half the spacetime supersymmetries.

The generalized curvature of this solution was shown \cite{Khuri,Khurinew}
to possess (anti) self-dual structure similar to that of the 't Hooft ansatz.
To see this we define a generalized curvature $\gcone$ in terms of the standard
curvature $\cone$ and $H_{\mu\alpha\beta}$:
\be
\gcone=\cone+{1\over 2}\left(\nabla_\sigma H^\mu{}_{\nu\rho}-
\nabla_\rho H^\mu{}_{\nu\sigma}\right)+
{1\over 4}\left(H^\lambda{}_{\nu\rho}H^\mu{}_{\sigma\lambda}
- H^\lambda{}_{\nu\sigma} H^\mu{}_{\rho\lambda}\right).
\ee
One can also define $\gcone$ as the Riemann tensor generated
by the generalized Christoffel symbols $\hat\Gamma^\mu_{\alpha\beta}$,
where $\hat\Gamma^\mu_{\alpha\beta}=\Gamma^\mu_{\alpha\beta}
-(1/2) H^\mu{}_{\alpha\beta}$.
The crucial observation for obtaining higher-loop and even exact solutions
is the following. For any solution given by (\ref{ansatz}),
we can express the generalized curvature in terms of the dilaton field as 
\be
\gcone=
\delta_{\mu\sigma}\nabla_\rho\nabla_\nu\phi
-\delta_{\mu\rho}\nabla_\sigma\nabla_\nu\phi
+\delta_{\nu\rho}\nabla_\sigma\nabla_\mu\phi
-\delta_{\nu\sigma}\nabla_\rho\nabla_\mu\phi 
\pm\epsilon_{\mu\nu\rho\lambda}\nabla_\sigma\nabla_\lambda\phi
\mp\epsilon_{\mu\nu\sigma\lambda}\nabla_\rho\nabla_\lambda\phi.
\ee
It easily follows that
\be
\gcone=\mp {1\over 2} \epsilon_{\rho\sigma}{}^{\lambda\gamma}
\hat R^\mu{}_{\nu\lambda\gamma}.
\ee
So the (anti) self-duality appears in the gravitational sector of the string
in terms of its generalized curvature.

We now turn to the exact heterotic solution. The tree-level supersymmetric
vacuum equations for the heterotic string are given by
\be
\delta\psi_M=\left(\nabla_M-{\textstyle {1\over 4}}H_{MAB}
\Gamma^{AB}\right)\epsilon=0
\ee
\be
\delta\lambda=\left(\Gamma^A\partial_A\phi-{\textstyle{1\over 6}}
H_{ABC}\Gamma^{ABC}\right)\epsilon=0
\ee
\be
\delta\chi=F_{AB}\Gamma^{AB}\epsilon=0
\ee
where $A,B,C,M=0,1,2,...,9$ and
where $\psi_M,\ \lambda$ and $\chi$ are the gravitino, dilatino and gaugino
fields. The Bianchi identity is given by
\be
dH={\alpha'\over 4} \left({\rm tr} R\wedge R-{1\over 30}{\rm Tr}
F\wedge F\right).
\ee
The $(9+1)$-dimensional Majorana-Weyl fermions decompose into
chiral spinors according to $SO(9,1)\supset SO(5,1) \otimes SO(4)$ for
the $M^{9,1}\to M^{5,1}\times M^4$ decomposition. Then (\ref{ansatz}) with
arbitrary dilaton and with constant chiral spinors $\epsilon_\pm$ solves the
supersymmetry equations with zero background fermi fields provided the YM gauge
field satisfies the instanton (anti) self-duality condition \cite{Strominger1}
\be
F_{\mu\nu}=\pm {1\over 2}\epsilon_{\mu\nu}{}^{\lambda\sigma}
F_{\lambda\sigma}.
\ee
In the absence of a gauge sector, the multi-fivebrane solution is identical to
the ``neutral" tree-level solution shown in (\ref{ansatz}). A perturbative ``gauge"
fivebrane solution was found in \cite{Strominger1}
An exact solution is obtained as follows. Define a generalized connection by
\be
\Omega^{AB}_{\pm M}=\omega^{AB}_M\pm H^{AB}_M 
\ee
in an SU(2) subgroup of the gauge group, and equate it to the gauge
connection $A_\mu$ \cite{Charap1,Charap2} so that the corresponding curvature $R(\Omega_{\pm})$
cancels against the Yang-Mills field strength $F$ and $dH=0$.
For $e^{-2\phi}\Box\ e^{2\phi}=0$ (or $e^{2\phi} = e^{2\phi_0} f$) the
curvature of the generalized connection can be written in
terms of the dilaton as in from which it follows that both $F$ and
$R$ are (anti) self-dual. This solution becomes exact since
$A_\mu=\Omega_{\pm\mu}$ implies that all the higher order corrections vanish
\cite{Callan1,Callan2}. The self-dual solution for the gauge connection is then
given by the 't Hooft ansatz. So the heterotic solution combines a YM instanton
in the gauge sector with an axionic instanton in the gravity sector. In
addition, the heterotic solution has finite action. Further arguments
supporting the exactness of this solution based on $(4,4)$ worldsheet
supersymmetry are shown in \cite{Callan1}. Note that at no point in this discussion
do we refer to the specific form of $f$, so that all of the above
arguments apply for an arbitrary solution of $f^{-1}\Box\ f=0$.

We now go back to the 't Hooft ansatz and solve the equation
$f^{-1}\Box\ f=0$. If we take $f$ to depend on all four coordinates we obtain
a multi-instanton solution
\be f_I=1+\sum_{i=1}^N{\rho_i^2\over |\vec x - \vec a_i|^2}
\ee
where $\rho_i^2$ is the instanton scale size and $\vec a_i$ the location in
four-space of the $i$th instanton. For $e^{2\phi} = e^{2\phi_0} f_I$,
and assuming no dimensions are compactified, we obtain
from (\ref{ansatz}) the neutral fivebrane of \cite{Luelem}  and the exact heterotic
fivebrane
of \cite{Callan1,Callan2} in $D=10$. The solitonic fivebrane tension
$\widetilde{T_6}$ is related to the fundamental string tension $T_2$
($=1/2\pi\alpha'$) by the Dirac quantization condition 
\be
\kappa_{10}^2 \widetilde{T_6} T_2 =n\pi
\ee
where $n$ is an integer and where $\kappa_{10}^2$ is the $D=10$
gravitational constant. This implies $\rho_i^2=e^{-2\phi_0}n_i\alpha'$,
where $n_i$ are integers. Near each source the solution is described by
an exact conformal field theory \cite{Callan1,Callan2,Rey,Khuri}.
\subsection{Monopoles, strings and domain walls in D$=$4}
Instead, let us single out a direction in the transverse four-space (say $x^4$)
and assume all fields are independent of this coordinate. Since all fields
are already independent of $x^5,x^6,x^7,x^8,x^9$, we may consistently assume
the $x^4,x^5,x^6,x^7,x^8,x^9$ are compactified on a six-dimensional torus,
where we shall take the $x^4$ circle to have circumference $Le^{-\phi_0}$
and the rest to have circumference $L$,
so that $\kappa_4^2=\kappa_{10}^2e^{\phi_0}/L^6$. Then the solution
for $f$ satisfying $f^{-1} \Box\ f=0$ has multi-monopole structure.

We may then modify the solution of the 't Hooft ansatz even further and choose
two directions in the four-space $(1234)$ (say $x^3$ and $x^4$) and assume all
fields are independent of both of these coordinates. We may now consistently
assume that $x^3,x^4,x^6,x^7,x^8,x^9$ are compactified on a six-dimensional
torus, where we shall take the $x^3$ and $x^4$ circles to have circumference
$Le^{-\phi_0}$ and the remainder to have circumference $L$, so that
$\kappa_4^2=\kappa_{10}^2 e^{2\phi_0}/L^6$. Then the solution
for $f$ satisfying $f^{-1} \Box\ f=0$ has multi-string structure

We complete the family of solitons that can be obtained from the solutions
of the 't Hooft ansatz by demanding that $f$ depend on only one coordinate,
say $x^1$. We may now consistently assume that $x^2,x^3,x^4,x^7,x^8,x^9$ are
compactified on a six-dimensional torus, where we shall take the $x^2$,
$x^3$ and $x^4$ circles to have circumference $Le^{-\phi_0}$ and the rest to
have circumference $\kappa_4^2=\kappa_{10}^2e^{3\phi_0}/L^6$. Then the solution
of $f^{-1} \Box\ f=0$ has domain wall structure.

 As for the fivebrane in $D=10$, the mass of the monopole, the mass per unit
length of the string and the mass per unit area of the domain wall saturate
a Bogomol'nyi bound with the topological charge. (In the case of the string
and domain, wall, however, we must  extrapolate the
meaning of the ADM mass to non-asymptotically flat spacetimes.)

\subsection{String/string duality}
Let us focus on the solitonic string configuration in the case of
a single source. In terms of the complex field
\be
T=T_1+iT_2=B_{34}+ie^{-2\sigma}=B_{34}+i\sqrt{{\rm det} g^S_{mn}}
\qquad m,n=3,4,6,7,8,9,
\ee
where $g^S_{MN}$ is the string $\sigma$-model metric, the solution takes the
form (with $z=x_1+x_2$)
\be
T={1\over 2\pi i}\ln {z\over r_0}
\ee
\be
ds^2=-dt^2 + dx_5^2 - {1\over 2\pi} \ln{r\over r_0} dz d\overline z
\ee
whereas both the four-dimensional (shifted) dilaton $\eta=\phi + \sigma$
and the four-dimensional two-form $B_{\mu\nu}$ are zero. In terms of the
canonical metric $g_{\mu\nu}$, $T_1$ and $T_2$, the relevant part of the
action takes the form
\be
S_4={1\over 2\kappa_4^2}\int d^4 x\sqrt{-g}
\left( R - {1\over 2T_2^2}g^{\mu\nu}\partial_\mu T \partial_\nu \overline T\right)
\ee
and is invariant under the $SL(2,R)$ transformation
\be
T\to {aT+b\over cT+d},\qquad ad-bc=1.
\ee
The discrete subgroup $SL(2,Z)$, for which $a$, $b$, $c$ and $d$ are
integers, is just a subgroup of the $O(6,22;Z)$ {\it target space duality},
which can be shown to be an exact symmetry of the compactified string theory
at each order of the string loop perturbation expansion.

This $SL(2,Z)$ is to be contrasted with the $SL(2,Z)$ symmetry of the
elementary four-dimensional solution of \cite{Dabholkar}. In their solution $T_1$ and $T_2$ are zero, but $\eta$ and $B_{\mu\nu}$
are non-zero. The relevant part of the action is
\be {S_4={1\over 2\kappa_4^2}\int d^4 x\sqrt{-g}
\left( R - 2g^{\mu\nu}\partial_\mu \eta \partial_\nu \eta
-{1\over 12} e^{-4\eta} H_{\mu\nu\rho} H^{\mu\nu\rho}
\right).}
\ee
The equations of motion of this theory also display an $SL(2,R)$ symmetry,
but this becomes manifest only after dualizing and introducing the axion
field $a$ via
\be
{\sqrt{-g}g^{\mu\nu}\partial_\nu a=
{1\over 3!}\epsilon^{\mu\nu\rho\sigma} H_{\nu\rho\sigma} e^{-4\eta}.}
\ee
Then in terms of the complex field
\be
S_1 + iS_2 =a + ie^{-2\eta} 
\ee
the Dabholkar {\it et al.} fundamental string solution may be written
\be
S={1\over 2\pi i} \ln {z\over r_0}
\ee
\be
ds^2=-dt^2 + dx_5^2 - {1\over 2\pi} \ln {r\over r_0} dz d\overline z.
\ee
Thus the two solutions are the same with the replacement $T\leftrightarrow
S$. It has been conjectured that this second $SL(2,Z)$ symmetry may also be a
symmetry of string theory \cite{Font:1990gx,Sen:1992fr,Senstrong}, but this is far from
obvious order by order in the string loop expansion since it involves a
strong/weak coupling duality $\eta\to - \eta$. What interpretation
are we to give to these two $SL(2,Z)$ symmetries: one an obvious symmetry of
the fundamental string and the other an obscure symmetry of the fundamental
string?

\subsection{Subsequent developments}
\label{subs}
\begin{itemize}
\item{String-string duality in $D=6$ implies $S$-duality in $D=4$}

$S$-duality in $D=4$ gauge theories refers to the conjectured $SL(2,Z)$ symmetry that acts on
the gauge coupling constant $e$ and theta angle $\theta$:
\be
S \rightarrow \frac{aS+b}{cS+d}
\la{sl2zs}
\ee
where $a,b,c,d$ are integers satisfying $ad-bc=1$ and where
\be
S=S_1+iS_2=\frac{\theta}{2\pi} + i\frac{4\pi}{e^2}
\la{S}
\ee
This is also called electric/magnetic duality because the integers $m$ and
$n$ which characterize the magnetic charges $Q_m=n/e$ and electric charges
$Q_e=e(m+n\theta/2\pi)$ of the particle spectrum transform as
\be
\left( \begin{array}{c}
m\\
n
\end{array}
\right)
\rightarrow
\left( \begin{array}{cc}
a&b\\
c&d
\end{array}
\right)
\left( \begin{array}{c}
m\\
n
\end{array}
\right)
\la{charges}
\ee
Such a symmetry would be inherently non-perturbative since, for $\theta=0$
and with $a=d=0$ and $b=-c=-1$, it reduces to the strong/weak coupling duality  
\[
{e}^2/4\pi \rightarrow4\pi/{e}^2
\]
\be
n\rightarrow m, m\rightarrow -n
\label{simple}
\ee
This in turn means that the coupling constant cannot get renormalized in
perturbation theory and hence that the renormalization group
$\beta$-function vanishes
\be
\beta(e)=0
\ee
This is guaranteed in $N=4$ supersymmetric Yang-Mills and also happens in
certain $N=2$ theories.  Moreover electric-magnetic duality 
follows by embedding these Yang-Mills theories in a superstring theory. 
In string theory the roles of the theta angle
$\theta$ and coupling constant $e$ are played by the VEVs of the the
four-dimensional axion field $a$ and dilaton field $\eta$: 
\be
<a>=\frac{\theta}{2\pi}
\ee
\be
{e}^2/4\pi=<e^{\eta}>=8G/\alpha'
\ee
Here $G$ is Newton's constant and $2\pi\alpha'$ is the inverse string tension.
Hence $S$-duality (\ref{sl2zs}) now becomes a transformation law for the
axion/dilaton field $S$: 
\be
S=S_1+iS_2={\rm a}+ie^{-\eta}
\la{Sfield}
\ee

The $S$-duality conjecture in string theory has its origins in supergravity \cite{Ferrara,Deser}. In
the late 70s and early 80s, it was realized that compactified supergravity
theories exhibit non-compact global symmetries  \cite{Cremmerjuliascherk,Cremmerjulia,Duff:1985bv} e.g
$SL(2,R)$, $O(22,6)$, $O(24,8)$, $E_7$, $E_8$, $E_9$, $E_{10}$.  In 1990 it was
conjectured \cite{Luduality2,Luduality3} that discrete subgroups of all these
symmetries   should be promoted to duality symmetries of either heterotic or 
Type
II superstrings. The case for $O(22,6;Z)$ had already been made. 
This is the well-established target space duality, sometimes called {\it
$T$-duality} \cite{Giveonreview}.  Stronger evidence for a strong/weak coupling
$SL(2,Z)$ duality in string theory was subsequently provided in
\cite{Font:1990gx,Rey,Duff:1992iv,Sen:1992fr,Liu,Khurinew,Schwarzsen,%
Senstrong,Khuristring},
stronger evidence for the combination of $S$ and $T$-duality into an $O(24,8;Z)$
in heterotic strings was provided in \cite{Luduality2,Duffclass} and
stronger evidence for their combination into a discrete $E_7$ in Type II 
strings
was provided in \cite{Hulltownsend}, where it was dubbed {\it $U$-duality}.  

Let us first consider $T$-duality and focus just on the moduli fields that arise
in  compactification on a $2$-torus of a $D=6$ string with dilaton $\Phi$, 
metric
$G_{MN}$ and $2$-form potential $B_{MN}$ with $3$-form field strength $H_{MNP}$.
Here the $T$-duality is just $O(2,2;Z)$.  Let us parametrize the compactified
($m,n=4,5$) components of string metric and 2-form as   
\be
G_{mn}=e^{\rho-\sigma}\left( \begin{array}{cc}
e^{-2\rho}+c^2&-c\\
-c&1
\end{array}\right)
\ee
and
\be
B_{mn}=b\epsilon_{mn}
\ee
The four-dimensional shifted dilaton $\eta$ is given by   
\be
e^{-\eta}=e^{-\Phi}\sqrt{det G_{mn}}=e^{-\Phi -\sigma}
\ee
and the axion field ${\rm a}$ is defined by 
\be
\epsilon^{\mu\nu\rho\sigma}\partial_{\sigma}{\rm a}=
\sqrt{-g}e^{-\eta}g^{\mu\sigma}g^{\nu\lambda}g^{\rho\tau}H_{\sigma\lambda\tau}
\ee
where $g_{\mu\nu}=G_{\mu\nu}$ and $\mu,\nu=0,1,2,3$. We further define the 
complex
Kahler form field $T$ and the complex structure field $U$ by  
\[
T=T_1+iT_2=b+ie^{-\sigma}
\]
\be
U=U_1+iU_2=c+ie^{-\rho}
\ee
Thus this $T$-duality may be written as
\be
O(2,2;Z)_{TU} \sim SL(2,Z)_T \times SL(2,Z)_U
\la{Tsplit}
\ee
where $SL(2,Z)_T$ acts on the $T$-field and $SL(2,Z)_U$ acts on the $U$-field
in the same way that $SL(2,Z)_S$ acts on the $S$-field in (\ref{sl2zs}).  In
contrast to $SL(2,Z)_S$, $SL(2,Z)_T \times SL(2,Z)_U$ is known to be not 
merely a
symmetry of the supergravity theory but an exact string symmetry order by order
in string perturbation theory. $SL(2,Z)_T$ does, however, contain a
minimum/maximum length duality mathematically similar to (\ref{simple}) 
\be
R \rightarrow \alpha'/R
\la{R}
\ee
where $R$ is the compactification scale given by
\be
\alpha'/R^2=<e^{\sigma}>.
\ee

Even before compactification, the Type IIB supergravity exhibits an $SL(2,R)$
whose  discrete subgroup has been conjectured to be a non-perturbative
symmetry of the Type IIB string \cite{Callan2,Hulltownsend}.  We shall
refer to this duality as $SL(2,Z)_X$ to distinguish it from the others.  
Combining
this with the known $T$-duality of the four dimensional theory obtained by
compactification on $T^6$ leads to the $E_7$.  So the explanation for 
$U$-duality
devolves upon the explanation for this $SL(2,Z)_X$.
%We shall return to this insection (\ref{triality}).  
 
Let us now investigate how both $N=4$ and $N=2$ exact electric/magnetic duality
follows from string theory.  As discussed above, there is a formal similarity
between this symmetry and that of $T$-duality.  It was argued in
\cite{Duffstrong} that these mathematical similarities between $SL(2,Z)_S$ and
$SL(2,Z)_T$ are not coincidental. Evidence was presented in favor of the idea
that the physics of the fundamental string in six spacetime dimensions may
equally well be described by a dual string and that one emerges as a soliton
solution of the other \cite{Luloop,Lublack,Senstrong,Duffminasian,Duffstrong,Harveystrominger}.
The string equations admits the singular {\it elementary} string solution
\cite{Dabholkar}  
\[
ds^2= (1-k^2/r^2)[-d\tau^2+d\sigma^2 + (1-k^2/r^2)^{-2}dr^2
+r^2d\Omega_{3}{}^2]
\]
\[
e^{\Phi}=1-k^2/r^2
\]
\be
e^{-\Phi}*H_3=2k^2\epsilon_3
\la{fund}
\ee
where
\be
k^2=\kappa^2 T/\Omega_3
\ee
$T=1/2\pi \alpha'$ is the string tension, $\Omega_3$ is the volume of $S^3$ and
$\epsilon_3$ is the volume form. It describes an
infinitely long string whose worldsheet lies in the plane $X^0=\tau,X^1
=\sigma$.  Its mass per unit length is given by 
\be
 M= T<e^{\Phi/2}>
\ee
and is thus heavier for stronger string coupling, as one would expect for
a fundamental string.  The string equations also admit the non-singular
{\it solitonic} string solution
\cite{Lublack,Duffminasian}
\[
ds^2= -d\tau^2+d\sigma^2 + (1-\tilde k^2/r^2)^{-2}dr^2 + r^2d\Omega_{3}{}^2
\]
\[
e^{-\Phi}=1-\tilde k^2/r^2
\]
\be
H_3=2\tilde k^2\epsilon_3
\la{sol}
\ee
whose tension $\tilde T=1/2\pi {\tilde \alpha}'$ is given by
\be
\tilde k^2=\kappa^2 \tilde T/\Omega_3
\ee
Its mass per unit length is given by
\be
\tilde {M}= \tilde T <e^{-\Phi/2}>
\ee
and is thus heavier for weaker string coupling, as one would expect
for a solitonic string. Thus we see that the solitonic string differs from the
fundamental string by the replacements 
\[
\Phi \rightarrow \tilde \Phi=-\Phi
\]
\[
G_{MN} \rightarrow \tilde G_{MN}=e^{-\Phi}G_{MN}
\] 
\[
H \rightarrow \tilde H=e^{-\Phi}*H
\]
\be
\alpha' \rightarrow \tilde \alpha'
\la{dual}
\ee
The Dirac quantization rule $eg=2\pi n$ ($n$=integer) relating the Noether
``electric'' charge
\be
e=\frac{1}{\sqrt{2}\kappa}\int_{S^3}e^{-\Phi}*H_3
\ee
to the topological ``magnetic'' charge
\be
g=\frac{1}{\sqrt{2}\kappa}\int_{S^3}H_3
\ee
translates into a quantization condition on the two tensions:
\be
2\kappa{}^2=n(2\pi)^3\alpha'\tilde \alpha'\,\,\,\,\,\,n=integer
\la{Dirac1}
\ee
where $\kappa$ is the six-dimensional gravitational constant. 
Both the string and dual string soliton solutions break half the
supersymmetries, both saturate
a Bogomol'nyi bound between the mass and the charge. These
solutions are the extreme mass equals
charge limit of more general two-parameter black string solutions
\cite{Horowitz1,Lublack}.

We now make the major assumption of string/string duality: the dual
string may be regarded as a fundamental string in its own right with a 
worldsheet
action that couples to the dual variables and has the dual tension
given in (\ref{dual}). It follows that the dual string equations admit the dual
string (\ref{sol}) as the fundamental solution and the
fundamental string (\ref{fund}) as the soliton solution.	When expressed 
in terms of
the dual metric, however, the former is singular and the latter non-singular. It
also follows from (\ref{Dirac1}) that in going from the fundamental string to
the dual string and interchanging  $\alpha'$ with 
${\tilde\alpha}'=2\kappa^2/(2\pi)^3\alpha'$, one also interchanges the roles of
worldsheet and spacetime loop expansions. Moreover, since the dilaton enters the
dual string equations with the opposite sign to the fundamental string, it was
argued in \cite{Luloop,Lublack,Duffminasian} that in $D=6$ the strong coupling 
regime
of the string should correspond to the weak coupling regime of the dual string:
\be
{\rm g}_6{}^2/(2\pi)^3 = <e^{\Phi}>=(2\pi)^3/{\tilde{\rm g}_6}^2
\la{coupling}
\ee
where ${\rm g_6}$ and $\tilde{\rm g}_6$ are the six-dimensional string
and dual string loop expansion parameters.

On compactification to four spacetime dimensions, the two theories
appear very similar, each acquiring an $O(2,2;Z)$ target space duality.	One's
first guess might be to assume that the strongly coupled
four-dimensional fundamental
string corresponds to the weakly coupled dual string, but in fact
something more subtle and
interesting happens: the roles of the $S$ and $T$ fields are
interchanged \cite{Duffstrong} so that the
strong/weak coupling $SL(2,Z)_S$ of the fundamental string emerges as a
subgroup of the target space
duality of the dual string:
\be
O(2,2;Z)_{SU} \sim SL(2,Z)_S \times SL(2,Z)_U
\la{Ssplit}
\ee
This {\it duality of dualities} is summarized in Table \ref{table}.
\begin{table}
$
\begin{array}{lll}
&Fundamental \, string&Dual \, string\\
&&\\
T-duality&O(2,2;Z)_{TU} &O(2,2;Z)_{SU}\\
&\sim SL(2,Z)_T \times	SL(2,Z)_U&\sim SL(2,Z)_S\times	SL(2,Z)_U\\
Moduli&T={\rm b}+ie^{-\sigma}&S={\rm a}+ie^{-\eta}\\
&{\rm b}=B_{45}&{\rm a}=\tilde B_{45}\\
&e^{-\sigma}=\sqrt{detG_{mn}}&e^{-\eta}=\sqrt{det \tilde{G}_{mn}}\\
Worldsheet \, coupling&<e^{\sigma}>=\alpha'/R^2&<e^{\eta}>={\rm g}^2/2\pi\\
Large/small \, radius &R\rightarrow \alpha'/R&{\rm g}^2/2\pi\rightarrow
2\pi/{\rm g}^2\\
S-duality&SL(2,Z)_S&SL(2,Z)_T\\
Axion/dilaton&S={\rm a}+ie^{-\eta}&T={\rm b}+ie^{-\sigma}\\
&d{\rm a}=e^{-\eta}*H&d{\rm b}=e^{-\sigma}\tilde{*} \tilde{H}\\
&e^{-\eta}=e^{-\Phi}\sqrt{detG_{mn}}&e^{-\sigma}=e^{\Phi}\sqrt{det
\tilde{G}_{mn}}\\
Spacetime \, coupling&<e^{\eta}>={\rm g}^2/2\pi&<e^{\sigma}>=\alpha'/R^2\\
Strong/weak \, coupling&{\rm g}^2/2\pi\rightarrow 2\pi/{\rm g}^2&R
\rightarrow \alpha'/R
\end{array}
$
\label{table}
\caption{Duality of dualities}
\end{table}
As a consistency check, we note that since $(2\pi R)^2/2\kappa^2=1/16\pi G$
the Dirac
quantization rule (\ref{Dirac1}) becomes (choosing $n$=1)
\be
8GR^2=\alpha'\tilde \alpha'
\la{Dirac2}
\ee
Invariance of this rule now requires that a strong/weak coupling
transformation on the fundamental
string ($8G/\alpha'\rightarrow \alpha'/8G$) must be accompanied by a
minimum/maximum length
transformation of the dual string ($\tilde \alpha'/R^2 \rightarrow
R^2/\tilde \alpha'$), and vice
versa.

The idea of an S-duality arising as a T-duality in going from $D$ to $D-2$ on $T^2$ later appeared in the Langlands programme in pure mathematics \cite{Langlands} with $D=6$ and in F-theory with $D=12$ \cite{Vafa} where  the S-duality is that of Type IIB.

\item{Four-dimensional Heterotic/Type IIA/Type IIB triality}

We have seen that in six spacetime dimensions, the heterotic string is dual to a Type IIA string. On further toroidal compactification to four spacetime dimensions, the heterotic string acquires an $SL(2, Z)_S$ strong/weak coupling duality and an $SL(2, Z)_T \times SL(2, Z)_U$ target space duality acting on the dilaton/axion, complex Kahler form and the complex structure fields $S,T,U$ respectively. Strong/weak duality in $D = 6$ interchanges the roles of $S$ and $T$ in $D = 4$ yielding a Type IIA string with fields $T, S, U.$  However, the target space symmetry of the heterotic theory also contains an $SL(2,Z)_U$ that acts on $U$, the complex structure of the torus. This suggests that, in addition to these $S$ and $T$ strings there ought to be a third $U$-string whose axion/dilaton field is $U$ and whose strong/weak coupling duality is $SL(2,Z)_U$. From a $D = 6$ perspective, this seems strange since we now interchange $G_{45}$ and $B_{45}$. Moreover, of the two electric field strengths which become magnetic, one is a winding gauge field and the other is Kaluza-Klein! So such a duality has no $D = 6$ Lorentz invariant meaning. In fact, this $U$ string is a Type IIB string, a result which may also be understood from the point of view of mirror symmetry: interchanging the roles of Kahler form and complex structure (which is equivalent to inverting the radius of one of the two circles) is a symmetry of the heterotic string but takes Type IIA into Type IIB \cite{Dine, Dai}. In summary, if we denote the heterotic, IIA and IIB strings by $H,A,B$ respectively and the axion/dilaton, complex Kahler form and complex structure by the triple $XYZ$ then we have a triality between the $S$-string ($H_{STU} = H_{SUT}$), the T-string ($B_{TUS} = A_{TSU}$) and the $U$-string ($A_{UST} = B_{UTS}$). We
note that D = 6 general covariance is a perturbative symmetry of the Type IIB string and therefore that the D = 4 Type IIB strings must have a perturbative $SL(2,Z)$ acting on the complex structure of the compactifying torus. Secondly we note that for both Type IIB theories, $ B_{TUS}$ and $B_{UTS}$, S is the complex structure field. Thus the T string has $SL(2, Z)_U \times SL(2, Z)_S$ and the U string has $SL(2, Z)_S \times SL(2, Z)_T$ as required. In this sense, four-dimensional string/string/string triality fills a gap left by six-dimensional string/string duality: although duality satisfactorily explains the strong/weak coupling duality of the D = 4 Type IIA string in terms of the target space duality of the heterotic string, the converse requires the Type IIB ingredient \cite{triality}.

 \begin{figure}[h]
\centering\includegraphics[scale=0.3]{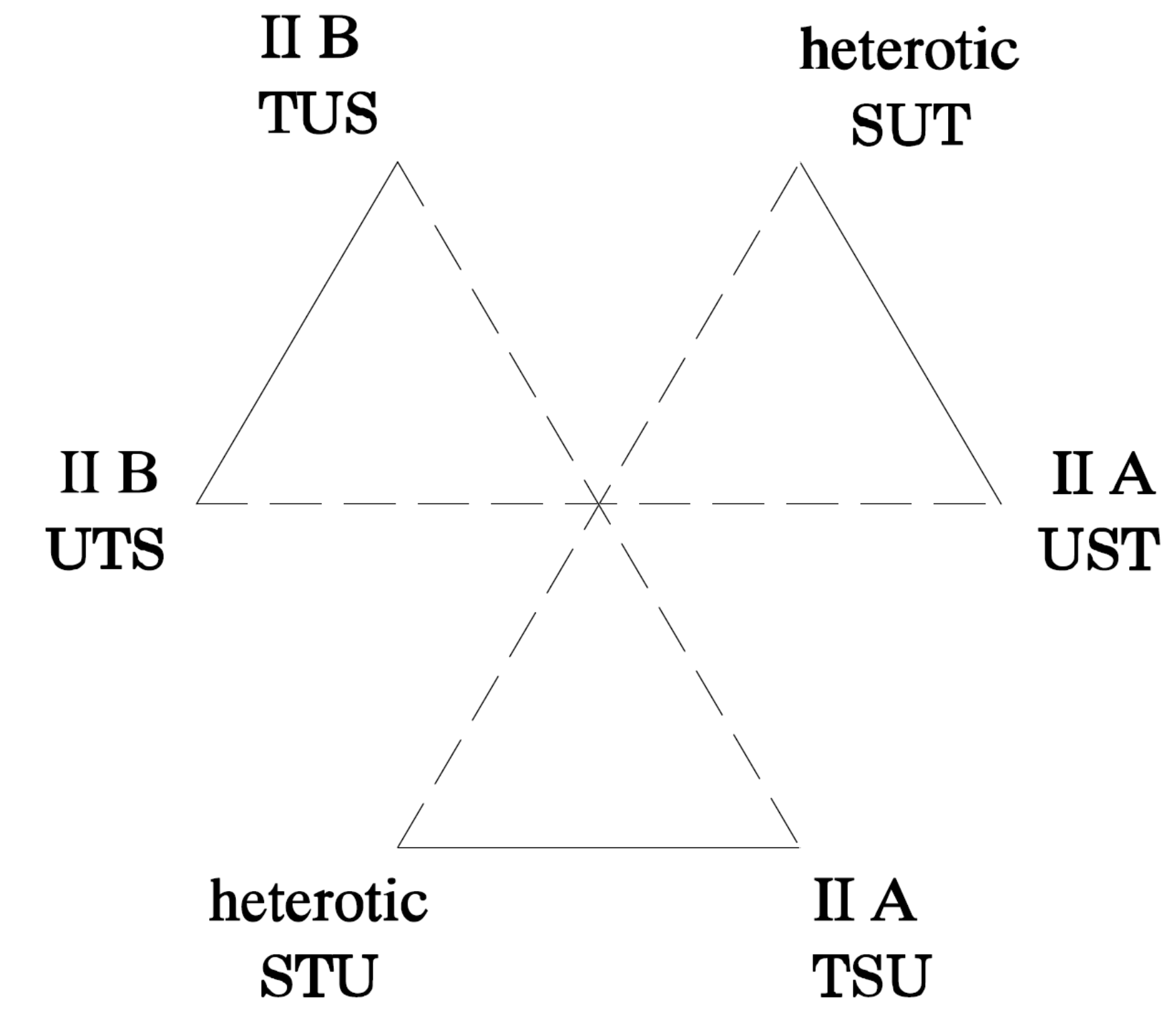}
\caption {Heterotic/Type IIA/Type IIB triality.  The solid lines
correspond to string/string dualities and the dashed lines
represent mirror transformations.}
\label{tri}
\end{figure}

\item{Triality and the STU model}

An interesting subsector of string compactification to four dimensions is provided by the STU model, introduced independently in \cite{Senvafa,triality}. This model has a low energy limit which is described by $N = 2$ supergravity coupled to three vector multiplets interacting through the special Kahler manifold  $[SL(2)/SO(2)]^3$.  (In the version of \cite{Senvafa}, the SL(2) are replaced by a subgroup denoted $\Gamma_0(2)$).  The three complex scalars are denoted by the letters S, T and U, hence the name of the model \cite{triality,stublack}. The remarkable feature that distinguishes it from generic N=2 supergravities coupled to vectors \cite{Kounas} is its S-T-U triality  \cite{triality}.  There are three different versions with two of the SL(2)s perturbative symmetries of the Lagrangian and the third a non-perturbative symmetry of the equations of motion. In a  fourth version all three are non-perturbative \cite{triality,stublack}. All four are on-shell equivalent. If there are in addition four hypermultiplets, the STU model is self-mirror.

  A general static spherically symmetric black hole solution depends on 4 electric and 4 magnetic charges denoted $q_0,q_1,q_2,q_3, p^0,p^1,p^2,p^3$, but the generating solution depends on just $8-3=5$ parameters.  The solution can usefully be embedded in
$N = 4 $ supergravity with symmetry $SL(2) \times SO(6, 22)$, the low-energy limit of the heterotic string compactified on $T^6$, where the charges transform as a ${\bf (2, 28)}$ and also in
$N = 8$ supergravity with symmetry $E_{7(7)}$, the low-energy limit of the Type IIA or Type IIB strings, compactified on $T^6$ or M-theory on $T^7$, where the charges transform as a ${\bf 56}$.
In both cases, remarkably, the same five parameters suffice to describe these $56$-charge black holes \cite{cveticyoum,cvetichull} after fixing the action of the isotropy subgroup $[SO(2)]^3$. The STU black hole entropy is a complicated function of the 8 charges \cite{stublack}:
\be
    (S/\pi)^2 = - (p\cdot q)^2 \\
    + 4\Big[(p^1q_1)(p^2q_2) + (p^1q_1)(p^3q_3) + (p^3q_3)(p^2q_2) \\
    + q_0p^1p^2p^2 - p^0q_1q_2q_3\Big]
    \label{stuentropy}
\ee
Some examples of supersymmetric black hole solutions \cite{Rahmfeld} are provided by the electric Kaluza-Klein black hole with $q = (1, 0, 0, 0)$ and $p= (0, 0, 0, 0)$; the electric winding black hole with $q = (0, 0, 0, -1)$ and $p= (0, 0, 0, 0)$; the magnetic Kaluza-Klein black hole with $q = (0,0,0,0)$ and $p= (0,-1,0,0)$; the magnetic winding black hole with $q = (0,0,0,0)$
and $p= (0, 0, -1, 0)$. By combining these 1-particle states, we may build up 2-, 3- and 4-particle bound states at threshold \cite{Rahmfeld,triality,stublack}. For example $q = (1,0,0,-1)$ and $p= (0,0,0,0)$; $q = (1,0,0,-1)$ and $p= (0,-1,0,0)$; $q = (1,0,0,-1)$ and $p= (0,-1,-1,0)$. The 1-, 2- and 3-particle states all yield vanishing contributions to the entropy. A non-zero value is obtained for the 4-particle example, however, which is just the Reissner-Nordstrom black hole.

\end{itemize}

\section{1996 Ten to eleven: It is not too late}

INTERNATIONAL SCHOOL OF SUBNUCLEAR PHYSICS - Director:  A. ZICHICHI
34th Course:  Effective Theories and Fundamental Interactions - Directors:  G. VENEZIANO - A. ZICHICHI
3 - 12 July 1996

\subsection{ M-theory: the theory formerly known as strings}
%\subsection{M-theory and dualities}
\noindent
Not so long ago it was widely believed that there were five different 
superstring 
theories each competing for the title of ``Theory of everything,''
that all-embracing theory that describes all physical phenomena. See 
Table \ref{Strings}. 

Moreover, on the $(d,D)$ brane-scan of 
supersymmetric extended objects with $d$ worldvolume dimensions 
moving in a spacetime 
of $D$ dimensions, all these theories occupied the same 
$(d=2, D=10)$ slot. See Table \ref{brane-scan}. The orthodox wisdom was 
that while $(d=2, D=10)$ 
was the Theory of Everything, the other branes on the scan were 
Theories of Nothing. 
All that has now changed. We now know that there are not five 
different theories at all but, together with $D=11$ supergravity, they 
form 
merely six different corners of a deeper, unique and more profound 
theory called ``M-theory'' where M stand for Magic, Mystery or 
Membrane. M-theory involves all of the other branes on the brane-scan, 
in particular 
the eleven-dimensional membrane $(d=3, D=11)$ and eleven-dimensional 
fivebrane $(d=6, D=11)$, thus 
resolving the mystery of why strings stop at ten dimensions while 
supersymmetry allows eleven \cite{Duff:1987qa}.

Although we can glimpse various corners of $M$-theory, the big 
picture 
still eludes us. Uncompactified $M$-theory has no dimensionless 
parameters, 
which is good from the uniqueness point of view but makes ordinary 
perturbation 
theory impossible since there are no small coupling constants to 
provide the expansion parameters.  A low 
energy, $E$, expansion is possible in powers of $E/M_{P}$, with 
$M_{P}$ the 
Planck mass, and leads to the familiar $D=11$ supergravity plus 
corrections of higher powers in the curvature. Figuring out what 
governs these corrections would go a long way in pinning down what 
$M$-theory really is. 

\begin{table}
\begin{center}
\vspace{.2in}
\begin{tabular}{l|c|c|c}
&\textbf{Gauge Group} & \textbf{Chiral?} & \textbf{Supersymmetry 
charges}\\ \hline 
\textbf{Type I}& $SO(32)$ & yes & 16\\ \hline 
\textbf{Type IIA} & $U(1)$ & no & 32\\ \hline
\textbf{Type IIB} & -- & yes & 32\\ \hline 
\textbf{Heterotic} & $E_8\times E_8$ & yes & 16\\ \hline 
\textbf{Heterotic} & $SO(32)$ & yes &16 \\
\hline
\end{tabular}
\end{center}
\caption{The Five Superstring Theories}
\la{Strings}
\end{table}

Why, therefore, do we place so much trust in a theory we cannot even 
define? First we know that its equations (though not in general its 
vacua) 
have the maximal number of 32 supersymmetry charges. This is a 
powerful 
constraint and provides many ``What else can it be?'' arguments in 
guessing what the theory looks like when compactified to $D<11$ 
dimensions. For example, when $M$-theory is compactified on a circle 
$S^{1}$ of radius $R_{11}$, it leads to the Type IIA string, with 
string 
coupling constant $g_{s}$ given by
\be
g_{s}=R_{11}^{3/2}   
\ee
We recover the weak coupling regime only when $R_{11}\rightarrow 0$, 
which 
explains the earlier illusion that the theory is defined in 
$D=10$. Similarly, if we compactify on a line segment $S^{1}/Z_{2}$, we recover the $E_{8} \times E_{8}$ 
heterotic string.  Moreover, although the corners of M-theory we 
understand best 
correspond to the weakly coupled, perturbative, regimes where the 
theory can be approximated by a string theory, they are 
related to one another by a web of dualities, some of which are 
rigorously established and some of which are still conjectural but 
eminently plausible. For example, if we further compactify Type IIA 
string on a 
circle of radius 
$R$, we can show rigorously that it is equivalent to the Type IIB 
string 
compactified on a circle or radius $1/R$. If we do the same thing for 
the $E_{8} \times E_{8}$ heterotic string we recover the $SO(32)$ 
heterotic string. These well-established relationships which remain 
within the 
weak coupling regimes are called {\it T-dualities}. The 
name {\it $S$-dualities} refers to the less well-established 
strong/weak coupling 
relationships. For example, the $SO(32)$ heterotic string is believed 
to be $S$-dual to the $SO(32)$ Type I string, and the Type IIB 
string to be self-$S$-dual. If we compactify more dimensions, other 
dualities can appear. For example, the heterotic string compactified on a 
six-dimensional torus $T^{6}$ is also believed to be self-$S$-dual. 
There is also the phenomenon of {\it duality of dualities} 
by which the $T$-duality of one theory is the $S$-duality of another. 
When M-theory is compactified on $T^{n}$, these $S$ and $T$ dualities 
are combined into what are termed $U$-dualities. 
All the consistency checks we have been able to 
think of (and after 20+ years there are dozens of them) have worked and 
convinced us that all these dualities are in fact valid. Of course we 
can compactify M-theory on more complicated manifolds such as the 
four-dimensional $K3$ or the six-dimensional Calabi-Yau spaces and 
these lead to a bewildering array of other dualities. For example: the 
heterotic string on $T^{4}$ is 
dual to the Type II string on $K3$; the heterotic 
string on $T^{6}$ is dual to the    
the Type II string on Calabi-Yau; the Type IIA string on a 
Calabi-Yau 
manifold is dual to the Type IIB string on the mirror Calabi-Yau 
manifold.
These more complicated compactifications lead to many more parameters 
in the theory, known to the mathematicians as {\it moduli}, but in 
physical uncompactified spacetime have the interpretation as 
expectation values of scalar fields \cite{Duff:1983vj}. Within string perturbation 
theory, these scalar fields have flat potentials and their 
expectation 
values are arbitrary. So deciding which topology Nature 
actually chooses and the values of the moduli within that topology is 
known as the {\it vacuum degeneracy problem}.

\subsection{String/string duality from M-theory}

Let us consider M-theory, with its fundamental membrane and solitonic fivebrane, on $R ^6 \times M^1 \times {\tilde M}^4$ where $M^1$ is a one-dimensional compact space of radius $R$ and ${\tilde M}^4$ is a four-dimensional compact space of volume V. We may obtain a fundamental string on $R^ 6$ by wrapping the membrane around $M^1$ and reducing on ${\tilde M}^4$. Let us denote the fundamental string sigma-model metrics in $D = 10$ and $D = 6$ by $G_{10}$ and $G_{6}$. Then from the corresponding Einstein Lagrangians
\be
\sqrt{-G_{11}}R_{11}=R^3\sqrt{-G_{10}}R_{10}=\frac{V}{R}\sqrt{-G_6}R_6
\ee
we may read off the strength of the string couplings in D = 10 \cite{DMW}
\be
\lambda_{10}{}^2=R^3
\ee
and D = 6
\be
\lambda_{6}{}^2=\frac{R}{V}
\ee

Similarly we may obtain a solitonic string on $R^6$  by wrapping the fivebrane around  ${\tilde M}^4$ and reducing on $M^1$.  Let us denote the solitonic string sigma-model metrics in $D = 7$ and $D = 6$ by ${\tilde G}_{7}$ and ${\tilde G}_{6}$. Then from the corresponding Einstein Lagrangians
\be
\sqrt{-G_{11}}R_{11}=V^{-3/2}\sqrt{-{\tilde G}_{7}}{\tilde R}_{7}=\frac{R}{V}\sqrt{-{\tilde G}_6}{\tilde R}_6
\ee
we may read off the strength of the string couplings in D = 7 \cite{DMW}
\be
{\tilde \lambda}_{7}{}^2=V^{-3/2}
\ee
and D = 6
\be
{\tilde \lambda}_{6}{}^2=\frac{V}{R}
\ee

Thus we see that the fundamental and solitonic strings are related by a strong/weak coupling:
\be
{\tilde \lambda}_6{}^2=1/\lambda_6{}^2
\ee

We shall be interested in $M^1 = S^ 1$ (in which case the fundamental string will be Type IIA) or $M^1= S^1/Z^2$ (in which case  the fundamental string will be heterotic $E_8 \times E_8$). Similarly, we will be interested in ${\tilde M}^4= T ^4$ (in which case the solitonic string will be Type IIA) or ${\tilde M}^4= K3$ (in which case the solitonic string will be heterotic). Thus there are four possible scenarios which are summarized in Table \ref{String/String dualities}. $(N_+, N_-)$ denote the $D = 6$ spacetime supersymmetries. In each case, the fundamental string will be weakly coupled as we shrink the size of the wrapping space $M^1$ and the dual string will be weakly coupled as we shrink the size of the wrapping space ${\tilde M}^4$.

In fact, there is in general a topological obstruction to wrapping the fivebrane around ${\tilde M}^4$ provided by
\be
\int{F_4}=2\pi m
\ee
where $F$ is the 4-form field strength of D = 11 supergravity, because the fivebrane cannot wrap around a 4-manifold that has $m \neq 0$. This is because the anti-self-dual 3-form field strength $T$ on the worldvolume of the fivebrane obeys
\be
dA_3=F_4
\ee
and the existence of a solution for $A_3$ therefore requires that $F_4$ must be cohomologically trivial. For M-theory on $R^6 \times S^1/Z^2 \times T^4$ this is no problem. However, for M-theory on $R^6 \times S^1/Z^2 \times K_3$, with instanton number $k$ in one $E_8$ and $(24 - k)$ in the other, the flux of $F_4$ over $K3$ is \cite{DMW}
\be
m=12-k
\ee
Consequently, the M-theoretic explanation of heterotic/heterotic duality requires $E_8 \times  E_8$ with the symmetric embedding $k = 12$. This has some far-reaching implications. For example, the duality exchanges gauge fields that can be seen in perturbation theory with gauge fields of a nonperturbative origin \cite{DMW}.

The dilaton ${\tilde \Phi} $, the string sigma-model metric ${\tilde G}_{MN}$ and 3-form field strength ${\tilde H}$ of the dual string are related to those of the fundamental string, $\Phi $, ${G}_{MN}$, and ${H}$ by the replacements
\[
\Phi \rightarrow {\tilde \Phi}=-\Phi
\]
\[
G_{MN} \rightarrow {\tilde G}_{MN}=e^{-\Phi}G_{MN}
\]
\be
H\rightarrow {\tilde H}= e^{-\Phi}*H
\ee
In the case of heterotic/Type IIA duality and Type IIA/heterotic duality, this operation takes us from one string to the other, but in the case of heterotic/heterotic duality and Type IIA/Type IIA duality this operation is a discrete symmetry of the theory. This Type IIA/Type IIA duality is hardly ever discussed in the literature in these terms, but we can recognise this symmetry as a subgroup of the SO(5, 5; Z) U-duality of the $D = 6$ Type IIA string.

\begin{table}
\begin{center}
\begin{tabular}{ccccccccccc}
$\textbf{$(N_+,N_-)$}$ &$\textbf{$M^1$}$&$\textbf{${\tilde M}^4$}$&Fundamental string&Dual string\\
&&&&\\
% \hline 
$\textbf{(1,0)}$& $S^1/Z_2$ & K3&heterotic&heterotic\\
%\hline 
$\textbf{(1,1)}$& $S^1$ & K3&Type IIA&heterotic\\
%\hline
$\textbf{(1,1)} $&$S^1/Z_2$&$T^4 $&heterotic&Type IIA \\
%\hline  
$\textbf{(2,2)} $&$S^1$&$T^4 $&Type IIA&Type IIA\\
%\hline 
\end{tabular}
\end{center}
\caption{String/String dualities}
\la{String/String dualities}
\end{table}

%\section{M=physics}
\subsection{Subsequent developments}
\begin{itemize}
\item{}
These include F-theory \cite{Vafa},  strong coupling expansion of Calabi-Yau compactifications \cite{Wittencalabi},  and string dynamics in six dimensions \cite{Seibergwittenphase,Duff:1996cf,Berkooz}.
\item{}
A more mathematical approach to M-theory may be found in a series of papers  involving Fiorenza, Huerta, Sati and Schreiber. See, for example, \cite{Huerta:2017utu,Sati:2018tvj,Fiorenza:2018ekd,Huerta:2018xyh,Fiorenza:2019usl}.
\end{itemize}
\section{2003 The status of local supersymmetry}

INTERNATIONAL SCHOOL OF SUBNUCLEAR PHYSICS - Director:  A. ZICHICHI
41st Course: From Quarks to Black Holes:  Progress in Understanding the Logic of Nature 
Directors:  G. t HOOFT - A. ZICHICHI
29 August - 7 September 2003

\subsection{Supersymmetry without Supermembranes: Not an option}
\label{compulsory}
{\it Gravity exists, so if there is any truth to supersymmetry then any
realistic supersymmetry theory must eventually be enlarged to a
supersymmetric theory of matter and gravitation, known as
supergravity.  Supersymmetry without supergravity is not an option,
though it may be a good approximation at energies below the Planck
Scale.}

Steven Weinberg, The Quantum Theory of Fields, Volume III, Supersymmetry

To paraphrase Weinberg:

{\it
Supergravity is itself only an effective nonrenormalizable theory
which breaks down at the Planck energies.  So if there is any truth to
supersymmetry then any realistic theory must eventually be enlarged to
superstrings which are ultraviolet finite.  Supersymmetry without 
superstrings is not an option.}

To paraphrase Weinberg again:

{\it
Superstring theory is itself only a perturbative theory which breaks down
at strong coupling.  So if there is any truth to supersymmetry then any
realistic theory must eventually be enlarged to the non-perturbative
M-theory, a theory involving higher dimensional extended objects: the
super $p$-branes\footnote{In my opinion, calling M-theory the strong coupling 
limit of string theory is a bit like calling string theory the high-energy limit of 
general relativity.}. Supersymmetry without M-theory is not an option.}

Yet two of the most basic questions of M-theory have until
now remained unanswered:

\subsection {What are the $D=11$ symmetries?}

In this lecture we argued that the equations of M-theory 
possess previously unidentified hidden spacetime (timelike and null) 
symmetries in addition to the well-known hidden internal (spacelike) 
symmetries.  For $11 \geq d \geq 3$, these coincide with the generalized 
structure groups discussed below and take the form ${\cal 
G}=SO(d-1,1) \times G(spacelike)$, ${\cal G}= ISO(d-1) \times 
G(null)$ and ${\cal G}=SO(d) \times G(timelike)$ with $1\leq d<11$.  
For example, $G(spacelike)=SO(16)$, $G(null)=[SU(8) \times 
U(1)] \times R^{56}$ and $G(timelike)=SO^*(16)$ when $d=3$.  The 
nomenclature derives from the fact that these symmetries also show up 
in the spacelike, null and timelike dimensional reductions of the 
theory.  However, we emphasize that we are proposing them as 
background-independent symmetries of the full unreduced and 
untruncated $D=11$ equations of motion, not merely their dimensional 
reduction.  Although extending spacetime symmetries, there is no 
conflict with the Coleman-Mandula theorem.  A more speculative idea \cite{holo} is 
that there exists a yet-to-be-discovered version of $D=11$ 
supergravity or $M$-theory that displays even bigger hidden symmetries 
corresponding to ${\cal G}$ with $d\leq 3$ which 
could be as large as $SL(32,R)$.

\begin{table}[t]
\begin{center}
\begin{tabular}{c|ccc}
$d/(11-d)$&$G(spacelike)$&$G(null)$&$G(timelike)$\\
\hline
11/0&$\{1\}$&$\{1\}$&$\{1\}$\\
10/1&$\{1\}$&$\{1\}$&$\{1\}$\\
9/2&$\SO(2)$&$\R$&$\SO(1,1)$\\
8/3&$\SO(3) \times \SO(2)$&$\ISO(2) \times \R$&$\SO(2,1) \times \SO(1,1)$\\
7/4&$\SO(5)$&$[\SO(3) \times \SO(2)] \times
\R6_{(3,2)}$&$ \SO(3,2)$\\
6/5&$\SO(5) \times \SO(5)$&$\SO(5) \times \R^{10}_{(10)}$&$\SO(5,\C)$\\
5/6&${\rm USp}(8)$&$[\SO(5) \times \SO(5)] \times
\R^{16}_{(4,4)}$&${\rm USp}(4,4)$\\
4/7&$\SU(8)$&${\rm USp}(8)\times \R^{27}_{(27)}$&$\SU^*(8)$\\
3/8&$\SO(16)$&$[\SU(8) \times \U(1)]\times
                        \R^{56}_{(28_{1/2},\overline{28}_{-1/2})}$&$\SO^*(16)$\\
\hline
2/9&$\SO(16) \times \SO(16)$&$\SO(16) \times \R^{120}_{(120)}$&$\SO(16,\C)$\\
1/10&$\SO(32)$&$[\SO(16)\times\SO(16)]\times\R^{256}_{(16,16)}$&$\SO(16,16)$\\
0/11&$\SL(32,\R)$&$\SL(32,\R)$&$\SL(32,\R)$
\end{tabular}
\end{center}
\caption{The generalized structure groups are given by ${\cal
G}=\SO(d-1,1) \times G(spacelike)$, ${\cal G}= \ISO(d-1) \times
G(null)$ and ${\cal G}= \SO(d) \times G(timelike)$.}
\label{gen}
\end{table}

\subsection{Counting supersymmetries  of M-theory vacua}

The equations of M-theory display the maximum number of supersymmetries
$N$=32, and so $n$, the number of supersymmetries preserved by a
particular vacuum, must be some integer between 0 and 32.  But are
some values of $n$ forbidden and, if so, which ones?  For quite some
time it was widely believed that, aside from the maximal $n=32$, $n$
is restricted to $0\leq n\leq 16$ with $n=16$ being realized by the
fundamental BPS objects of M-theory: the M2-brane, the M5-brane, the
M-wave and the M-monopole.  The subsequent discovery of intersecting
brane configurations with $n=0$, 1, 2, 3, 4, 5, 6, 8, 16 lent credence
to this argument.  On the other hand, it has been shown
that all values $0\leq n \leq 32$ are in principle allowed by the
M-theory algebra and examples of 
vacua with $16< n < 32$ have indeed 
since been found.  

In M-theory vacua with vanishing 4-form $F_{(4)}$, one can invoke the
ordinary Riemannian holonomy $H \subset SO(10,1)$ to account for
unbroken supersymmetries $n=1, 2, 3, 4, 6, 8, 16, 32$.  To explain the
more exotic fractions of supersymmetry, in particular $16<n<32$, we
need to generalize the notion of holonomy to accommodate non-zero
$F_{(4)}$.
 We show that the number of supersymmetries 
 preserved by an M-theory vacuum is given by the number of singlets 
 appearing in the decomposition of the 32-dimensional representation of 
 ${\cal G}$ under ${\cal G} \supset {\cal H}$ where ${\cal G}$ are 
 generalized structure groups that replace $SO(1,10)$ and ${\cal H}$ 
 are generalized holonomy groups.  In general we require the maximal 
 ${\cal G}$, namely $SL(32,R)$, but smaller ${\cal G}$ appear in 
 special cases such as product manifolds.

\subsection{Subsequent developments}
\begin{itemize}
\item{Generalized holonomy}

Generalized holonomy is developed further in \cite{hidden,holo,lps,alex}.  We conjectured, albeit on flimsy evidence, that the number of vacuum supersymmetries allowed by M-theory is restricted to
\[n = 0,1,2,3,4,5,6,8,10,12,14,16,18,20,22,24,26,28,32\]
Interestingly enough, a Godel universe with $n = 14$ was subsequently discovered \cite{harmark} which completes this list. 
Furthermore: $n = 31$ has now been ruled out for both Type IIB \cite{gran} and Type IIA \cite{bandos}. $n = 30$ has now been ruled out for M-theory \cite{gran2}. The class of M-theory plane waves found in \cite{gran2} has $ = 16, 20, 26$ but not $n = 28$, although plane wave solutions with $n = 28$ do appear in Type IIB \cite{bena}.  Backgrounds with $n > 24$ are necessarily (locally) homogeneous. See \cite{OF} where it is also conjectured that 24 is the minimum number which guarantees this. See \cite{Gran} for a recent review.
\item{LHC}

According to much of the popular media (and even some phenomenologists and experimentalists), the failure to find supersymmetric particles at the LHC signals the demise of supersymmetry. See, for example,

https://www.economist.com/science-and-technology/2016/11/12/a-bet-about-a-cherished-theory-of-physics-may-soon-pay-out

https://www.forbes.com/sites/startswithabang/2017/10/06/five-brilliant-ideas-for-new-physics-that-need-to-die-already/\#4472764857b7

https://www.scientificamerican.com/article/is-supersymmetry-dead/

https://www.theguardian.com/science/2013/jun/16/has-physics-gone-too-far

Since the super in superstring and supermembrane refers to supersymmetry, this failure to detect any superpartners is also said to cast doubt on string and M-theory, but string and M-theory are compatible with supersymmetry becoming evident only at much higher energies. In fact, they are silent about what energies supersymmetry would reveal itself. This is a valid criticism of our current state of knowledge but it is not a ``fudge'', as some journalists have claimed.  One would expect to see super-particles at the LHC only if supersymmetry is the solution to the ``gauge-hierarchy problem''. This is an extra assumption, favoured by some particle physicists, but it is not intrinsic to supersymmetry. Many of those same journalists think that supersymmetry was invented to solve the gauge hierarchy problem when in fact it precedes it.

When in the 1970s, encouraged by Abdus Salam and Chris Isham at Imperial College,  and by visits of Stanley Deser, I embarked on a career devoted to quantum gravity (a force forty orders of magnitude weaker than the others) I was well aware that this meant a departure from the kind of close association of theory and experiment traditionally enjoyed by particle physicists. We were in it for the long haul and empirical confirmation, if it came at all, was likely to be indirect. Nevertheless I thought it worthwhile given that the unification of gravity and quantum theory is the most important unresolved quandary in science.
Strange then that many journalists seem to regard this as new problem unique to string/M theory and/or supersymmetry.
I, along with many others belonging to the ``hep-th'' wing of theoretical physics, was attracted to global and local supersymmetry because they are respectively the square root of special and general relativity and hence provide a natural framework for incorporating gravity. There were three outstanding issues in quantum gravity in the 1970s:
 (1) Ultraviolet divergences and non-renormalizabilIty (2) The microscopic origin of the Bekenstein-Hawking black hole entropy (3) The black hole information paradox. Supersymmetry in the form of string theory has since provided an answer to (1), supersymmetry in the form of M-theory has since provided an answer to (2) and supersymmetry in the form of AdS/CFT has (according to Hawking and others) since provided an answer to (3). Moreover, I know of no other theory that provides adequate answers to any of (1), (2) or (3). Supersymmetry is still alive and kicking.

\end{itemize}

\section{2010 Black holes, qubits and quantum information}

INTERNATIONAL SCHOOL OF SUBNUCLEAR PHYSICS - Director: A. ZICHICHI 48th Course: What is Known and Unexpected at LHC 
Directors: G. t HOOFT - A. ZICHICHI 29 August - 7 September 2010

\subsection{Three qubits: Alice, Bob and Charlie}

Remarkably, there is  a correspondence between the measure of tripartite entanglement of three qubits and the entropy $S$ of the 8-charge STU black hole of Section  \ref{subs}.  Both are given by Cayley's hyperdeterminant 
\cite{Duff:2006uz}.

The three qubit system (where $A, B, C=0,1$)  is described by the state
\[
 |\Psi\rangle = a_{ABC}|ABC\rangle 
 \]
 \[
= a_{000}|000\rangle+a_{001}|001\rangle+a_{010}|010\rangle +a_{011}|011\rangle 
\]
\be
+a_{100}|100\rangle+a_{101}|101\rangle+a_{110}|110\rangle+a_{111}|111\rangle 
\ee
The tripartite entanglement of Alice, Bob and Charlie is given by the ``3-tangle''  \cite{tangle}
\beq
\tau_{ABC}=4|{\rm Det}~a_{ABC}|
\eeq
where ${\rm Det}~a_{ABC}$ is  Cayley's hyperdeterminant quartic in the hypermatrix $a_{ABC}$
\[
{\rm Det}~a_{ABC}=-\frac{1}{2}\epsilon^{A_{1}A_{2}}\epsilon^{B_{1}B_{2}}\epsilon^{A_{3}A_{4}}\epsilon^{B_{3}B_{4}}\epsilon^{C_{1}C_{4}}\epsilon^{C_{2}C_{3}}
{a}_{A_{1}B_{1}C_{1}}{a}_{A_{2}B_{2}C_{2}}{a}_{A_{3}B_{3}C_{3}}{a}_{A_{4}B_{4}C_{4}}
\]
\be
=  a_{000}^2 a_{111}^2 + a_{001}^2 a_{110}^2 +a_{010}^2 a_{101}^2 + a_{100}^2 a_{011}^2
\ee
\[
-2(a_{000}a_{001}a_{110}a_{111}+a_{000}a_{010}a_{101}a_{111}
+ a_{000}a_{100}a_{011}a_{111}+a_{001}a_{010}a_{101}a_{110}
\]
\be
+ a_{001}a_{100}a_{011}a_{110}+a_{010}a_{100}a_{011}a_{101})
+ 4 (a_{000}a_{011}a_{101}a_{110} + a_{001}a_{010}a_{100}a_{111})
\label{hyperdet}
\ee
The hyperdeterminant is invariant under $SL(2)_{A} \times SL(2)_{B} \times SL(2)_{C}$, with $a_{ABC}$ transforming as a $(2,2,2)$, and under a discrete triality that interchanges A, B and C. 

\subsection{STU black holes}
By identifying the 8 charges with the 8 components of the three-qubit hypermatrix $a_{ABC}$, 
\be
(p^0, p^1, p^2, p^3, q_0, q_1, q_2, q_3)=
(a_{000}, -a_{001}, -a_{111}, -a_{011}, a_{110}, a_{101}, a_{101}, a_{010})
\ee
 one finds \cite{Duff19}
 \be
S=\pi\sqrt{|{\rm Det}~a_{ABC}|}=\frac{\pi}{2}\sqrt{\tau_{ABC}}
\ee
This turns out to be just the tip of an iceberg and further papers  \cite{Kallosh:2006zs,Levay:2006kf,Duff:2006ue,Levay:2006pt,Levay:2007nm,Duff:2007wa,PR,wrap,BDL} have written a more complete dictionary, which translates a variety of phenomena in one language to those in the other. For example, in the N = 2 theory  the 3-qubit entanglement classification, is matched by the black hole classification with either 1/2 or 0 fraction of supersymmetry preserved. By embedding in the N = 8 theory, we can  include the finer supersymmetry preserving distinctions.

There is, in fact, a quantum information theoretic interpretation of the 56 charge N = 8 black hole in terms of a Hilbert space consisting of 7 copies of the 3-qubit Hilbert space \cite{Duff:2006ue,Duff:2008eei}. It relies on $[SL(2)]^7$ being a subgroup of $E_{7(7)}$ and admits the interpretation, via the Fano plane, of a tripartite entanglement of seven qubits, with the entanglement measure given by Cartans quartic $E_{7(7)}$ invariant. Remarkably, however, because the generating solution depends on the same 5 parameters as the STU model, its classification of states will exactly parallel that of the usual 3-qubits. Indeed, the Cartan invariant reduces to Cayley's hyperdeterminant in a canonical basis \cite{Kallosh:2006zs}. Nevertheless, we still do not know whether there are any physical reasons underlying these mathematical coincidences.

\subsection{Wrapped branes as qubits}

In the same spirit we consider the configurations of intersecting D3-branes, whose wrapping around the six compact dimensions $T^6$ provides the microscopic string-theoretic interpretation of the charges, and associate the three-qubit basis vectors $ |ABC\rangle$, ($A,B,C=0$ or $1$) with the corresponding 8 wrapping cycles \cite{wrap}. 

Thus our microscopic analysis of the black hole has provided an explanation for the appearance of the qubit two-valuedness (0 or 1) that was lacking in the previous macroscopic treatments \cite{Duff:2006uz,Kallosh:2006zs,Levay:2006kf,Duff:2006ue,Levay:2006pt,Duff:2007wa,Levay:2007nm}: the brane can wrap one circle or the other in each $T^2$.  The number of qubits is three because of the six extra dimensions of string theory. 
\begin{table}
\begin{center}
\renewcommand{\arraystretch}{1.6}
\renewcommand{\tabcolsep}{4pt}
\begin{tabular}{||cc|cc|cc||c|c|c|}
\hline
4&5&6&7&8&9&$\rm{macro~charges}$&\rm{micro~charges}&$|ABC>$\\
\hline
 x & o & x & o & x  & o &  $ p^0$& 0&$|000>$\\
 o & x & o & x & x  & o &  $q_1$&0& $|110>$\\
 o & x & x &  o & o & x &  $q_2$&$-N_3 {\rm sin}\theta {\rm cos}\theta$&$|101>$\\
 x & o & o & x & o & x  & $ q_3$&$N_3 {\rm sin}\theta {\rm cos}\theta$&$|011>$\\
\hline
 o & x & o & x & o& x&$q_0$& $N_0+N_3 {\rm sin}^2\theta$&$|111>$\\
 x & o & x &  o & o & x  &$-p^1$&$ -N_3 {\rm cos}^2\theta$&$|001>$\\
 x & o & o & x & x & o &$-p^2$&$-N_2$&$|010>$\\
 o & x & x & o & x  & o&$ -p^3$& $-N_1$&$|100>$ \\
\hline
\end{tabular}
\caption{Three qubit interpretation of the 8-charge D=4 black hole from  four D3-branes wrapping around the lower four cycles of $T^6$ with wrapping numbers $N_0,N_1,N_2,N_3$ . Note that they intersect over a string at angle $\theta$.}
\label{intersect3}
\end{center}
\end{table}

To wrap or not to wrap: that is the qubit.

In particular, we relate a well-known fact of quantum information theory, that the most general real three-qubit state can be parameterized by four real numbers and an angle, to a well-known fact of string theory, that the most general $STU$ black hole can be described by four D3-branes intersecting at an angle.

\subsection{Subsequent developments}

\begin {itemize}
\item{Quantum information}

Falsifiable predictions in the fields of high-energy physics or cosmology are hard to come by, especially for ambitious attempts, such as string/M-theory, to accommodate all the fundamental interactions. In the field of quantum information theory, however, the work described in this lecture has shown that the stringy black hole/qubit correspondence can reproduce well-known results in the classification of two and three qubit entanglement. In \cite{4qubit} this correspondence was taken one step further to predict new results in the less well-understood case of four-qubit entanglement that can in principle be tested in the laboratory.

\item{It from bit?}

Looking at the hep-th arXiv in 2018, we see that quantum information has become a dominant theme that has attracted the attention of leading string theorists, for example \cite{Maldainfo,Sussinfo,Witteninfo}. However, we cannot claim much credit for this since these developments have not followed the kind of black hole/qubit correspondence discussed above. A different kind of black hole/qubit correspondence, namely ER=EPR has been very influential \cite{ER} as has the holographic derivation of entanglement entropy \cite{RT1,RT2}.

\end{itemize}
\section{\bf 2016 M-physics}
\label{M}

\subsection{Oxford English Dictionary: M-theory}
\la{OED2}
\indent
{\bf M-theory}, n.  Particle Physics.

Brit.  $\epsilon$m$\theta$ieri , U.S. $\epsilon$m$\theta$ieri, $\epsilon$m$\theta$eri

[< M (app. representing MEMBRANE n.) + THEORY n.1]

A unified theory involving branes that subsumes eleven-dimensional supergravity and the
five ten-dimensional superstring theories.

\smallskip

Quot. 1996 is from a paper received for publication earlier (23 Oct. 1995) than quot.
1995 (17 Dec.).

{\bf 1995} Re: Confinement: Massive Gauge Bosons in sci.physics (Usenet newsgroup)
17 Dec., String theorists are a mathematically sophisticated crew, so I'm sure they would
enjoy an abstract description of the M-theory (as it's called) from which one could
then derive all its varied guises.

{\bf 1996} J. H. Schwarz  in Physics Lett. B. 367 97/1
If one assumes the existence of a fundamental theory in eleven dimensions
(let's call it the M theory), this provides a powerful heuristic basis for
understanding various results in string theory. [Note] This name was suggested by E. Witten.

{\bf 1998} Sci. Amer. Feb. 59/2 Despite all these successes, physicists are glimpsing only small
corners of M-theory; the big picture is still lacking.

{\bf 2002} U.S. News \& World Rep. 6 May
59/1 M-theory..holds that our universe may occupy just part of a many-dimensional
mega-universe. In that picture, it could be shadowed by another universe on a
different brane- M-theory jargon for 3-D membrane.

\subsection{Where M stands for\ldots}

More M-etymology:

``Recent results indicate that if one assumes the existence of a
fundamental theory in eleven dimensions (let's call it the `M-theory'
[This name was suggested by E. Witten]), this provides a powerful
heuristic basis for understanding various results in string theory.''
J. Schwarz, hep-th/9510101.

``As it has been proposed that the eleven-dimensional theory is a
supermembrane theory but there are some reasons to doubt that
interpretation, we will non-committally call it M-theory, leaving for
future the relation of {\it M} to membranes.'' P. Horava and E. Witten,
hep-th/9510209

``For instance, the eleven-dimensional `M-theory' (where M stands for
magic, mystery or membrane, according to taste) on $X \times S^{1}$,
with $X$ any ten-manifold, is equivalent to Type IIA on $X$, with a
Type IIA string coupling constant that becomes small when the radius
of $S^{1}$ goes to zero.'' E. Witten, hep-th/9512219

\subsection{Subsequent developments}
\begin{itemize}
\item{M-theory and string theory}

What is the future of branes? I will finish on an optimistic note borrowed from Scientific American \cite{sciam} (and Isaac Newton):

{\it Edward Witten is fond of imagining how physics might develop on another planet, where major discoveries such as general relativity, quantum mechanics and supersymmetry are made in a different order than on Earth. In a similar vein, I would like to suggest that on planets more logical than ours, branes in 11 dimensions would have been the starting point from which 10-dimensional string theory was subsequently derived. Indeed, future terrestrial historians may judge the late 20th century as a time when theorists were like children playing on the seashore, diverting themselves in now and then finding a smoother pebble or prettier shell in superstrings, whilst the great ocean of M-theory lay all undiscovered before them.}
\end{itemize}
\section{2017 Thirty years of Erice on the Brane}

INTERNATIONAL SCHOOL OF SUBNUCLEAR PHYSICS - Director: A. ZICHICHI
55th Course: Highlights from LHC and the other Frontiers of Physics
Directors: G. 't HOOFT  A. ZICHICHI 
14 - 23 June 2017
\subsection{Nino}
\begin{figure}
\centering\includegraphics[scale=0.25]{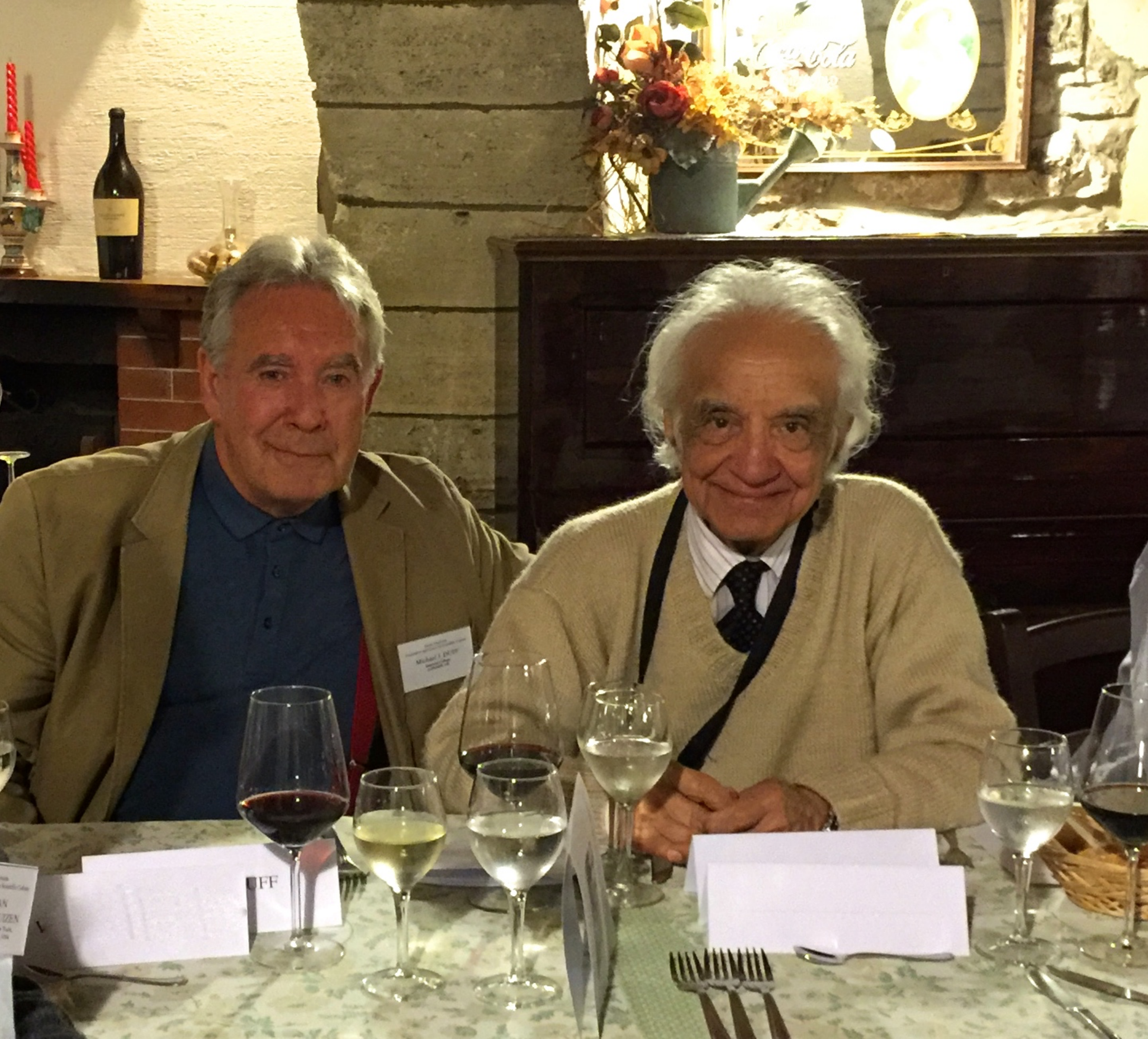}
\caption{Nino and the author}
\label{nino}
\end{figure}

See Figure \ref{nino}.
\subsection{Subsequent developments}
This paper.
\section{Acknowledgements}

First and foremost I am indebted to Nino Zichichi for his support and encouragement of brane research over these thirty years. I am also grateful to my colleagues in the Theory Group at Imperial College for many useful conversations, to Philip Candelas  for his hospitality in the Mathematical Institute, Oxford University and to Marlan Scully for his hospitality in the Institute for Quantum Science and Engineering, Texas A\&M University. Special thanks to Leron Borsten and Jianxin Lu for correcting errors and useful suggestions.  I acknowledge the Leverhulme Trust for an Emeritus Fellowship and the Hagler Institute for Advanced Study at Texas A\&M for a Faculty Fellowship. The work  is supported in part by the STFC under rolling grant ST/P000762/1. %%%%%%%%%%%%%%%%%%%%%%%%%%%%%%%%%%%%%%%%%%%%%%%%%%%%%%%%%%%

\end{document}